\documentclass[pra,twocolumn,aps,groupedaddress,showpacs,nofootinbib]{revtex4}
\usepackage{amsmath,amsfonts,amssymb}
\usepackage{wrapfig}
\usepackage{graphicx}
\usepackage{bbm,bm,epsfig}
\usepackage{graphics}
\usepackage{hyperref}

\newcommand{\be}{\begin{eqnarray}}
\newcommand{\ee}{\end{eqnarray}}
\newcommand{\eins}{\mbox{$1 \hspace{-1.0mm}  {\bf l}$}}
\newcommand{\id}{\mbox{$1 \hspace{-1.0mm}  {\bf l}$}}

\newcommand{\sz}{\sigma_z}
\newcommand{\C}{{\mathbb{C}}}

\begin{document}



\title{Quantum simulation of interacting high--dimensional systems: the influence of noise}

\author{Wolfgang D\"ur$^{1,2}$, Michael J. Bremner$^{1,2}$ and Hans J. Briegel$^{1,2}$}

\affiliation{$^1$ Institut f{\"u}r Theoretische Physik,
Universit{\"a}t Innsbruck,
Technikerstra{\ss}e 25, A-6020 Innsbruck, Austria\\
$^2$ Institut f\"ur Quantenoptik und Quanteninformation der
\"Osterreichischen Akademie der Wissenschaften, Innsbruck,
Austria.}

\date{\today}

\begin{abstract}

We consider the simulation of interacting high-dimensional systems
using pairwise interacting qubits. The main tool in this context
is the generation of effective many-body interactions, and we
examine a number of different protocols for obtaining them. These
methods include the usage of higher-order processes (commutator
method), unitary conjugation or graph state encoding, as well as
teleportation based approaches. We illustrate and compare these
methods in detail and analyze the time cost for simulation. In the
second part of the article, we investigate the influence of noise
on the simulation process. We concentrate on errors in the
interaction Hamiltonians and consider two generic noise models,
(i) timing errors in pairwise interactions and (ii) noisy pairwise
interactions described by Master equations of Lindblad form. We
analyze and compare the effect of noise for the different
simulation methods and propose a way to significantly reduce the
influence of noise by making use of entanglement purification
together with a teleportation based protocol.
\end{abstract}



\maketitle

\tableofcontents


\section{Introduction}

Quantum computers are an example of a wide class of
quantum-control systems that are capable of simulating the
Hamiltonian dynamics of any finite-dimensional system. In the
field of quantum information theory such systems are called universal
quantum simulators, a name first coined by Feynman a quarter of a
century ago to describe a class of physical systems whose dynamics
can be manipulated in order to mimic the dynamics of any other
system \cite{Feynman82a}. Feynman suggested that such systems
might be able to overcome the seemingly insurmountable problem of
efficiently simulating quantum mechanical systems. Meanwhile it has been shown that
universal quantum computers, and all systems that can
efficiently simulate quantum computers, are capable of efficiently
simulating the dynamics of all finite-dimensional $k$-local
Hamiltonians \cite{Lloyd97} and all sparse Hamiltonians
\cite{Childs04, Berry05}. These classes of Hamiltonians include
the set of all local spin systems and all Hamiltonians that can be
efficiently mapped to such systems.

More specifically, we say that a quantum-control system is a
universal quantum simulator on a set of $n$ subsystems if, under
ideal conditions, it is capable of generating any unitary
operation on that set of subsystems. There are many physical
systems that can satisfy this condition given the right forms of
control, for instance an ``always-on'' two-body Hamiltonian that
entangles a set of $n$ qubits can be thought of as a universal
quantum simulator if is supplemented by arbitrarily fast
single-qubit unitary control \cite{Dodd02a}. These systems can
also be considered to be universal quantum computers because they
can efficiently simulate the quantum circuit model (see
\cite{Bremner03a, Bremner04a, Wocjan02b, Wocjan02c, Bennett01a,
Dur01a, Haselgrove03a, Vidal01b, Jane03,Bullock04,Murg04} for
other examples involving Hamiltonians manipulated by single-qubit
control). This form of quantum simulation is often referred to as
Hamiltonian simulation, as it involves the manipulation of a fixed
system Hamiltonian. Such universal quantum simulators are
appealing from an experimental perspective because they have the
prospect of utilizing ``global'' operations which can sometimes be
more easily manipulated and created than quantum gates. Other,
more direct approaches are concerned with simulation of specific
interaction Hamiltonians (e.g. certain model Hamiltonians for high
$T_c$ superconductivity) using quantum optical systems such as
neutral atoms stored in an optical lattice
\cite{Jaksch98,Jaksch05,Micheli05}. This form of quantum
simulator, while not universal, are more experimentally feasible
in the short term.

Although many universal quantum simulators are equivalent to a
full scale universal quantum computer, it is expected and hoped
for that even without using complex quantum error correction or
fault tolerant methods, one can simulate specific (relevant)
Hamiltonians with sufficiently high accuracy, and gain in this way
new insight in the corresponding systems. In particular, quantum
simulators operating on a few tens of spins can be expected to be
realized on a much shorter timescale than a fault tolerant
universal quantum computer that needs to operate on hundreds of
thousand spins.

With this in mind, in this article we undertake a study into the
effects of noise on a number of different simulation protocols.
The aims of our investigation are twofold:

\begin{itemize}
\item[(i)] We introduce and investigate different methods to
generate many--body interactions from two--body interactions with
help of local control operations.

\item[(ii)] We study the influence of noise in two--body
interaction and local control operations on the simulation
process.
\end{itemize}

We concentrate on the simulation of many-body interaction
Hamiltonians from two-body interaction Hamiltonians. Despite the
fact that many-body interactions are often neglected in the
theoretical description of quantum systems as they mostly appear
as higher order processes, such terms play a pivotal role in the
theory of quantum simulation. This becomes particularly apparent
when one uses a system of qubits to simulate higher-dimensional
systems. For instance, if one wanted to simulate any eight
dimensional spin system with three qubits then one would have to
be able to simulate three-qubit interaction terms. A more extreme
example of this is shown in \cite{Ortiz01, Verstraete05} where it
is demonstrated that Fermionic systems can be simulated using
interacting spin systems. Such simulations utilize a
generalization of the Jordan-Wigner transformation \cite{Jordan28}
to demonstrate how Fermionic Hamiltonians can be mapped to a spin
Hamiltonian which contains many-body interaction terms. Another
application of many-body interaction Hamiltonians is in quantum
error correcting protocols for adiabatic quantum computing.
Although it is known that two--local Hamiltonians are universal
for adiabatic quantum computing \cite{Kempe04}, such schemes are
not fault--tolerant. For existing error correction protocols it
seems that, in order to correct errors in an adiabatic algorithm
that uses a two-local Hamiltonian, we are required to use
many-body interaction Hamiltonians \cite{Jordan05}.

In this paper we particularly focus on the effects of noise on
protocols for simulating many-body interaction terms that are
tensor products of the Pauli matrices. Such simulations, while
having a simple form, are interesting because they can often be
simulated via protocols that can be highly parallelized and they
can be used to form the building blocks of more sophisticated
protocols. In addition, such Hamiltonians can transform local
noise into highly non-local noise. Our approach in this article is
to examine the effects of noise on the entangling operations used
in these protocols. The noise models that we study are quite
general and can applied to a variety of physical implementations
and, in principle, the analysis that we perform can be extended to
more complex simulations.

More generally, the challenge ahead in the study of quantum
simulators is to more precisely identify where the limitations
lie. This problem has two clear pathways, the first being to
identify the \emph{algorithmic constraints} on quantum simulation.
For instance, it is known that simulating the static properties of
quantum systems seems to be a computationally difficult task
because the problem of identifying whether or not a state of a
two-local Hamiltonian is the ground state of that Hamiltonian is
\texttt{QMA}-complete \cite{Kempe04} (the quantum analogue of
\texttt{NP}-complete). Thus it is thought that this task is not
efficiently solvable by a quantum computer or any quantum
simulator that is computationally equivalent to a quantum
computer. In spite of such results, there are currently few proven
restrictions on the class of systems that \emph{are} efficiently
solvable on a quantum computer or an equivalent quantum simulator.
In particular, much work needs to be done to identify physical
systems that can be efficiently simulated by a universal quantum
simulator and which quantities of these systems can be efficiently
extracted. For example, recent work \cite{Brown06} has highlighted
that there are subtle algorithmic conditions that must be
satisfied in order to ensure that a quantum simulation protocol
will achieve any speed-up over a classical simulation.

The second pathway is to identify the \emph{physical restrictions}
to our capacity to perform quantum simulations, which is the main
focus of this article. If a quantum computer endures too much
noise \cite{Harrow03}, its dynamics can be classically simulated,
the same is true for all quantum simulators though the amount of
noise that is too much for a general quantum simulator is not yet
known. One way of approaching the problem of how much noise can be
tolerated by a quantum simulator is to address the problem of how
to eradicate noise in a simulation. Quantum computers can be made
fault-tolerant through the use of error-correcting codes. The
fault tolerance theorem tells us that that a noisy quantum
computer employing quantum error correction can simulate an ideal
quantum computer without too much additional overhead given that
amount of noise is below a particular threshold and that the
errors that occur are of the right kind (for a summary see e.g.
chapter 10 of \cite{Nielsen00a}). The key message of the threshold
theorem is that it is in principle possible to build quantum
simulation devices that can overcome the effects of noise and that
any fault-tolerant quantum computing architecture is an example of
such a device.

Currently, it is not known how far the theory of fault-tolerant
quantum computing can be extended. The first versions of the
fault-tolerance theorem applied only to architectures that
implement the circuit model of quantum computing, however, the
theorem has since been extended to one-way models of quantum
computing \cite{Rausendorf03, Nielsen04, Aliferis06a, Dawson06a,
Rausendorf06a, Rausendorf07a}. As of yet, there is not a complete
understanding of how to develop a theory of fault-tolerance for
systems that are controlled by adiabatic evolutions and for
general time-varying Hamiltonian evolutions. That said, in
\cite{Aharanov05} the authors demonstrated that the
fault-tolerance theorem can be extended to account for systems
enduring certain forms of non-Markovian noise by considering a
model of quantum computation where circuits are simulated by a
``local'' time-varying Hamiltonian, proving that such Hamiltonian
control systems are also fault-tolerant. In addition to this there
have been promising recent developments towards developing a
theory of quantum error correction for adiabatic models of quantum
computing \cite{Jordan05}.

So far we have discussed quantum simulators in the context of
fault-tolerant quantum computers, yet it is far from clear as to
whether quantum simulators have to be fault-tolerant to be useful.
Current fault-tolerance thresholds for quantum computers with
reasonable noise models lie between $10^{-5}$ and $10^{-3}$. While
theoretical advances may raise these thresholds, current belief is
it will not be possible to build any quantum computer that
satisfies a fault-tolerant threshold for many years. As such, it
is worthwhile to ask what can be done with systems that contain
too much noise to be made fault-tolerant and yet by many measures
are still highly quantum \cite{Virmani05}? Such systems can
simulate classical computations efficiently as well as a limited
class of quantum systems that one may not be able to classically
simulate. With the present paper, where we investigate the
influence of noise on the simulation of many-body interactions, we
aim to shed some light on the possibility of using quantum
simulators in such an intermediate regime.

The paper is organized as follows. In part I, we investigate
several methods to generate many--body interactions. In Sec.
\ref{commutator} we describe the standard commutator method that
makes use of the Lie-Trotter product expansion. In Sec. \ref{GSE}
we introduce a method that is based on unitary conjugation or
graph state encoding, while we consider teleportation based
methods in Sec. \ref{TBM}. Part II of the article is concerned
with a detailed study of the influence of noise in interactions on
the simulation process. In Sec. \ref{DMF} we discuss distance
measures and fidelity of noisy processes. In Sec.
\ref{Noisemodel1} we describe our first noise model, where noisy
interactions are described by master equations of Lindblad form.
We consider a second noise model taking random fluctuations in
interaction time into account in Sec. \ref{Noisemodel2}. In
Sections \ref{commutatorNOISE}, \ref{GSENOISE} and \ref{TBMNOISE}
we investigate in detail the influence of noise on the simulation
process for commutator method, graph state encoding and
teleportation based method respectively. In Sec. \ref{TBMNOISE} we
also discuss the usage of entanglement purification to
significantly reduce the influence of noise. The results for the
different methods are compared in Sec. \ref{compare}, and we
summarize in Sec. \ref{conclusions}.

\section*{\Large PART I: Methods to generate many--body interaction Hamiltonians}

We will consider $m$ systems, each of dimension $d$, with
associated Hilbert space ${\cal H}=(\C^d)^{\otimes m}$. In the
case of $d=2$, i.e. qubits, we make use of the Pauli-matrices
which we denote by $\sigma_0\equiv \eins$, $\sigma_1\equiv
\sigma_x$,$\sigma_2\equiv \sigma_y$,$\sigma_3\equiv \sigma_z$.
When it is clear from the context, we will often omit tensor
products, i.e. we identify $\sigma_i^{(A)} \otimes \sigma_j^{(B)}
\equiv \sigma_i^{(A)}\sigma_j^{(B)}$. We will also set $\hbar=1$ in the following.

We assume that the $d$--level systems interact pairwise, and we
will be interested in methods to generate effective many--body
interactions involving up to $n \leq m$ of these systems. For
simplicity, we will mainly consider $d=2$, i.e. qubits. The
simulation of arbitrary many--body Hamiltonians of $m$ qubits (or,
equivalently, Hamiltonians of higher dimensional systems with
$D=2^m$) can be achieved if one is capable of generating
\begin{itemize}
\item[(i)] a specific many--body interaction Hamiltonian $H=\sigma_z^{\otimes
n}$ for all $n\leq m$;
\item[(ii)] fast local unitary control of the
individual qubits.
\end{itemize}
Hence we will concentrate on the following in methods to generate
a basic $m$--body interaction for some fixed $m$.

Given (i) and (ii) are fulfilled, standard techniques from
Hamiltonian simulation can be applied, where intermediate fast
local unitary operations are used to manipulate the basic
Hamiltonian $H$ and generate an arbitrary desired effective
Hamiltonian \be H'=\sum_k \lambda_k H_k, \label{Hprime} \ee where
$H_k$ are $m$--body interaction Hamiltonians consisting of Pauli
matrices. Here, one makes use of the facts that any $m$--body
Hamiltonian can be represented in the Pauli basis, and
Hamiltonians $H_k$ consisting of Pauli--matrices can effectively
generated from $H$ via unitary conjugation, i.e. \be U_k e^{-i t
H} U_k^\dagger = e^{-i t U_kHU_k^\dagger}, \ee where local unitary
operations $U_k$ suffice. Using the identity \be \lim_{M\to
\infty} \left(\prod_k e^{-i H_k t/M}\right)^M = e^{-i t \sum_k
H_k}, \ee we find for sufficiently short times $t=\delta t$, \be
\prod_k e^{-i\delta t \lambda_k H_k} &=& \eins -i \delta t \sum_k \lambda_k H_k + O(\delta t^2) \nonumber\\
&\approx& e^{-i\delta t \sum_k \lambda_k H_k}, \ee that is,
sequences of applications of the standard Hamiltonian $H$,
together with intermediate local unitary operations $U_k^\dagger
U_{k+1}$, generate (up to higher order corrections in $\delta t$)
a unitary operation that is generated by an effective Hamiltonian
$H'$ (Eq. \ref{Hprime}). Note that more complicated sequences
allow for the simulation of the effective Hamiltonian $H'$ with
higher accuracy, with corrections appearing only in higher order
$\delta t$.

\section{Commutator method}\label{commutator}

\subsection{Three--body interactions}

A somewhat standard approach to simulating a three body
interaction from a given two--body interaction is to apply a
sequence of time evolutions, generated by different Hamiltonians,
each for a short time $\delta t$. The sequence is chosen in such a
way that all first order terms in $\delta t$ --when performing a
Taylor expansion-- cancel, and only higher order terms in $\delta
t$ remain. These higher order terms include products of different
two--body Hamiltonians, and can hence correspond to effective
many--body interactions. To be more precise, consider the
following sequence of time evolutions, generated by the
Hamiltonians $H_k$ applied for time $\delta t$\be U_{\rm
tot}=e^{iH_4\delta t}e^{iH_3\delta t}e^{iH_2\delta t}e^{iH_1\delta
t}. \label{manybody} \ee For small $\delta t$, one can Taylor
expand this expression and obtains \be
U_{\rm tot}=&& \eins + i\delta t \sum_{j=1}^{4}H_j \nonumber \\
&-& \frac{\delta t^2}{2} \left(\sum_{j=1}^4H_j^2 + 2\sum_{l=2}^4 \sum_{j=1}^{l-1} H_lH_j \right) \nonumber \\
&+& O(\delta t^3) \ee Considering the case \be H_4=-H_2,
\hspace{1cm} H_3=-H_1, \ee one obtains \be
U_{\rm tot} &=& \eins + \delta t^2(H_1H_2-H_2H_1) + O(\delta_t^3) \nonumber \\
&=& e^{i 2\delta t^2 \times (-i/2)[H_1,H_2]}+O(\delta t^3). \ee
The sequence of interactions described by some properly chosen
two--body Hamiltonians hence correspond --up to higher order
corrections-- to an evolution which is described by an Hamiltonian
that is essentially given by \be H_{\rm eff}=-i/2[H_1,H_2], \ee
which can be an effective three--body interaction.

{\bf Example:} For a system of three qubits and \be
\label{Hamiltonians12}
H_1=\sigma_z^{(A)}\sigma_x^{(B)},\nonumber\\
H_2=\sigma_y^{(B)}\sigma_z^{(C)}, \ee we have that \be
e^{[H_1,H_2] \delta t^2}+ O(\delta t^3) &\approx& \eins^{(ABC)}+i2
\delta t^2
\sigma_z^{(A)}\sigma_z^{(B)}\sigma_z^{(C)} \nonumber \\ & & + O(\delta t^3)\nonumber \\
&\approx& e^{i \delta t' H_{\rm eff}} + O(\delta t'^{3/2}). \ee
That is, if we evolve a system as indicated in Eq.
(\ref{manybody}) using only two--body interactions with
$H_4=-H_2,H_3=-H_1$ for a total time of $4 \delta t$, then the
resulting evolutions is ---up to higher order corrections in
$\delta t$--- the same as the one resulting from a three--body
Hamiltonian \be H_{\rm
eff}=\sigma_z^{(A)}\sigma_z^{(B)}\sigma_z^{(C)}, \ee applied for a
time \be \delta t'=2 \delta t^2. \ee Note that there is a dilation
factor of $2/\delta t$, i.e. the required physical time $t$ to
implement an effective three--body interaction for a time $t'$ is
given by $t=4 \sqrt{\delta t'/2}$, since the effective three body
interaction only appears in second order in $\delta t$. We have
assumed that $H_1,H_2$ as well as $-H_1,-H_2$ can be implemented.
The simulation of $H_{\rm eff}$ is only correct in first order in
$\delta t'$, and unwanted terms appear already in order $\delta
t'^{3/2}$. This corresponds to a reduced accuracy as compared to
standard Hamiltonian simulation schemes for two--body
interactions, where the desired Hamiltonian is correctly produced
up to first order, and undesired terms (errors) appear only in
second order. We will refer to this kind of errors as Taylor
expansion errors, which are independent from errors in
interactions and local control operations which will discussed in
detail in Sec. \ref{Errors}. The Taylor expansion errors need to
be taken into account when using the effective $m$--body
Hamiltonian to simulate other Hamiltonians via Hamiltonian
simulation techniques.

\subsection{Many--body interactions}\label{Manybodyinteractions}

In principle, many--body interactions Hamiltonians for arbitrary
number of qubits $m$ can be generated in a recursive way using
above method. For instance, one of the two Hamiltonians, say
$H_1$, is replaced by an effective $m-1$ body Hamiltonian.
Together with an appropriate two--body Hamiltonian a new effective
$m$--body Hamiltonian can be produced. Note, however, that there
is a dilation factor of $\delta t/2$ in each of these simulation
processes. That is, the implementation of a $m$--body Hamiltonian
for time $t'$ requires a physical time $t = O(t'^{2^{-(m-1)}})$,
if basic two--body Hamiltonians are used.

The time cost can be reduced when using an alternative method,
where two $n$--body interactions are used to generate a
$m=(2n-1)$--body interaction. For instance, the generation of a
$5$--body interaction has a time cost of $O(\delta t^{1/4})$, as
compared to $O(\delta t^{1/8})$ when using the first method. The
desired $m$--body Hamiltonian appears in $O(\delta t^{m-1})$ and
corrections appear in $O(\delta t^{m})$, where $m=5$ in our
example. Rewriting this in the new effective time $\delta t_m$,
i.e. one realizes a $m$--body Hamiltonian for time $\delta t_m$,
one finds \be \delta t_m = O(\delta t^{m-1}), \ee and corrections
appear in \be O(\delta t_m^{m/(m-1)}). \ee That is, \be
U_{\rm tot} &=& \eins + i O(\delta t^{m-1}) H_{\rm eff} + O(\delta t^m) \nonumber\\
&=& \eins + i \delta t_m H_{\rm eff} + O(\delta t_m^{m/(m-1)}) \nonumber\\
&\approx& e^{i\delta t_m H_{\rm eff}}. \ee To ensure that the
Taylor expansion is a good approximation at all instances of the
protocol, all involved times, in particular $\delta t=O(\delta
t_m^{1/(m-1)})$ (which corresponds to the physical time for which
two--body interactions are applied), need to be sufficiently
small. This limits the possible values of $\delta t_m$, which is
important when considering the simulation of general $m$--body
Hamiltonians from the basic one. In particular, the total time
required to simulate a Hamiltonian for time $t_{\rm tot}$ with
Hamiltonian simulation techniques (i.e. generating an effective
$m$--body Hamiltonian for time $\delta t_m$, and using this
Hamiltonian together with intermediate local unitary operations to
simulate other Hamiltonians for larger times by repeating this
process $t_{\rm tot}/\delta t_m$ times) requires a total time of
\be t_{\rm tot}/\delta t_m \times O(\delta t_m^{1/(m-1)}) = t_{\rm
tot} O(\delta t_m^{-(m-2)/(m-1)}). \ee The dilation factor (or
time cost) $O(\delta t_m^{-(m-2)/(m-1)})$ approaches $O(\delta
t_m^{-1})$ for large $m$ and can be significant. Recall that
$\delta t_m^{1/(m-1)} \ll 1$ needs to be fulfilled, which implies
that the dilation factor, for large $m$, will typically be of
order $10^{m}$ or larger.

{\bf Example:} To make above considerations more concrete,
consider the generation of an effective $5$--body interaction
Hamiltonian for time $\delta t''$, \be U_{\rm tot}(\delta t'')
\approx e^{-i\delta t''H_{\rm eff}}, \ee where we use the notation
$\delta t''\equiv \delta t_5, \delta t'\equiv \delta t_3$. This
could take place as follows: (i) use two--body interactions
$U_{AB}(\pm \delta t),U_{BC}(\pm \delta t)$ generated by two--body
Hamiltonians $\pm H_{AB},\pm H_{BC}$ between systems $AB$, $BC$
for time $\delta t=\sqrt{\delta t'/2}$ to produce an effective
three--body interaction $U_{ABC}(\delta t')$ generated by the
effective three--body Hamiltonian $H_{ABC} = -i/2[H_{BC},H_{AB}]$.
That is, \be
U_{ABC}(\delta t') &=& U_{AB}(-\delta t)U_{BC}(-\delta t)U_{AB}(\delta t)U_{BC}(\delta t) \nonumber\\
&\approx& e^{i \delta t' H_{ABC}}. \ee Use the same method to
produce a three--body interaction $U_{CDE}(\delta t')$ generated
by the effective three--body Hamiltonian $H_{CDE} =
[H_{DE},H_{CD}]$. (ii) Use these three--body interactions
$U_{ABC}(\delta t')$, $U_{CDE}(\delta t')$ to produce an effective
$5$--body interaction $U_{\rm tot}(\delta t'')$ generated by the
effective $5$--body Hamiltonian $H_{\rm eff}= -i/2[H_{CDE},H_{ABC}]$
for time \be \delta t'' = \sqrt{\delta t'/2} = (\delta
t/8)^{\frac{1}{4}}. \ee That is, \be
U_{\rm tot}(\delta t'') &=& U_{ABC}(-\delta t')U_{CDE}(-\delta t')U_{ABC}(\delta t')U_{CDE}(\delta t) \nonumber\\
&\approx& e^{i \delta t'' H_{\rm eff}}. \ee The choice \be
H_{AB}=\sigma_z^{(A)}\sigma_x^{(B)}, & &  H_{BC}=\sigma_y^{(B)}\sigma_y^{(C)}, \nonumber\\
H_{CD}=\sigma_x^{(C)}\sigma_y^{(D)}, & &
H_{DE}=\sigma_x^{(D)}\sigma_z^{(E)}, \ee leads to an effective
$5$--body Hamiltonian \be H_{\rm eff}=\sigma_z^{\otimes 5}. \ee

\section{Unitary conjugation and graph state encoding}\label{GSE}

A second method to generate effective many--body interaction
Hamiltonians from basic two--body interaction Hamiltonians is by
unitary conjugation. That is, before [after] the evolution with
respect to a single-- or two--body Hamiltonian $H$, a (possibly
non--local) unitary operation $U$ [$U^\dagger$] is applied. The
resulting evolution is given by \be U e^{-i t H} U^\dagger = e^{-i
t UHU^\dagger}, \ee i.e. by the transformed Hamiltonian
$H'=UHU^\dagger$. If $U$ is itself a non--local unitary operation
(e.g. generated by two--body interactions), then the resulting
effective Hamiltonian $H'$ can contain many--body terms. The
unitary conjugation can be viewed as a change of basis, where
single-- and two--body terms of an interaction Hamiltonian act
effectively as many--body terms in the new basis. In the
following, we will concentrate a specific family of basis changes,
associated with graph states. The corresponding unitary operations
$U$ can be efficiently implemented using basic two--body
interactions, and allow for the systematic construction of the
desired many--body Hamiltonians via a proper choice of the graph.
This gives a powerful tool for the construction of many--body
interaction Hamiltonians.

Consider as a first example a system of two qubits, and a unitary
operation $U=U_{PG}$,
\be
U_{PG}=diag(1,1,1,-1),
\ee
i.e. a phase gate. A phase gate can e.g. be induced by a simple
two--body Hamiltonian \be H=(\id-\sigma_z)\otimes(\id -\sigma_z),
\ee applied for time $t=\pi/4$, which is locally equivalent to an
Ising Hamiltonian $H=\sigma_z\otimes\sigma_z$. With this choice of
$U$, operators of the form $\sigma_i\otimes \id$ are transformed
to $U_{PG}(\sigma_i\otimes \id) U_{PG}^\dagger = \sigma_i \otimes
\sigma_{f(i)}$, with $f(i) = 0$ if $i=0,3$ and $f(i)= 3$ if
$i=1,2$. That is, a single--body Hamiltonian acting on the first
qubit,
\be H=\sum \lambda_k \sigma_k\otimes \id, \ee is transformed into
a two--body Hamiltonian \be
H'&=&U_{PG} H U_{PG}^\dagger  \\
&=& \lambda_0 \sigma_0\otimes \sigma_0 + \lambda_1 \sigma_1\otimes
\sigma_3 + \lambda_2 \sigma_2\otimes \sigma_3 + \lambda_3
\sigma_3\otimes \sigma_0.\nonumber \ee We remark that the
phase--gate corresponds to the unitary transformation from a
product basis $|k_1k_2\rangle$, $k_i=0,1$ in the $x$--basis (i.e.
$k_i=0$ corresponds to $|0\rangle_x =
\frac{1}{\sqrt{2}}(|0\rangle_z + |1\rangle_z)\equiv
\frac{1}{\sqrt{2}}(|0\rangle +|1\rangle)$, and $|1\rangle_x =
\frac{1}{\sqrt{2}}(|0\rangle - |1\rangle)$) into a Graph--state
basis $|G,k_1k_2\rangle$, where $G$ is the graph with vertices
$1,2$ and a single edge, $E=\{1,2\}$. That is, \be
|G,k_1k_2\rangle=U_{PG} |k_1k_2\rangle.
\ee

\subsection{Graph--state encoding}

More generally, for systems of $N$ qubits one can define an
encoding into a graph-state basis corresponding to an arbitrary
graph $G$. In this case, the graph $G=(V,E)$, is a set $V$ of $N$
vertices connected by edges E, corresponds to an interaction
pattern (specified by $E$) between the qubits (that are associated
with vertices). That is, \be |G,{\bm
k}\rangle=\prod_{\{\alpha,\beta\}\in E} U_{PG}^{(\alpha,\beta)}
|{\bm k}\rangle, \ee where ${\bm k}$ is a binary vector of length
$N$. We define the neighborhood $N_\alpha$ of a given vertex
$\alpha$ as the set of all vertices connected to $\alpha$ in the
graph, $N_\alpha=\{\beta| \{\alpha,\beta\} \in E\}$. We use the
notation
\be \sigma_i^{N_\alpha} = \otimes_{\alpha \in N_\alpha}
\sigma_i^{(\alpha)}, \ee to refer to an operator acting on all
qubits in the neighborhood of $\alpha$. Using a such graph state
encoding, the corresponding unitary operation is given by \be
U=\prod_{\{\alpha,\beta\} \in E} U_{PG}^{(\alpha,\beta)}. \ee Note
that the phase gates commute and can hence be implemented in
parallel. It is now straightforward to determine the effect of
unitary conjugation with such a $U$. Pauli operators acting on
qubit $\alpha$ are transferred as follows \cite{Hein06}\be
\label{trafoH_Graph}
U\sigma_0^{(\alpha)}U^\dagger &=& \sigma_0^{(\alpha)}, \nonumber \\
U\sigma_1^{(\alpha)}U^\dagger &=& \sigma_1^{(\alpha)}\sigma_3^{N_\alpha}, \nonumber \\
U\sigma_2^{(\alpha)}U^\dagger &=& \sigma_2^{(\alpha)}\sigma_3^{N_\alpha}, \nonumber \\
U\sigma_3^{(\alpha)}U^\dagger &=& \sigma_3^{(\alpha)}. \ee We have
that $\sigma_1^{(\alpha)}$ and $\sigma_2^{(\alpha)}$ are
transformed into effective $m$--body interaction terms, where $m$
is given by the local degree of the graph (i.e. the number of
neighbors of $\alpha$) plus one. The action of
$\sigma_3^{(\alpha)}$ remains local. The transformation of
two--body terms follows immediately from Eq. (\ref{trafoH_Graph}),
i.e. \be U \sigma_i^{(\alpha)}\otimes \sigma_j^{(\beta)} U^\dagger
= (U \sigma_i^{(\alpha)} U^\dagger)  (U \sigma_j^{(\beta)}
U^\dagger). \ee Note that $(U \sigma_i^{(\alpha)} U^\dagger)$ and
$(U \sigma_j^{(\beta)} U^\dagger)$ can contain terms that act on
the same qubit which need to be multiplied. Depending on the
neighborhood relation of qubits $\alpha$ and $\beta$, and on the
kind of Pauli operators, the resulting total operator can have
support on up to $|N_\alpha|+|N_\beta|+2$ (i.e. acts
non--trivially on this number of qubits), but may also be the
identity. For instance, \be
U \sigma_1^{(\alpha)}\otimes \sigma_1^{(\beta)} U^\dagger &=& \sigma_1^{(\alpha)}\sigma_1^{(\beta)}\sigma_3^{N_\alpha}\sigma_3^{N_\beta}\nonumber \\
&=&\sigma_1^{(\alpha)}\sigma_1^{(\beta)}\sigma_3^{N_\alpha \oplus
N_\beta}, \ee where $\oplus$ denotes the XOR (exclusive or
operation), i.e. $N_\alpha \oplus N_\beta \equiv (N_\alpha\cup
N_\beta) \setminus (N_\alpha\cap N_\beta)$.

To illustrate the transformation rules, we consider a simple
example of three qubits arranged as an open chain, i.e. with edges
$E=\{\{1,2\},\{2,3\}\}$ and corresponding unitary
$U=U_{PG}^{(1,2)}U_{PG}^{(2,3)}$. The two-qubit operator
$\sigma_1^{(1)}\sigma_3^{(2)}$ transforms to a single qubit
operator $\sigma_x^{(1)}$, as $U\sigma_1^{(1)}U^\dagger =
\sigma_1^{(1)}\sigma_3^{(2)}$ and $U\sigma_3^{(2)}U^\dagger =
\sigma_3^{(2)}$. Similarly, a two qubit operator
$\sigma_1^{(1)}\sigma_1^{(3)}$ is transformed to a two--qubit
operator $\sigma_1^{(1)}\sigma_1^{(3)}$, while
$\sigma_1^{(1)}\sigma_3^{(3)}$ transforms to a three--body
operator $\sigma_1^{(1)}\sigma_3^{(2)}\sigma_3^{(3)}$. In this
case, a three--body operator can also be obtained from a single
qubit operator, e.g. $\sigma_1^{(2)}$ transforms into
$\sigma_3^{(1)}\sigma_1^{(2)}\sigma_3^{(3)}$.

\subsection{Example 1: Three--body interaction Hamiltonian exhibiting a quantum phase transition}

We consider now the generation of a three--body interaction
Hamiltonian with transversal magnetic field, where we assume a
linear chain of qubits with periodic boundary conditions. We have
\be \label{HGSE} H'=\sum_\alpha
(-\sigma_3^{(\alpha-1)}\sigma_1^{(\alpha)}\sigma_3^{(\alpha+1)} +
B \sigma_1^{(\alpha)}). \ee This Hamiltonian has been considered
in Ref. \cite{Pachos04a,Pachos04b,Kay05} in
the context of optical lattices in a triangular configuration,
where three--body processes may lead in certain parameter regimes
to an interaction of this form. This Hamiltonian exhibits a
quantum phase transition. More importantly, it has a finite energy
gap above its unique ground state, a finite classical correlation
length but despite of this an diverging entanglement length \cite{Po04}. We
remark that the ground state of this system for $B=0$ is a one--dimensional
cluster state, i.e. a graph state corresponding to a closed linear
chain.

We now show how to generate the Hamiltonian $H'$ using only
two--body interaction and a simple graph state encoding. To this
aim, we consider a graph with edges $(2\alpha,2\alpha+1)$. That
is, every second qubit is connected with its right neighbor.
We use this
graph for our graph state encoding, i.e. define the unitary
operation $U$ as \be \label{UGSE} U=\prod_\alpha
U_{PG}^{(2\alpha,2\alpha+1)}. \ee It is now straightforward to
check the following transformation rules \be
U\sigma_3^{(2\alpha-1)}\sigma_1^{(2\alpha)}U^\dagger &=& \sigma_3^{(2\alpha-1)}\sigma_1^{(2\alpha)}\sigma_3^{(2\alpha+1)}, \nonumber\\
U\sigma_1^{(2\alpha-1)}\sigma_3^{(2\alpha)}U^\dagger &=& \sigma_3^{(2\alpha-2)}\sigma_1^{(2\alpha-1)}\sigma_3^{(2\alpha)},\nonumber\\
U\sigma_3^{(2\alpha)}\sigma_1^{(2\alpha+1)}U^\dagger &=& \sigma_1^{(2\alpha)},\nonumber\\
U\sigma_1^{(2\alpha)}\sigma_3^{(2\alpha+1)}U^\dagger &=&
\sigma_1^{(2\alpha+1)}. \ee That is, the two--body Hamiltonian \be
H_1= \sum_\alpha
-(\sigma_3^{(2\alpha-1)}\sigma_1^{(2\alpha)}+\sigma_1^{(2\alpha-1)}\sigma_3^{(2\alpha)})
\ee transforms into \be H_1'=\sum_\alpha
-\sigma_3^{(\alpha-1)}\sigma_1^{(\alpha)}\sigma_3^{(\alpha+1)},
\ee and gives the desired three--body interaction terms, while \be
H_2= B \sum_\alpha
(\sigma_3^{(2\alpha)}\sigma_1^{(2\alpha+1)}+\sigma_1^{(2\alpha)}\sigma_3^{(2\alpha+1)})
\ee transforms to \be H_2'=B \sum_\alpha \sigma_1^{(\alpha)} \ee
and provides the transversal magnetic field in $X$--direction. In
total, the two--body Hamiltonian $H=H_1+H_2$ is transformed via
unitary conjugation with $U$ (\ref{UGSE}) into the desired
Hamiltonian $H'$ (\ref{HGSE}). That is, for all times $t$, $U
e^{-it (H_1+H_2)}U^\dagger = e^{-it H'}$. We remark that one can
in addition achieve a magnetic field in $Z$--direction via a
single--body Hamiltonian $\sum_\alpha \sigma_3^{(\alpha)}$ which
is not changed by unitary conjugation.

We emphasize that the method proposed here to generate the
three--body Hamiltonian may be significantly simpler and easier to
implement in optical lattices than the original proposal of Ref.
\cite{Pachos04a,Pachos04b,Kay05}, as it does not rely on
higher--order processes. We require, however, individual
addressability of the individual qubits in some form. The
encoding-- and decoding pattern (i.e. the operation $U$) can
directly be generated from the available two--body interaction
with is essentially given by a pairwise Ising Hamiltonian
$\sigma_z\otimes \sigma_z$ resulting e.g. from controlled cold
collisions. The same interaction, together with fast local unitary
control operations, can also be used to generate the Hamiltonian
$H=H_1+H_2$ using Hamiltonian simulation techniques. That is, the
simulation of the evolution with respect to the three-body
Hamiltonian $H'$ for time $t_{\rm tot}$ takes place as follows:
\begin{enumerate}
\item Use the basic two--body interactions to generate $U$; \item
Simulate the evolution with respect to the Hamiltonian $H=H_1+H_2$
for time $\delta t$ by using the basic two--body interaction
together with fast local unitary operations. Here, standard
Hamiltonian simulation techniques are applied, and the desired
process is approximated up to corrections $O(\delta t^2)$. \item
Repeat the process $t_{\rm tot}/\delta t$ times. For $\delta t$
sufficiently small the total evolution will be a high--fidelity
approximation of the evolution generated by $H$ applied for time
$t_{\rm tot}$. \item Use the basic two--body interactions to
generate $U^\dagger =U$.
\end{enumerate}
The total evolution after steps 1-4 corresponds to $e^{-it_{\rm
tot}H'}$ as desired.

\subsection{Example 2: Interacting $d$--dimensional systems}

We now consider the simulation of Hamiltonians corresponding to
(interacting) $d$--dimensional systems. To illustrate this
approach, we first consider a single $d$--dimensional system with
$d=2^4=16$. In contrast to the previous example, we will make use
of {\em auxiliary systems} here. That is, we consider a system of
five qubits $1,2,3,4,A$ and the associated graph $G$ with edges
$\{k,A\}$, $k=1,2,3,4$, i.e. each of the qubits 1,2,3,4 (which
represent the 16--dimensional system) is connected to qubit $A$.
Here, qubit $A$ serves as auxiliary system and is prepared in an
eigenstate of $\sigma_x$ with eigenvalue +1, i.e. in the state
$|0\rangle_x=(|0\rangle+|1\rangle)_A/\sqrt{2}$. This ensures that
any effective Hamiltonian that acts on qubit $A$ as either the
identity $\eins^{(A)}$ or $\sigma_x^{(A)}$, leaves this qubit
invariant, i.e. qubit $A$ is decoupled from the evolution in this
case. However, Hamiltonians involving qubit $A$ can be used to
manipulate and trigger the effective Hamiltonian acting on system
qubits $1$ to $4$. The unitary operation $U$ corresponding to the
graph state encoding is given by \be U=\prod_{k=1}^{4}
U_{PG}^{(k,A)}. \ee This leads to the following transformation
rules of single-- and two--body operators under unitary
conjugation, where $k,k_i \in \{1,2,3,4\}$ \be
U\sigma_1^{(\alpha)}\sigma_3^{(A)}U^\dagger &=& \sigma_1^{(\alpha)},\nonumber\\
U\sigma_2^{(\alpha)}\sigma_3^{(A)}U^\dagger &=& \sigma_2^{(\alpha)},\nonumber\\
U\sigma_3^{(\alpha)}U^\dagger &=& \sigma_3^{(\alpha)},\nonumber\\
U\sigma_1^{(\alpha_1)}\sigma_1^{(\alpha_2)}U^\dagger &=& \sigma_1^{(\alpha_1)}\sigma_1^{(\alpha_2)},\nonumber\\
U\sigma_1^{(\alpha_1)}\sigma_2^{(\alpha_2)}U^\dagger &=& \sigma_1^{(\alpha_1)}\sigma_2^{(\alpha_2)},\nonumber\\
U\sigma_2^{(\alpha_1)}\sigma_2^{(\alpha_2)}U^\dagger &=& \sigma_2^{(\alpha_1)}\sigma_2^{(\alpha_2)},\nonumber\\
U\sigma_3^{(\alpha_1)}\sigma_3^{(\alpha_2)}U^\dagger &=& \sigma_3^{(\alpha_1)}\sigma_3^{(\alpha_2)},\nonumber\\
U\sigma_3^{(\alpha_1)}\sigma_3^{(A)}U^\dagger &=& \sigma_3^{(\alpha_2)}\sigma_3^{(\alpha_3)}\sigma_3^{(\alpha_4)}\sigma_1^{(A)},\nonumber\\
U\sigma_1^{(A)}U^\dagger &=&
\sigma_3^{(1)}\sigma_3^{(2)}\sigma_3^{(3)}\sigma_3^{(4)}\sigma_1^{(A)}.
\ee Note that the list contains only operators that leave the
ancilla qubit $A$ unaltered. As can be seen, all single qubit
terms, as well as specific two, three and four qubit interaction
terms on the system qubits can be obtained. This allows one, in
principle, to use Hamiltonian simulation techniques to generate
any interaction Hamiltonian acting on the four system qubits.
Note, however, that the generation of effective single qubit
unitary operations (lines 1-3) requires in part two--body
interactions. As a Hamiltonian simulation scheme operating on
effective Hamiltonians may require such effective intermediate
single--qubit unitaries applied for some time $t=O(1)$, the time
cost to simulate certain many--body interaction Hamiltonians for
time $\delta t$ will be of order $t$, i.e. a dilation factor of
$t/\delta t$.

In a similar way, two $d$--level systems of this kind can be
simulated and coupled pairwise. Consider in addition to qubits
$1-4,A$ corresponding to a first $d=16$--dimensional system a
second set of five qubits $1'-4',A'$ corresponding to a second
$d=16$--dimensional system. We consider for system $1'-4',A'$ a
similar graph state encoding, i.e. the the total graph has
additional edges $(k',A')$, $k'=1',2',3',4'$ and $U$ changes
accordingly. Again, qubit $A'$ serves as auxiliary system and is
prepared in state $|0\rangle_x$, and manipulation of the second
$d$--level system works in exactly the same way as for system 1
(qubits 1-4). In addition, coupling between the two $d$--level
systems can e.g. be achieved via the following two--body
interactions \be
U\sigma_3^{(\alpha_1)}\sigma_3^{(\alpha_2')}U^\dagger &=& \sigma_3^{(\alpha_1)}\sigma_3^{(\alpha_2')},\nonumber\\
U\sigma_3^{(\alpha_1)}\sigma_1^{(A')}U^\dagger &=& \sigma_3^{(\alpha_1)}(\otimes_{\alpha'=1}^{4}\sigma_3^{(\alpha')}) \sigma_1^{(A')},\\
U\sigma_1^{(A)}\sigma_3^{(\alpha_2')}U^\dagger &=& (\otimes_{\alpha=1}^{4}\sigma_3^{(\alpha)})\sigma_3^{(\alpha_2')} \sigma_1^{(A)},\nonumber\\
U\sigma_1^{(A)}\sigma_1^{(A')}U^\dagger &=&
(\otimes_{\alpha=1}^{4}\sigma_3^{(\alpha)})(\otimes_{\alpha'=1'}^{4'}\sigma_3^{(\alpha')})
\sigma_1^{(A)}\sigma_1^{(A')}.\nonumber \ee The generalization to
$N$ $d$--dimensional systems that interact pairwise is
straightforward.

\subsection{Example 3: 2D setup with 4--body plaquette interaction}

We consider now the simulation of a Hamiltonian that contains
four--body terms and corresponds to a 2D--setup where qubits are
arranged on a square lattice. In particular, we consider
interactions between qubits on the same plaquette $\square$, i.e.
a Hamiltonian of the form \be \label{plaquette1} H'=\sum_\alpha
\sigma_3^{(\alpha,\alpha)}\sigma_3^{(\alpha,\alpha+1)}\sigma_3^{(\alpha+1,\alpha)}\sigma_3^{(\alpha+1,\alpha+1)}.
\ee Such a Hamiltonian can be generated by considering $N$ qubits
arranged on a rectangular array, and $N$ additional auxiliary
qubits that are placed in the center of each square $\square$ and
prepared in the eigenstate of $\sigma_x$, $|0\rangle_x$. The graph
is such that each of the auxiliary qubits $A_\alpha$ is connected
to all system qubits in the corresponding square, i.e. to qubits
$(\alpha,\alpha),(\alpha,\alpha+1),(\alpha+1,\alpha),(\alpha+1,\alpha+1)$.
Hence, single-qubit operations on auxiliary qubit $A_\alpha$
transforms to a four--qubit operation on the plaquette,
\be
U\sigma_1^{(A_\alpha)}U^\dagger =
\sigma_3^{(\alpha,\alpha)}\sigma_3^{(\alpha,\alpha+1)}\sigma_3^{(\alpha+1,\alpha)}\sigma_3^{(\alpha+1,\alpha+1)}
\sigma_1^{A_\alpha},
\ee
while leaving the auxiliary qubit
unchanged. The simple Hamiltonian $H=B \sum_\alpha
\sigma_1^{(A_\alpha)}$, which may e.g. be generated by applying a
homogenous magnetic field in $X$--direction, hence transforms to
the desired Hamiltonian $H'$ (\ref{plaquette1}). The construction
can be easily generalized to higher dimensions (cubic lattice) and
other geometries (e.g. hexagonal lattices).

\subsection{Time cost and universal quantum simulation}

\subsubsection{Fixed graph state encoding: }
The examples considered so far assume that the encoding (or basis
change) is fixed throughout the whole simulation process. This
implies that the time cost to simulate a many--body Hamiltonian
$H=\sum_k \lambda_k H_k$ for time $t_{\rm tot}$, where each of the
terms $H_k$ can be obtained by a unitary conjugated single-- or
two--body Hamiltonian is given by \be \pi/2 + t_{\rm
tot}\sum_k\lambda_k. \ee Here, we assume that the Hamiltonians
$H_k$ can be obtained directly via unitary conjugation (as in
example 1). That is, one has a constant overhead in simulation
time, as the basis changes that need to be applied at the
beginning and the end of the simulation process each have time
cost $\pi/4$ (phase gates that can be applied in parallel). There
is no additional time overhead to simulate many--body
Hamiltonians, as the basic single-- and two--body Hamiltonians are
directly transformed to the required form. If the method is only
used to generate a basic $m$--body Hamiltonian for time $\delta
t$, then the time cost is essentially $\pi/(2\delta t)$, i.e.
essentially the same as in the commutator method (Sec.
\ref{commutator}). However, here the implementation of the
many--body Hamiltonian is {\em exact}, and the choice of time
$\delta t_m$, the evolution time of the Hamiltonian, is not
limited by requirements on the desired accuracy or the validity of
approximations.

If, as discussed in example 2, the generation of the effective
many--body Hamiltonian also makes use of Hamiltonian simulation
techniques --in particular effective single--body unitary
operations generated by two--body interactions--, one encounters
an additional increase of time cost. Here, however, the graph
state encoding remains fixed throughout the simulation process.

\subsubsection{Variable graph state encoding}
\label{variableGSE} Another possibility is to vary the graph state
encoding. Assume we wish to generate an interaction Hamiltonian
\be H=\sum_k \lambda_k H_k, \ee for time short time $\delta t$,
where $H_k$ are interaction Hamiltonians of different kind, e.g.
various $m$--body terms with different $m$. While a particular
graph state encoding is specially suitable to implement a specific
Hamiltonian $H_k$
---and in fact such a graph state encoding can be chosen in a
constructive way---, it might be complicated (or impossible) to
generate all desired terms using the same encoding. However, one
may change the encoding during the simulation process. That is,
generate $e^{-i\delta t \sum_k \lambda_k H_k}$ via a sequence of
different unitary conjugations and simple single-- and two--body
Hamiltonians $\tilde H_k$, i.e.
\be
\prod_k U_k e^{-i \delta t \lambda_k
\tilde H_k} U_k^\dagger \approx e^{-i \delta t \sum_k \lambda_k U_k
\tilde H_k U_k^\dagger} + O(\delta t^2).
\ee
This is similar to standard Hamiltonian simulation, except that
here the intermediate unitary operations $U_k$ are not local, but
correspond to some graph-state encoding.

Note that in such a set--up, it is even sufficient to restrict to
{\em local} Hamiltonians $\tilde H_k$, which are transformed via
different graph--state encodings (generated by two--body
interaction) to the desired many--body interaction Hamiltonian.
That is, we assume that two--body interactions between any pair of
qubits and local control operations are available. In this case,
the proof of universality is very easy. In fact, it is sufficient
to show that a proper choice of graph state encoding allows one to
generate a $m$--body interaction term \be \label{basic}
\otimes_{\alpha\in S} \sigma_{j_\alpha}^{(\alpha)} \ee on any
subset $S$ of parties, where $j_\alpha \in \{1,2,3\}$. Such a
basic term can be manipulated by means of standard Hamiltonian
simulation techniques (i.e. intermediate local unitaries, which
can be generated from single qubit (i.e. local) Hamiltonians and a
graph state encoding corresponding to the trivial, empty graph,
i.e. $U_k=\eins$) to generate effectively any Hamiltonian acting
on the subset of $S$ of qubits for some (small) time $\delta t$.
As the subset $S$ is arbitrary, the same method allows one to
generate arbitrary Hamiltonians, containing terms acting on
different subsets $S$. If a Hamiltonian consists of $L$ basic
terms of the form (\ref{basic}) with coefficients $\lambda_k$,
then the time cost to simulate such a Hamiltonian is given by
$L\pi/2+\sum\lambda_k \delta t$. A term of the form (\ref{basic})
is most easily generated by choosing a graph state encoding
corresponding a graph with edges $(\alpha,\alpha_j)$, where
$\alpha$ is some fixed qubit $\in S$, and $\alpha_j$ runs over all
remaining qubits in $S$ (i.e. qubit $\alpha$ is connected with all
other qubits in $S$). The corresponding unitary $U$ transforms
single--qubit terms acting on qubit $\alpha$ into an interaction
term involving all qubits in $S$, \be
U\sigma_1^{(\alpha)}U^\dagger = \sigma_1^{(\alpha)}
\otimes_{\alpha_j \in S, \alpha_j\not = \alpha}
\sigma_3^{(\alpha_j)}, \ee which can be transformed via local
unitary operations to the desired form (\ref{basic}).

The method used in this proof of principle demonstration is not

optimal, as a new graph state encoding is required for each
individual term. As already demonstrated in two examples above, a
clever choice of graph state encoding allows one to generate many
terms using the same encoding, and hence a significant reduction
of the time cost can be achieved. In particular, if the
Hamiltonian corresponds to some translational invariant set--up or
exhibits symmetries, a proper choice of graph state encoding
--where the graph has the same symmetry properties-- allows in
many cases to generate all terms of the same form (connected e.g.
by the translation operator) simultaneously. For instance, in
example 3, two different graph state encodings, corresponding to
the graph discussed in example 3 and an empty graph, are
sufficient to generate an interaction Hamiltonian with arbitrary
4--qubit interaction terms. This allows, for example, to simulate
a ring exchange Hamiltonian \cite{Hermele04,Buchler05} \be
H=\sum_{\square} (b_1^\dagger b_2 b_3^\dagger b_4 + b_1 b_2^\dagger b_3 b_4^\dagger) ),
\ee where the sum runs over all plaquettes $\square$ and opposite
corners are labelled as $1,3$ $[2,4]$ respectively. Note that
generalizations to other geometries or higher dimensions are
straightforward.

\section{Teleportation based methods}\label{TBM}

A third method to generate many--body interaction terms is by
using a teleportation based approach. That is, entangled states
are used as a resource to generate by means of teleportation an
appropriate multi--qubit unitary operation, corresponding to a
time evolution with respect to a certain many--body Hamiltonian
$H$. In the following we will discuss two teleportation based
methods that allow one to generate time evolutions with respect to
a Hamiltonian of the form \be \label{Hmultiparticle}
H=\otimes_{\alpha \in S} \sigma_3^{(\alpha)}, \ee for an arbitrary
time $t$ in a deterministic way, where $S$ is some set of $|S|$
qubits. We will consider $|S|=m$ in the following. This
corresponds to the implementation of a unitary operation of the
form $U_\varphi\equiv \exp(-i\varphi H)$ with $H$ given by Eq.
\ref{Hmultiparticle} and $\varphi=t$.

Different methods to generate specific, non--local unitary
operations are known. Except in special cases (e.g. for a CNOT
gate), either a sequence of operations using a set of entangled
states, or entangled states of higher dimension (including some
auxiliary particles) are required to make the process
deterministic. There exist a one to one correspondence between
arbitrary quantum states $E$ (described by a density operator) and
completely positive maps ${\cal E}$ \cite{dePillis67,
Jamiolkowski72,Cirac00a}. The state $E$ (possibly shared between
many parties) can be used as a resource to implement by means of
local operations and classical communication (essentially a
teleportation process) the map ${\cal E}$ on an arbitrary input
state probabilistically (i.e. for certain measurement outcomes in
the teleportation process). In many relevant cases, e.g. for gates
of the form $U_\varphi$, either $U_\varphi$ or $U_\varphi^\dagger$
is implemented with probability $p=1/2$. In this case, the
implementation of $U_\varphi$ can be made deterministic by using a
sequence of teleportation processes \cite{Cirac00a} corresponding
to unitary operations $U_\varphi, U_{2\varphi}, U_{4\varphi},
etc.$, where additional unitaries are only applied if the previous
unitary was not implemented successfully. Alternatively, a
different entangled state consisting of additional auxiliary
particles can be used for teleportation, where the measurement
basis of auxiliary particles are determined by the measurement
outcomes of the teleportation process. In \cite{Raussendorf03a,
Browne06} such a method is demonstrated for the purposes
simulating Hamiltonians that are diagonal in the computational
basis. This approach is similar to the measurement based one--way
quantum computation using cluster states \cite{Raussendorf01a}. We
will discuss both methods in detail in the following.

Cirac et al \cite{Cirac00a} demonstrated that the Jamio{\l}kowski
isomorphism defines a teleportation protocol that allows us to use
entangled states as a resource for generating entangling unitary
evolutions. In this section we  revisit their work and demonstrate
several related protocols that allow us to use entangled states to
generate entangling unitary operations.

\subsection{The Jamio{\l}kowski Isomorphism}
\label{subsec:Jisomorphism}

We now review some of the features of the Jamio{\l}kowski
isomorphism. This review is incomplete and discusses only those
features of the isomorphism that we feel are necessary to
understand the teleportation protocols which will be discussed
later in this section. For a more complete treatment we refer the
reader to \cite{Dur01b, Dur05a}.

The Jamio{\l}kowski isomorphism is a mapping between the set of
linear operators and the matrix algebra defined on a composite
Hilbert space. More precisely, given the composite Hilbert space
${\cal H}_{A'}\otimes {\cal H}_A$ and a matrix $E$ defined on this
composite space then the isomorphism is given by
\begin{equation}
\mathcal{E}(M) = d_A^2
\text{tr}_{A\bar{A}}\big[E^{A'A}P_\Phi^{A\bar{A}}M^{\bar{A}}\big],
\end{equation}
where $H^{\bar{A}}$ has the same dimension as $H^A$, $M$ is any
matrix defined in the matrix algebra of $H^A$ (and subsequently is
also defined for $H^{\bar{A}}$), and $P_\Phi^{A\bar{A}}$ is a
projector representing the maximally entangled state
\begin{equation}
|\Phi\rangle =
\frac{1}{\sqrt{d_A}}\sum_{i=1}^{d_A}|i\rangle^{\bar{A}}
|i\rangle^A
\end{equation}
where $d_A$ is the dimension of system $A$, and denote $P_{\Phi}=|\Phi\rangle\langle\Phi|$.
The inverse mapping is given by
\begin{equation}
E = \mathcal{E}^{\bar{A}}\otimes \mathcal {I}(P_\Phi^{\bar{A}A}),
\end{equation}
where $\mathcal{I}$ is the identity map.

It should be clearly noted that the Jamio{\l}kowski isomorphism is
defined for the set of all linear operators and the set of all
matrices for the composite Hilbert space ${\cal
H}_{\bar{A}}\otimes {\cal H}_A$. This set of physical maps is a
subset of this set of linear operators. Likewise, the set of
quantum density matrices is a subset of all matrices on this
composite Hilbert space. It turns out, that we can define a
teleportation-like protocol that helps us to understand the
circumstances for which the isomorphism is physically relevant.

\begin{figure}[th]
\includegraphics[width=5cm]{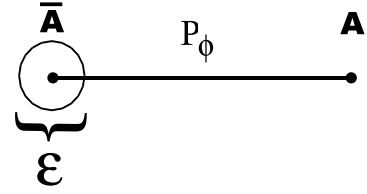}
\caption{\label{fig:Inv} In order to obtain the state $E$ the completely positive map (CPM)
$\mathcal{E}$ is applied to system $\bar{A}$ of the joint system
of $A$ and $\bar{A}$, which is prepared in the maximally entangled
state $P_\Phi^{\bar{A}A}$. (Figure taken from Ref. \cite{Dur05a})}
\end{figure}

\begin{figure}[th]
\includegraphics[width=8cm]{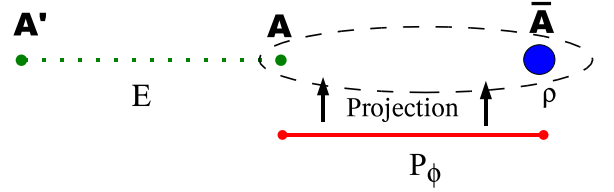}
\caption{\label{fig:Iso} (Color online) Given the state $E$ on the
composite system $A'$ and $A$, the CPM $\mathcal{E}$ is evaluated
for an arbitrary input state $\rho$ by taking $\rho$ as an input
at system $\bar{A}$. Then the joint system $A\bar{A}$ is measured
in a Bell basis containing the maximally entangled state
$P_\Phi^{A\bar{A}}$. With probability $\frac{1}{d_A^2}$ the
desired output state $\mathcal{E}(\rho)$ is then obtained at
system $A'$. (Figure taken from Ref. \cite{Dur05a})}
\end{figure}

If we have a maximally entangled state, $|\Phi\rangle^{\bar{A}A}$,
and we act with a trace-preserving completely positive map,
$\mathcal{E}\otimes \mathcal{I}$, on the system then clearly the
resulting state, $E$ is a quantum state. See figure \ref{fig:Inv}.
We shall now see how the state $E$ can be used to
probabilistically teleport the operation $\mathcal{E}(\rho)$ onto
some arbitrary state $\rho$. Consider the composite system
$A'A\bar{A}$ with Hilbert space ${\cal H}_{A'}\otimes {\cal
H}_A\otimes {\cal H}_{\bar{A}}$. Prepare $E$ on the composite
system $A'A$ and system $\bar{A}$ is in some arbitrary state,
$\rho$,
\begin{equation}
\rho_{A'A\bar{A}} = E\otimes \rho.
\end{equation}
Then, if we then perform a projective measurement on
$\rho_{A'A\bar{A}}$ in a maximally entangled basis which includes
$|\Phi\rangle_{A\bar{A}}$, the Jamio{\l}kowski isomorphism tells
us directly that there is a $\frac{1}{d_A^2}$ probability that the
outcome of this measurement will leave system $A'$ in the state
$\mathcal{E}(\rho)$. Thus, we say that with probability
$\frac{1}{d_A^2}$, the operation ${\cal E}(\rho)$ is ``teleported
to system $A'$". Such a process is outlined in figure
\ref{fig:Iso}.

\subsubsection{The Jamio{\l}kowski isomorphism for many-body systems}

We now turn to the case when $A$
is a composite system. Following the description in \cite{Dur01,Dur05a}
we will consider the case where $A$ and $A'$ are many-body systems
that have Hilbert spaces of the same type. That is we can write,
\begin{equation}
{\cal H}^A = {\cal H}^{A_1}\otimes ...\otimes {\cal H}^{A_N},
\end{equation}
where each constituent Hilbert-space has dimension $d_{A_i}$ and
${\cal H}^A\simeq {\cal H}^{A'}$. If we choose the projector
$P_\Phi^{\Bar{A}A}$ to be the tensor product of projectors onto
the two-qudit maximally entangled states
$|\Phi\rangle^{\Bar{A}_iA_i}$ then we can have a straightforward
way of extending the isomporphism to the many-body setting and
retain the interpretation in terms of a teleportation protocol
\cite{Dur01b}. More precisely $P_\Phi^{\Bar{A}A}$ is given by,
\begin{equation}
P_\Phi^{\Bar{A}A}=P_\Phi^{\Bar{A}_1A_1}\otimes ... \otimes
P_\Phi^{\Bar{A}_NA_N},
\end{equation}
where
\begin{equation}
|\Phi\rangle =
\frac{1}{\sqrt{d_{A_i}}}\sum_{k=1}^{d_{A_i}}|k\rangle^{\Bar{A}_i}|k\rangle^{A_i}.
\end{equation}

\begin{figure}[th]
\includegraphics[width=6cm]{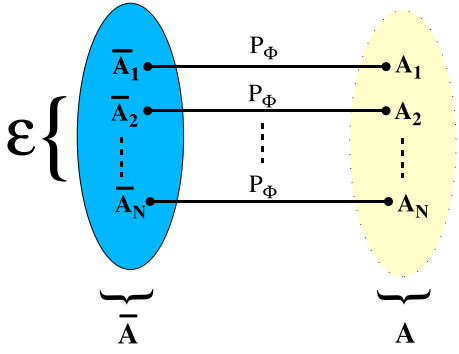}
\caption{\label{fig:MultiInv} (Color online) In order to obtain
the state $E$ the CPM $\mathcal{E}$ is applied to the subsystems
$\bar{A}_i$ of the composite system
$\bar{A}=(\bar{A}_1,\ldots,\bar{A}_N)$, which are (locally)
prepared in the maximally entangled states
$P_\Phi^{\bar{A}_iA_i}$. (Figure taken from Ref. \cite{Dur05a})}
\end{figure}

\begin{figure}[th]
\includegraphics[width=8.5cm]{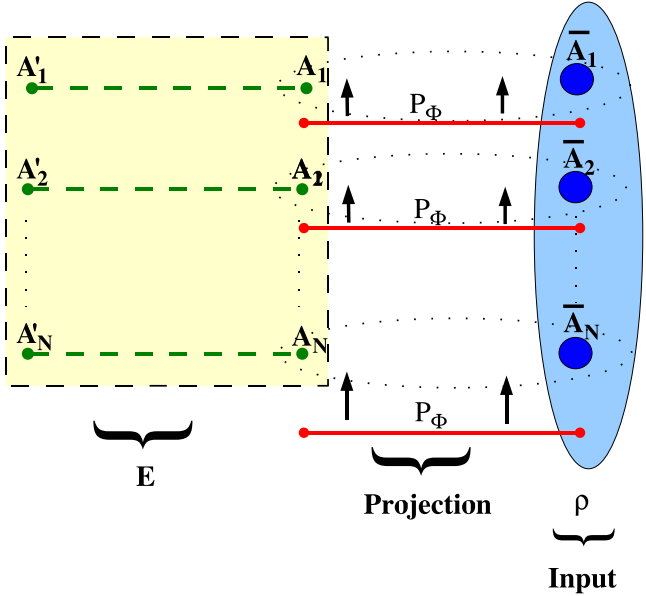}
\caption{\label{fig:MultiIso} (Color online) Given the state $E$
on the composite system $A'=(A'_1,\ldots,A'_N)$ and
$A=(A_1,\ldots,A_N)$, the CPM $\mathcal{E}$ is evaluated for an
arbitrary multipartite input state $\rho$ by taking $\rho$ as an
input at system $\bar{A}=(\bar{A}_1,\ldots,\bar{A}_N)$. Then the
joint systems $A_i\bar{A}_i$ are (locally) measured in a Bell
basis containing the maximally entangled state
$P_\Phi^{A_i\bar{A}_i}$. With probability $\frac{1}{d_A^2}$ the
desired output state $\mathcal{E}(\rho)$ is then obtained at
system $A'$. (Figure taken from Ref. \cite{Dur05a})}
\end{figure}

The only difference between teleportation protocols in the
many-body case and the case that we have already discussed is that
in this many-body scenario, the teleportation protocol requires
$N$ projective measurements to be made between the systems $A$ and
$\Bar{A}$ instead of one. The definition of $E$ in the many-body
case is depicted in figure \ref{fig:MultiInv}. The many-body
teleportation protocol for the implementation of ${\cal E}(\rho)$
on system $A'$ demonstrated in figure \ref{fig:MultiIso}.

\subsection{Teleportation-based gates using weakly entangled states}

As we have seen, the Jamio{\l}kowski isomorphism shows us how to
create a teleportation protocol that probabilistically generates
quantum operations when given certain entangled quantum states.
For the purposes of this paper we are interested in simulation
protocols that generate unitary operations of the form $U(\alpha)
= e^{-i\alpha \sigma_z^{\otimes n}}$. Such a protocol was described in
\cite{Cirac00a} where the authors demonstrated a non-deterministic
simulation protocol based on the Jamio{\l}kowski isomorphism that
succeeds with a higher success probability than the protocol that
we discussed in the previous subsection. In the next section we
will demonstrate how this protocol can be made deterministic
through the use of an additional ancillary system.

\subsubsection{Quantum control and the Jamio{\l}kowski isomorphism}

In order to understand the protocol demonstrated in
\cite{Cirac00a}, it helps to have an understanding of the way in
which control operations on a state, $E$, are transferred to the
associated operation $\mathcal{E}$ under the Jamio{\l}kowski
isomorphism. A list of control rules is given in \cite{Dur05a},
here we primarily interested in two of these rules:
\begin{enumerate}
\item Given matrices $B_1, C_1$ that can be enacted on system $A'$
and $B_2, C_2$ on system $A$, then the matrix,
\begin{equation}
\label{eq:localmanipE} E'=B_1^{A'}\otimes
B_2^{A}E^{\mathcal{E}}C_1^{A'}\otimes C_2^{A}
\end{equation}
is isomorphic to the map,
\begin{equation}
\mathcal{E}'(M) = B_1\mathcal{E}(B_2^TMC_2^T)C_1
\end{equation}
where the superscript $T$ denotes transposition.

\item $\mathcal{E}$ is separable with respect to $A_k$ and $A_l$
iff $E$ is separable with respect to $A_k$ and $A_l$. That is the
CPM corresponding to state $E\otimes F$ is $\mathcal{E}\otimes
\mathcal{F}$, where ${\cal E}$ and ${\cal F}$ are the maps
isomorphic to $E$ and $F$.
\end{enumerate}

The first rule can be extended to demonstrate that if control is
locally limited to parties $A$ and $A'$ then the allowable set of
control operations are limited to local operations with one-way classical communication. The only allowable
classical communication must flow from part $A$ to party $A'$ as
the manipulations performed by party $A$ occur prior to
implementing the map $\mathcal{E}$ whereas those performed by $A$
occur after. To see this, consider the following mixture of
matrices,
\begin{equation}
E'=\sum_j Q_j^{A'}\otimes R_j^{A}E(Q_j^{A'}\otimes
R_j^{A})^\dagger.
\end{equation}
Using Eq. \ref{eq:localmanipE} we see that the corresponding CPM
is,
\begin{equation}
\label{eq:localmanipEfinal} \mathcal{E}'=\sum_j Q_j
\mathcal{E}(R_j^TMR_j^*)Q_j^\dagger.
\end{equation}
If the matrices $\{Q_j\}$ and $\{R_j\}$ represent measurements and
unitary operations, then the operation $R_j$ occurs before the
operation $Q_j$. Thus the set of operations that can physically be
performed on the state $E$ are restricted to,
\begin{equation}
E' = \sum_{ij}Q_{ij}^{A'}\otimes R_{j}^AE(Q_{ij}^{A'}\otimes
R_{j}^A)^\dagger,
\end{equation}
where the CPM $\mathcal{R}{\rho}=\sum_j R_j \rho R_j^\dagger$ is
bi-stochastic trace-preserving CPM \footnote{Thus the map
$\tilde{\mathcal{R}}=\sum_j R_j^T\rho R_j^{*}$ is also a
trace-preserving CPM}, and if $j$ corresponds to a measurement
outcome at party $A$ the operation $Q_j(\rho) = \sum_i B_{ij}\rho
B_{ij}^\dagger$ is a trace-preserving CPM. It is interesting to
note that in the case where all operations are either unitary or
projective measurements then the bi-stochasticity of $\mathcal{R}$
is assured. This can be seen by noting that for any unitary
operator $U$, $(U^{*})^\dagger =U^{T} = (U^{*})^{-1}$ and all
projectors onto pure states are real and Hermitian.

\subsubsection{Using weakly entangled states to generate phase gates}
\label{subsubsec:weaklyenttelep}

In this section we review the non-deterministic simulation
protocol for generating unitary operations of the form $U(\alpha)
= e^{-i\alpha \sz^{\otimes n}}$ that was presented in
\cite{Dur01b}.

Firstly, we shall give the two-qubit protocol that appeared in
\cite{Cirac00a} in which it was shown how to generate the unitary
operator, $e^{-i\alpha \sz\otimes \sz}$, on any given input state.
We will then discuss why this protocol works and how to extend it
to the more general many-qubit case.

\begin{enumerate}
\item Prepare the state,
\begin{eqnarray}
|\alpha\rangle & = & \cos{\alpha}
|\Phi^{+}\rangle^{A_1A'_1}|\Phi^{+}\rangle^{A_2A'_2} \nonumber \\
& & -
i\sin{\alpha}|\Phi^{-}\rangle^{A_1A'_1}|\Phi^{-}\rangle^{A_2A'_2},
\end{eqnarray}
where $\alpha = \pi/2^N$, $|\Phi^+\rangle =
1/\sqrt{2}(|00\rangle+|11\rangle)$ and $|\Phi^-\rangle =
1/\sqrt{2}(|00\rangle-|11\rangle)$. The state, $|\alpha\rangle$,
is prepared between parties $A$ and $A'$.

\item Introduce party $\bar{A}$ which is in the desired input
state $\rho$.

\item Perform a Bell basis measurements between $A_1\bar{A}_1$ and
$A_2\bar{A}_2$. Note that we can write the Bell basis as
\begin{equation}
|\Psi_{i_1,i_2}\rangle = \eins \otimes
\sigma_{i_1,i_2}|\Phi^{+}\rangle,
\end{equation}
where $\sigma_{1,1}=\eins$, $\sigma_{1,2}=\sz$,
$\sigma_{2,1}=\sigma_y$, and $\sigma_{2,2}=\sigma_x$.

\item If the measurement on $A_1\bar{A}_1$ produces the outcome
$|\Psi_{i_1,i_2}\rangle$, apply $\sigma_{i_1,i_2}$ to qubit
$A'_1$. If the measurement on $A_2\bar{A}_2$ produces the outcome
$|\Psi_{j_1,j_2}\rangle$, apply $\sigma_{j_1,j_2}$ to qubit
$A'_2$.

\item If $i_1=j_1$ then $U(\alpha)=e^{-i\alpha \sz\otimes \sz}$
has been applied to system $A'$ and the procedure terminates. This
occurs with probability $1/2$.

\item If $i_1\neq j_1$ then we have applied $U(-\alpha)$ to system
$A'$. This occurs with probability $1/2$. In this case, we
re-label system $A'$ as system $\bar{A}$ and then prepare the
state $|2\alpha\rangle$ on $AA'$ and repeat the procedure.

The probability of success, that is that $U(\alpha)$ has been
applied to system $A'$, upon repetition is $1/2$. With probability
$1/2$ the unitary $U(-3\alpha)$ is applied to system $A'$.

\item For the $k$th iteration we prepare the state
$|2^{k-1}\alpha\rangle$. After $N$ steps the procedure must
necessarily terminate as $U(-(2^N-1)\alpha)=U(\alpha)$ up to an
irrelevant global phase.

\end{enumerate}

Why does this procedure work? It is a variation of the
teleportation protocol that we saw in the previous section. The
state $|\alpha\rangle$ can be prepared by the following operation
on the pair maximally entangled states
$|\Phi^{+}\rangle^{A_1A'_1}|\Phi^{+}\rangle^{A_2A'_2}$,
\begin{equation}
|\alpha\rangle =e^{-i\alpha \eins\otimes \sz\otimes \eins\otimes
\sz}|\Phi^{+}\rangle^{A_1A'_1}|\Phi^{+}\rangle^{A_2A'_2}.
\end{equation}
Looking at the definition of the the Jamio{\l}kowski isomorphism
and noting that $A$, $A'$, and $\bar{A}$ are each qubits then it
is clear that the state $|\alpha\rangle\langle\alpha|=E$ and it's
corresponding CPM is $\mathcal{E}\equiv U(\alpha)$. According to
the teleportation protocol of the previous section, in order to
apply $U(\alpha)$ to an unknown state we need to perform
Bell-basis measurements on systems $A$ and $\bar{A}$. For the
state $|\alpha\rangle$ given above, a successful run of the
protocol coincides with measuring systems $A_1\bar{A}_1$ and
$A_2\bar{A}_2$ to be in the state $|\Phi^+\rangle$, this occurs
with probability $1/16$.

In the protocol that we have presented in this section, we use
local unitary operations and one-way classical communication to
increase this success probability to $1/2$ for any given instance
of the protocol. The key to this improved success probability is
that the Pauli correction operators we apply, $\sigma_{i_1,i_2}$
and $\sigma_{j_1,j_2}$, allow us to manipulate the state $\alpha$.

A Bell basis can be generated by applying a single-qubit Pauli
operator to the state $|\Phi^+\rangle$,
\begin{equation}
|\Psi_{i_1,i_2}\rangle = \eins \otimes
\sigma_{i_1,i_2}|\Phi^{+}\rangle.
\end{equation}
This equation tells us that performing a measurement in this basis
and getting the outcome $|\Psi_{i_1,i_2}\rangle$ is is equivalent
to applying the projector $|\Phi^{+}\rangle$ followed by the Pauli
operation $\sigma_{i_1,i_2}$. Interestingly, it is irrelevant
which qubit the operation $\sigma_{i_1,i_2}$ is applied to. In
Step 3 of this protocol we perform Bell measurements between
systems $A$ and $\bar{A}$. If the outcome of these measurements
are all $|\Phi^{+}\rangle$ we have directly applied the gate
$U(\alpha)$. Alternatively, we can see from Equations
(\ref{eq:localmanipE}) and (\ref{eq:localmanipEfinal}) that we
will have applied the map
$\mathcal{E}\big((\sigma_{i_1,i_2}\otimes \sigma_{j_1,j_2})\rho(
\sigma_{i_1,i_2}\otimes \sigma_{j_1,j_2})\big)$. In the case where
$i_1=j_1$ we can apply the corrections of given in Step 4 of the
protocol above and succeed in applying $U(\alpha)$. These
corrections work because they commute with the unitary
$U(\alpha)$. In the case where $i_1\neq j_1$, the effect of
applying the correction is to apply $U(-\alpha)$ to $\rho$, as
\begin{equation}
U(\alpha)\sigma_{i_1,i_2}\otimes \sigma_{j_1,j_2}
=\sigma_{i_1,i_2}\otimes \sigma_{j_1,j_2} U(-\alpha).
\end{equation}

This protocol works half of the time because $U(\alpha)$ commutes
with half of the set of two-qubit Pauli product operators
$\{\sigma_j \otimes \sigma_k|j,k = 0,...,3\}$. In the next section
we demonstrate how to make a similar protocol that
deterministically applies $U(\alpha)$ to $\rho$ through the use of
an ancillary qubit.

As has been shown in \cite{Dur01b}, one can generalize this protocol to produce evolutions by
the $n-qubit$ unitary $U(\alpha)= e^{-i\alpha \sz^{\otimes n}}$.
In this case the probability of a success in any instance of the
protocol is still $1/2$ as in the two-qubit case. In essence the
protocol is the same. Prepare the state $|\alpha\rangle$ using $n$
Bell pairs, perform $n$ Bell-basis measurements, apply $n$ Pauli
correction operations. In each instance the correction operations
are the same as those for the two-qubit case. That is, if
$|\Psi_{i_1,i_2}\rangle$ is measured, apply the correction
$\sigma_{i_1,i_2}$ on the appropriate qubit. Like in the two-qubit
case, success is in an instance of the protocol is achieved in the
case when tensor product of all of the Pauli correction operations
that are applied commutes with the operator $\sigma_z^{\otimes n}$, which
occurs with a success probability of $\frac{1}{2}$.

\subsection{Teleportation-based gates using GHZ-type states}

In the previous section we saw how we can use teleportation as a
primitive to stochastically generate unitary operations of the
form $U=e^{-i\alpha \sz^{\otimes n}}$. It has been demonstrated
\cite{Aliferis04a, Gottesman99a} that gate-teleportation protocols
like this are always deterministic if the entangling operation to
be teleported onto $\rho$ is an element of the Clifford group
(defined as the normalizer of the Pauli group). If an operation is
not in this group, as is the case with $U(\alpha)$, then the
necessary correction operations in a deterministic protocol are
\emph{entangling}.

In this section we shall demonstrate that by performing
teleportation using GHZ-type states as a primitive we can
deterministically ``teleport'' the non-Clifford operations like
$U(\alpha)$. In essence, by utilizing prior entanglement with
ancillary qubits we can avoid applying \emph{entangling}
correction operations to the teleportation protocol. The
simulation protocol that we demonstrate here is essentially the
same as those which was demonstrated in the context of graph-state
based simulation in \cite{Raussendorf03a, Browne06}. However, here
we use the language of teleportation in order to highlight the
connection with the work presented in \cite{Cirac00a, Dur01b}.

Consider the state
\begin{equation}
|GHZ\rangle = \frac{1}{\sqrt{2}}(|+\rangle^{\otimes
n}|0\rangle_E+|-\rangle^{\otimes n}|1\rangle_E),
\end{equation}
this state is an $n+1$ qubit GHZ-state. The GHZ state is a
two-colorable graph state. It is easy to see that this state can
be generated from phase gates by applying a phase gate between
qubit $E$ and each of the other $n$ qubits in the system. It has
recently been shown that such states can be purified
\cite{Mu01,Ma01,Aschauer05a,Aschauer05}.

\subsubsection{Simulation protocol}

In Section \ref{subsubsec:weaklyenttelep} we saw a teleportation
protocol that used weakly entangled states as a resource to
generate $U(\alpha)$ that had a probability of success in any
given iteration of 1/2. This protocol failed when the Pauli
correction operations did not commute with $U(\alpha)$. When this
occurred, the result of the protocol was to apply the operator
$U(-\alpha)$ to the state $\rho$ because the correction operation
commutes through $U(\alpha)$ to generate $U(-\alpha)$.
Interestingly, in a failing instance of the protocol, if we could
``go back in time'' and change the weakly entangled state
$|\alpha\rangle$ to $|-\alpha\rangle$ we would have applied the
operator $U(\alpha)$ to $\rho$.

In this section we demonstrate an alternate method for generating
the states $|\alpha\rangle$ or $|-\alpha\rangle$, that is the
states corresponding to the maps $U=e^{-i\alpha \sz\otimes \sz}$
and $U=e^{i\alpha \sz\otimes \sz}$. This method allows us to
generate these states after performing Bell measurements between
systems $\bar{A}$ and $A$.

Consider the following $GHZ$-like state,
\begin{eqnarray}
|\kappa\rangle & = &\frac{1}{\sqrt{2}}(
|\Phi^{+}\rangle^{A_1A'_1}|\Phi^{+}\rangle^{A_2A'_2}|0\rangle^E \nonumber \\
& & +\eins\otimes \sz\otimes \eins\otimes
\sz|\Phi^{+}\rangle^{A_1A'_1}|\Phi^{+}\rangle^{A_2A'_2}|1\rangle^E),
\end{eqnarray}
where the superscript $E$ denotes the state of an ancillary
system. We define the following measurement basis for system $E$:
\begin{eqnarray}
\{|m \rangle ,|m_\bot\rangle\}  & = & \{\cos(\alpha)|0\rangle
+i\sin(\alpha)|1\rangle, \nonumber \\ && i\sin(\alpha)|0\rangle
+\cos(\alpha)|1\rangle\} \\
\{|-m \rangle ,|-m_\bot\rangle\}  & = & \{\cos(\alpha)|0\rangle
-i\sin(\alpha)|1\rangle, \nonumber \\ && i\sin(\alpha)|0\rangle
-\cos(\alpha)|1\rangle\}.
\end{eqnarray}
If we re-write the state $|\kappa\rangle$ in terms of the $\{|m
\rangle ,|m_\bot\rangle\}$ we find,
\begin{eqnarray}
|\kappa\rangle & = &\frac{1}{\sqrt{2}}\big[\big(
\cos(\alpha)|\Phi^{+}\rangle^{A_1A'_1}|\Phi^{+}\rangle^{A_2A'_2} \nonumber \\
& & - i\sin(\alpha)\eins\otimes \sz\otimes \eins\otimes
\sz|\Phi^{+}\rangle^{A_1A'_1}|\Phi^{+}\rangle^{A_2A'_2}\big)|m\rangle^E\nonumber
\\
& & +\big( \cos(\alpha)\eins\otimes \sz\otimes \eins\otimes
\sz|\Phi^{+}\rangle^{A_1A'_1}|\Phi^{+}\rangle^{A_2A'_2} \nonumber \\
& &
+i\sin(\alpha)|\Phi^{+}\rangle^{A_1A'_1}|\Phi^{+}\rangle^{A_2A'_2}\big)|m_\bot\rangle^E\big].
\end{eqnarray}
Which can be simplified to,
\begin{eqnarray}
\label{eq:altalpha} |\kappa\rangle & =
&\frac{1}{\sqrt{2}}\big[e^{-i\alpha
\sz^{A'_1}\sz^{A'_2}}|\Phi^{+}\rangle^{A_1A'_1}|\Phi^{+}\rangle^{A_2A'_2}|m\rangle^E\nonumber
\\
& & + \sz^{A'_1}\sz^{A'_2}e^{-i\alpha \sz^{A'_1}\sz^{A'_2}
}|\Phi^{+}\rangle^{A_1A'_1}|\Phi^{+}\rangle^{A_2A'_2}|m_\bot\rangle^E\big]
\nonumber \\
& = & \frac{1}{\sqrt{2}}(|\alpha\rangle |m\rangle +
\sz^{A'_1}\sz^{A'_2}|\alpha\rangle |m_\bot\rangle).
\end{eqnarray}
Eq. \ref{eq:altalpha} demonstrates that a measurement on
$|\kappa\rangle$ of system $E$ in the $\{|m \rangle
,|m_\bot\rangle\}$ basis will either generate the state
$|\alpha\rangle$, if the outcome is $m$, or the state
$\sz^{A'_1}\sz^{A'_2}|\alpha\rangle$, given $m_\bot$ as the
outcome. By performing a similar analysis it is possible to see
that a measurement in the basis $\{|-m \rangle ,|-m_\bot\rangle\}$
produces either the state $|-\alpha\rangle$ or the state
$\sz^{A'_1}\sz^{A'_2}|-\alpha\rangle$. Given that we can always
perform Pauli correction operations on system $A'$ after this
measurement, we see that a measurement on the ancilla system $E$
can generate the states $|\alpha\rangle$ or $|-\alpha\rangle$.

With this technique in mind, we shall now outline our new
protocol.

We first consider the two-qubit case. We want to deterministically
generate an evolution by the unitary $U=e^{-i\alpha \sz\otimes
\sz}$ on the unknown two-qubit state $\rho$ for any $\alpha$.

\begin{enumerate}
\item Prepare the resource state $|\kappa\rangle$

\item Introduce party $\bar{A}$ which is in the desired input
state $\rho$.

\item Perform Bell basis measurements between $A_1\bar{A}_1$ and
$A_2\bar{A}_2$.

\item If the measurement on $A_1\bar{A}_1$ produces the outcome
$|\Psi_{i_1,i_2}\rangle$, apply $\sigma_{i_1,i_2}$ to qubit
$A'_1$. If the measurement on $A_2\bar{A}_2$ produces the outcome
$|\Psi_{j_1,j_2}\rangle$, apply $\sigma_{j_1,j_2}$ to qubit
$A'_2$.

\item If $i_1 = j_1$, then measure system $E$ in the $\{|m \rangle
,|m_\bot\rangle\}$ basis. If the outcome is $m$ the procedure
terminates. If the outcome is $m_\bot$ then apply a $\sz$
operation to qubits $A'_1$ and $A'_2$ to terminate the procedure.

\item If $i_1 \neq j_1$, then measure system $E$ in the $\{|-m
\rangle ,|-m_\bot\rangle\}$ basis. If the outcome is $m$ the
procedure terminates. If the outcome is $-m_\bot$ then apply a
$\sz$ operation to qubits $A'_1$ and $A'_2$ to terminate the
procedure.

\item Upon termination, system $A'$ will have undergone the
evolution
\begin{equation}
\mathcal{E}(\rho^{A'})= U(\alpha)\rho U(\alpha)^\dagger.
\end{equation}

\end{enumerate}

Essentially, the procedure that we have outlined here is the same
as that for weakly entangled states. The key difference is that in
the case of the protocol for weakly entangled states, we
constructed $|\alpha\rangle$ first. Under that protocol, we are
restricted to manipulations that are local operations with one-way classical communication. In this
protocol, we keep systems $A$ and $A'$ entangled with system $E$
until the final step. The effect of measuring system $E$ is to
collapse the total composite system into the desired state, either
$|\alpha\rangle$ or $|-\alpha\rangle$.

This protocol is easily extended to generate evolutions of the
form $U=e^{-i\alpha \sz^{\otimes n}}$.  Firstly prepare the state
\begin{eqnarray}
|\kappa\rangle & = &
\frac{1}{\sqrt{2}}(|\Phi^{+}\rangle^{A_1A'_1}...|\Phi^{+}\rangle^{A_nA'_n}
|0\rangle \nonumber \\ &&+
\sz^{A'_1}...\sz^{A'_n}|\Phi^{+}\rangle^{A_1A'_1}...|\Phi^{+}\rangle^{A_nA'_n}
|1\rangle).
\end{eqnarray}
Then perform the same Bell measurements and Pauli corrections as
above. If the operations applied after the measurement in Step 4
commute with the operator $\sz^{A'_1}...\sz^{A'_n}$, then measure
in the $\{|m \rangle ,|m_\bot\rangle\}$ basis and apply the
operator $\sz^{A'_1}...\sz^{A'_n}$ if the outcome is $m_\bot$. If
they do not, measure in the $\{|-m \rangle ,|-m_\bot\rangle\}$
basis and apply $\sz^{A'_1}...\sz^{A'_n}$ if the outcome is
$m_\bot$.

\section*{\Large PART II: Influence of noise}\label{Errors}

As we discussed in the Introduction, building a fault-tolerant quantum computer is a challenging task that may not be accomplished in the near future. 
In the meantime it
is hoped that non-trivial quantum simulation experiments could
still be performed. In particular, the demonstration of an
imperfect quantum simulation that could still outperform the
best-known simulation algorithms on a classical computer would be
a significant experimental achievement.

In Part II of this paper, we discuss the influence of noise and
imperfections on the simulation of many-body Hamiltonians using
the different methods described in Part I. We focus our attention
on protocols for simulating Hamiltonian evolutions of the form $H_\alpha
= e^{-i\alpha \sigma_z^{\otimes m}}$. As we saw in Part I, while
Hamiltonians such as $H_\alpha$ have a particularly simple form, they
can be used alongside the Lie-Trotter formula to simulate the
dynamics of any finite-dimensional Hamiltonian. Our focus is on
these more simple Hamiltonians because in many cases the effects
of noise on these protocols can be analytically studied. This
gives us insight into the types of errors that might occur in more
sophisticated simulation protocols. In particular, the following
analysis could be used to derive error bounds on the measurement
statistics of more complex simulations that use Hamiltonians like
$H_\alpha$ as a building block.

We begin Part II by discussing general methods for classifying
noise processes. We then go on to introduce the noise models that
we consider, local dephasing and depolarizing noise and errors
induced by random errors in the evolution time of entangling
Hamiltonians. We then set about integrating these noise models
into the protocols that we discussed in Part I and examining the
effects analytically, where possible, and computationally. Mostly,
we look at the effect of introducing noise to the entangling
operations in these protocols and consider local operations to be
perfect. This is done because for the purposes of quantum
simulation we are concerned with the way that the noise is spread
throughout the system as a result of the entangling operations in
each protocol. In addition, we often focus on regimes
where the effect of local noise on single-qubit operations can be
incorporated into the noise model for the entangling operations
without much loss of generality.

\section{Distance measures and fidelity}\label{DMF}

We will use two different measures to quantify the quality of the
simulation process. Time evolution with respect to a specific
many--body Hamiltonian $H$ for a time $t$ corresponds to a unitary
operation $U_t$ (which is the desired evolution), while time
evolution governed by a master equation of Lindblad form
(describing the noisy evolution), leads to a completely positive
map ${\cal E}_t$. We consider the case where the operation is
acting on $m$ qubits and determine how close ${\cal E}_t$ is to
$U_t$.

We use two devices for comparing the closeness of $\mathcal{E}_t$
to $U_t$, Jamio{\l}kowski fidelity \cite{Gilchrist04a} and the
\emph{local noise equivalent}. The mathematical detail of both of
these devices are discussed in more detail below. It is worth
noting here that both the Jamio{\l}kowski fidelity and the local
noise equivalent have useful operational interpretations when
being used to describe a simulation protocol.

%

\subsection{Jamio{\l}kowski fidelity}

The Jamio{\l}kowski fidelity $F(U_t,\mathcal{E}_t)$ measures the
closeness of one quantum operation to another. This fidelity is
based on the Jamio{\l}kowski isomorphism that was introduced in
Section \ref{subsec:Jisomorphism} that relates quantum operations
to quantum states. The Jamio{\l}kowski fidelity measures the
overlap between two quantum operations by measuring the overlap
between their corresponding states as defined by the
Jamio{\l}kowski isomorphism. This will be more rigorously defined
below.

There are many ways that one might wish to go about defining
fidelities for comparing quantum operations. The Jamio{\l}kowski
isomorphism is useful because it is easy to calculate, it
satisfies some important mathematical properties, and it has
several clear operational interpretations which are closely linked
to the theory of quantum computation and simulation
\cite{Gilchrist04a}.

The first operational interpretation of the Jamio{\l}kowski
fidelity is in terms of a \emph{sampling} quantum computation.
Imagine that we were to implement a quantum operation, $U$, on
some initial state $|x\rangle$ in order to attain measurement
outcomes, $y$, on the output state. Say that we vary the input $x$
and sample in order to attain the joint probability distribution
$p_x(y)$. Such a scenario is the way that one might perform any
number of physically relevant simulation protocols. Say that the
ideal distribution, that which occurs if $U$ is implemented
perfectly, is $p_x(y)$ and the actual experimentally achieved
distribution is $q_x(y)$, which occurs as a result of implementing
and imperfect operation, $\mathcal{E}$. Then it can be shown
\cite{Gilchrist04a} that the Jamio{\l}kowski fidelity provides a
lower bound on the overlap between the two probability
distributions. That is,
\begin{equation}
F(U,\mathcal{E})\leq B(p_x(y),q_x(y))
\end{equation}
where $B(p_x(y),q_x(y))$ is the Bhattacharya overlap between the
probability distributions and the problem instances are chosen
uniformly at random. The Bhattacharya overlap is a measure for
comparing probability distributions, while it doesn't have a
succinct operational interpretation, it can be used to bound the
Kolmogorov distance between two probability distributions.

Let us now consider a different simulation protocol, one that is
based on a \emph{function} computation. Imagine that we desire to
use a unitary process, $U$, to simulate an evolution from the
state $|x\rangle$ to the state $|f(x)\rangle$. We then perform
measurements on the output state $|f(x)\rangle$ in order to
determine the property of that state given by $f(x)$. If we
instead performed the evolution $\mathcal{E}$ and output the state
$\mathcal{E}(|x\rangle\langle x|)$, then the average probability,
$p_e$, of getting an incorrect value of $f(x)$ given a uniform
distribution over $x$ is bounded by,
\begin{equation}
p_e \leq 1-F(U,\mathcal{E}).
\end{equation}
More succinctly, if an experiment is designed to simulate some
system in order to deterministically evaluate some property of a
state, then the Jamio{\l}kowski fidelity provides a bound on the
average probability of an error.

We now move on to the more mathematically oriented properties of
the Jamio{\l}kowski fidelity. There are a number of ways of
defining a fidelity for a quantum operation that are based on the
Jamio{\l}kowski isomorphism. In the case where we are always
comparing an ideal unitary operation to a CP map it is convenient
to use the following definition: \be F({\cal
E}_t,U_t)=F(E_t,|\Psi_t\rangle)=\langle \Psi_t|E_t|\Psi_t\rangle,
\ee where \be
|\Psi_t\rangle &=& \eins^{(A)} \otimes U_t^{(B)} |\Phi\rangle, \nonumber\\
E_t &=& {\cal E}_t^{(B)} |\Phi\rangle_{AB}\langle \Phi|, \ee and
$|\Phi\rangle =
1/\sqrt{2^m}\sum_{k=0}^{2^m-1}|k\rangle_A|k\rangle_B$ is a
maximally entangled state of $2m$ qubits, where we labelled the
basis of $m$ qubits by $\{|0\rangle,|1\rangle,\ldots
|2^m-1\rangle\}$, i.e. identified the $m$ qubits with a
$2^m$--dimensional system. The operation acts on one part of the
maximally entangled state. This fidelity was called the
Jamio{\l}kowski process fidelity in \cite{Gilchrist04a}, as it can
be only requires elements of an operation's process matrix in
order to be calculated. In this paper we simply refer to it as the
Jamio{\l}kowski fidelity.

It is worth noting that the Jamio{\l}kowski fidelity is linearly
related to the \emph{average} fidelity, $\bar{F}(U,\mathcal{E})$,
between a unitary operation $U$ and a CP map $\mathcal{E}$,
\begin{equation}
\bar{F}(U,\mathcal{E})= \int d\psi \langle\psi|U^\dagger
\mathcal{E}(|\psi \rangle\langle \psi|)U|\psi\rangle,
\end{equation}
where the integral is over the pure states $|\psi\rangle$. For a
system of dimension $d$, the average fidelity can be calculated
from the Jamio{\l}kowski fidelity via,
\begin{equation}
\bar{F}(U,\mathcal{E})=\frac{F(U,\mathcal{E})d+1}{d+1}.
\end{equation}

\subsection{Simple bounds on the Jamio{\l}kowski fidelity for simulation protocols}

The Jamio{\l}kowski fidelity, as defined for this paper, can be
used to generate several distance measures on the the set of all
quantum operations. One such measure, $D(\mathcal{E}_t,U_t)$, will
be used in this paper \cite{Gilchrist04a},
\begin{equation}
D(\mathcal{E}_t,U_t) = \sqrt{1-F(\mathcal{E}_t,U_t)}.
\end{equation}
This distance measure is a metric on the space of quantum
operations that is stable (that is remains unchanged) under the
addition of ancillary quantum systems. Importantly, this measure
also satisfies the following chaining inequality,
\begin{equation}
\label{eq:chaining} D(U_2\circ
U_1,\mathcal{E}_2\circ\mathcal{E}_1)\leq
D(U_1,\mathcal{E}_1)+D(U_2,\mathcal{E}_2),
\end{equation}
where $U_{1,2}$ are unitary operations and $\mathcal{E}_{1,2}$ are
trace-preserving quantum maps. This chaining inequality basically
tells us that if we compare two sequences of quantum operations,
then the distance between the two sequences will not diverge more
rapidly than the distance between each individual step. This
inequality turns out to be useful in our context, as quantum simulation
protocols can often be cast in terms of sequences of unitary operations.

The chaining inequality allows us to obtain an upper bound on the
distance (and correspondingly a lower bound on the fidelity)
between two control sequences. Given an ideal unitary sequence $U
= U_m\circ U_{m-1} \circ \dots \circ U_{1}$ and the corresponding
noisy sequence $\mathcal{E} = \mathcal{E}_m\circ \mathcal{E}_{m-1}
\circ \dots\circ \mathcal{E}_{1}$ then the chaining inequality
tells us,
\begin{equation}
D(U,\mathcal{E})\leq \sum_{j=1}^m D(U_j,\mathcal{E}_j)\leq m
D_{max},
\end{equation}
where $D_{max}=\max_j D(U_j,\mathcal{E}_j)$.

Consider now the case where we wish
to compare an ideal operation $U$ with a noisy operation of the
form,
\begin{equation}
\mathcal{E}=U\circ \mathcal{E}_U = \mathcal{E}_U \circ U.
\end{equation}
where $\mathcal{E}_U$ commutes with $U$. Using the definition of
$F(U,\mathcal{E})$ it is easy to see that,
\begin{equation}
\label{eq:commnoisefid} F(U,\mathcal{E}) = F(I,\mathcal{E}_U).
\end{equation}
Correspondingly, we find
\begin{equation}
\label{eq:commnoisedist} D(U,\mathcal{E}) = D(I,\mathcal{E}_U).
\end{equation}
While this seems quite obvious, such circumstances occur readily
in physical situations. For instance a system with Hamiltonian
proportional to $\sz^{\otimes m}$ undergoing \emph{dephasing},
\emph{timing errors} (as we will see later), or alternatively for
$U$ and $\mathcal{E}$ that are both generated by \emph{fast
control pulses} (as is the case when using the Lie-Trotter
summation formula).

Each of the simulation protocols discussed in this paper use
sequences of unitary operations followed by measurements. The
ideal unitary operations that appear in each protocol are either
local unitary operations (often referred to as LU operations) or
are applications of the entangling Hamiltonian $\sz\otimes \sz$
between pairs of the qubits in this system for some time $t$, that
is the unitary $U_t$. In the case where we can consider all LU
operations to be perfect it is very easy to establish upper
(lower) bounds on the Jamio{\l}kowski distance (fidelity) using
the the chaining inequality.

A generic control sequence that appears in this paper looks
something like,
\begin{equation}
U = (LU)_m\circ U_t \circ (LU)_{m-1} \circ U_t \circ \dots \circ
(LU)_{2}\circ U_t \circ (LU)_{1},
\end{equation}
where $(LU)_j$ denotes some local unitary operations and $U_t$ is
an fixed entangling unitary that could in principle be applied
between any set of qubits in the system. If we consider noise that
only occurs as a result of the entangling operation $U_t$ then we
can replace $U_t$ in the above sequence by some quantum map
$\mathcal{E}_t$ that describes the noisy entangling operation and
our noisy control sequence looks like,
\begin{equation}
\mathcal{E} = \hat{(LU)_m}\circ \mathcal{E}_t \circ \hat{(LU)_{m-1}} \circ
\mathcal{E}_t \circ \dots \circ \hat{(LU)_{2}}\circ \mathcal{E}_t \circ
\hat{(LU)_{1}}.
\end{equation}
Comparing these two sequences with the chaining inequality we
find,
\begin{eqnarray}
D(U,\mathcal{E}) & \leq & D((LU)_m,(LU)_m)+ D(U_t,\mathcal{E}_t)
\nonumber \\ & & + D((LU)_{m-1},(LU)_{m-1}) + D(U_t,\mathcal{E}_t)
+\dots .\nonumber \\
\end{eqnarray}
If we note that for all $j$ $D((LU)_j,(LU)_j)=0$ then the above
expression can be simplified to,
\begin{equation}
\label{eq:simpledistbound} D(U,\mathcal{E}) \leq k
D(U_t,\mathcal{E}_t),
\end{equation}
for a sequence involving $k$ implementations of the map
$\mathcal{E}_t$. Thus the problem of establishing an upper bound
on the distance between $U$ and $\mathcal{E}$ is reduced to
finding the distance between $U_t$ and $\mathcal{E}_t$. The
corresponding fidelity bound is:
\begin{equation}
\label{eq:simpleFbound} F(U,\mathcal{E})\geq 1-k^2
D(U_t,\mathcal{E}_t)^2
\end{equation}

This bound is not the tightest bound that can always be
established analytically. Later in this paper we shall see similar
constructions that are optimized depending on on the simulation
model that appears and the noise that is being modelled. Better
bounds can be found by grouping the basic operations in each
simulation control sequence that commute with each other before
applying the chaining inequality. For instance, if all noise terms
and unitary operations in a control sequence commute with one
another the fidelity can be calculated precisely. Conversely, if
each local unitary operation that appears does not commute with
$U_t$ then it is not possible to find a better bound on the
fidelity without resorting to further approximations.

%


While the Jamio{\l}kowski fidelity is in general a suitable
distance measure for noisy quantum operations, one has to be a bit
careful about its interpretation. To be precise, when considering
unitary operations $U_t$ with small $t$, i.e. time evolution with
respect to the desired Hamiltonian for a small time $t$, the
operation will be very close to the identity. The influence of
noise, as resulting from a master equation description we consider
here, also increases with time $t$. For small $t$, the influence
of noise is hence very small, and also the noisy operation will be
close to the identity. This will lead to a fidelity close to 1,
also in cases where the noise dominates over the Hamiltonian part
of the evolution. In particular, it can happen that for small $t$
the fidelity is close to one, but the noisy operation is not
capable of creating entanglement (i.e. the corresponding operator
$E_t$ is separable \cite{Cirac00a}).

\subsection{Gate fidelity for long--time evolutions}

To allow for a better comparison, we thus also consider unitary
gates generated by time evolutions for long times of $O(1)$. The
ideal evolution is given by the unitary gate $U_t$, which is
obtained by the time evolution with respect to the desired
Hamiltonian $H$ for time $t$, where e.g. $t=\pi/4$. The ideal
evolution is compared with a sequence of basic noisy evolutions
that approximate the desired gate. To be precise, the noisy
evolution ${\cal E}_{\delta t}$ for (possibly small) times $\delta
t$ is applied $n$ times, where $n=t/\delta t$, to approximate the
desired unitary operation $U_t$. The Jamio{\l}kowski fidelity
\be
F_t({\cal E}_{\delta t}) \equiv F(\prod_{k=1}^n{\cal E}_{\delta
t},U_t),
\ee
is a suitable quantity to measure the quality of
${\cal E}_{\delta t}$ and compare such noisy evolutions with
different times $\delta t$. Also from a practical point of view,
$F_t$ is a useful quantity, as in many cases the goal is to
generate a certain unitary gate, and $F_t$ measures directly with
which fidelity this is possible. It turns out that the optimal
strategy is often not to generate very precise approximation of
the desired evolutions for short times $\delta t$ and apply a
sequence of many of such evolutions, but to chose some fixed
intermediate time $\delta t'$ which only require a moderate
repetition of these basic evolutions.

\section{Noise model 1: Master equation}\label{Noisemodel1}

We will describe the influence of noise by a master equation of
Lindblad form that is derived under certain assumptions, e.g. the
Markov approximation. To be precise, we model the interaction
between systems by a master equation of the form \be
\frac{\partial}{\partial t} \rho  = \mathcal{H}\, \rho +
\mathcal{L}\, \rho \ee where we have separated the evolution into
two parts, a Hamiltonian part, \be \mathcal{H}\rho  := -i[H,
\rho],
\ee corresponding to the {\it ideal unitary process} in question,
and a noise part described by some Liouvillian superoperator
$\mathcal{L}$. The formal solution of this master equation can be
written as \be \label{formalsolutionME} \rho(t)=e^{({\mathcal
H+{\mathcal L}})t} \rho(0), \ee which can, for small $t=\delta t$,
be Taylor expanded
\be \label{expansionME}
\rho(\delta t) \approx & & \rho(0) \nonumber \\ &+& \delta t({\mathcal H}+{\mathcal L})\rho(0) \nonumber \\
&+& \frac{\delta t^2}{2} ({\mathcal H}^2+{\mathcal L}^2+\mathcal {LH+HL})\rho(0) \nonumber \\
&+& O(\delta t^3). \ee

In the following, we will often assume a coupling of the
individual particles to independent thermal reservoirs, that is
noise acts independently on particles. The Liouvillian ${\cal L}$
is in this case given by a sum of terms corresponding to different
particles, \be {\cal L}=\sum_\alpha{\cal L}^{(\alpha)}.
\ee For qubits, we describe the coupling of a single particle to a
thermal reservoir by a quantum-optical noise term \cite{Br93}, \be
\label{Lindblad}
{\cal L}^{(\alpha)}\rho &=& -\frac{B}{2}(1-s)[\sigma_+^{(\alpha)}\sigma_-^{(\alpha)}\rho + \rho\sigma_+^{(\alpha)}\sigma_-^{(\alpha)} - 2 \sigma_-^{(\alpha)}\rho\sigma_+^{(\alpha)}] \nonumber\\
&& -\frac{B}{2} s [\sigma_-^{(\alpha)}\sigma_+^{(\alpha)}\rho + \rho\sigma_-^{(\alpha)}\sigma_+^{(\alpha)} - 2 \sigma_+^{(\alpha)}\rho\sigma_-^{(\alpha)}] \nonumber\\
&& -\frac{2C-B}{8} [2\rho -2
\sigma_z^{(\alpha)}\rho\sigma_z^{(\alpha)}], \ee with
$\sigma_\pm^{(\alpha)} \equiv 1/2(\sigma_x^{(\alpha)} \pm i
\sigma_y^{(\alpha)})$ act on particle $\alpha$ and $2C\geq B$.
While parameters $B,C$ give the decay rate of inversion and
polarization, $s\in [0,1]$ depends on the temperature $T$ of the
bath. More precisely $s= \text{lim}_{t\rightarrow \infty} \langle
\frac{\mathbf{1}+\sigma_z}{2}\rangle_t$, where $s=1/2$ corresponds
to $T=\infty$. For $B=0$ and $C=\gamma$, this corresponds to
dephasing noise, while for $s=1/2$ and $B=C\equiv \kappa$, we
obtain white noise, corresponding to a depolarizing channel.
Without interaction, i.e. for $H=0$, the corresponding time
dependent completely positive maps obtained by integrating the
master equation are given by a dephasing channel, \be
\label{dephasing} {\cal D}^{(\alpha)}\rho= p(t)\rho +
\frac{1-p(t)}{2}\left( \rho + \sigma_3^{(\alpha)} \rho
\sigma_3^{(\alpha)} \right)\hspace{0.2cm}, \ee with $p(t) =
e^{-\gamma t}$ or a depolarizing channel, \be \label{Depol} {\cal
M}^{(\alpha)}\rho= p(t)\rho + \frac{1-p(t)}{4} \sum_{j=0}^3
\sigma_j^{(\alpha)} \rho \sigma_j^{(\alpha)}, \ee with $p(t) =
e^{-\kappa t}$ respectively.

\subsection{Local noise equivalent}\label{subsec:LNE}

We also use a third quantity to determine the quality of the
simulation process, which is the {\em local noise equivalent}
(LNE). Here, the noisy process described by ${\cal E}_t$ is
compared with a (virtual) noisy process $\tilde {\cal E}_t$ that
corresponds to an evolution with respect to a superoperator given
by \be {\cal H} + \sum_{\alpha=1}^{m}{\cal L}^{(\alpha)} \ee for
time $t$, where the Hamiltonain part ${\cal H}\rho = -i[H,\rho]$
corresponds to the ideal evolution, while the noise part is
described by individual ({\em local}) coupling of qubits to
thermal baths (see Eq. \ref{Lindblad}). In particular, we consider
the cases of dephasing noise ($B=0$, $C=\gamma$) and white noise
($s=1/2$, $B=C\equiv \kappa$). The parameter $\gamma$ [$\kappa$]
determines the strength of coupling of each qubit to the
reservoir, and is adjusted in such a way that the processes ${\cal
E}_t$ and $\tilde {\cal E}_t = \exp[({\cal H} +
\sum_{\alpha=1}^{m}{\cal L}^{(\alpha)})t]$ lead to the same
Jamio{\l}kowski fidelity with respect to the ideal unitary
evolution $U_t$, i.e. \be \label{LNE} F({\cal E}_t,U_t) =
F(\tilde{\cal E}_t,U_t) \ee

The coupling strength $\gamma$ [$\kappa$] for which Eq. \ref{LNE}
is fulfilled hence give the local noise equivalent, i.e. the
amount of local noise that would lead to same fidelity after an
evolution for time $t$. This provides a suitable measure for the
quality of a noisy evolution. We remark that typically ${\cal
E}_t$ and $\tilde {\cal E}_t$ will not be identical. Nevertheless,
if the initial process containing two--body interactions is
described by single qubit dephasing/depolarizing noise with some
parameter $\gamma_0$ [$\kappa_0$], a LNE $\gamma >\gamma_0 [\kappa
> \kappa_0]$ can be considered as increase of noise level, where
it is implicitly assumed by such a comparison that generation of a
many--body interaction is as difficult as the generation of a
two--body interaction (which might, however, not be the case). As
we will see in the following, some methods typically lead to a LNE
which is much larger than the initial $\gamma_0,\kappa_0$.

In certain cases, the LNE can be determined analytically from the
Jamio{\l}kowski fidelity. Consider e.g. the case where the ideal
(desired) evolution is given by a Hamiltonian $H=\sigma_z^{\otimes
m}$, and we consider dephasing noise on individual qubits
described by a Liouvillian ${\cal L}^{(\alpha)}\rho=
-\gamma/2(\rho - \sigma_z^{(\alpha)}\rho\sigma_z^{\alpha)})$. In
this case, the Jamio{\l}kowski fidelity of the process
$\tilde{\cal E}_t = \exp[({\cal H} + \sum_\alpha{\cal
L}^{(\alpha)})t]=\exp({\cal H}t)\exp(\sum_\alpha{\cal
L}^{(\alpha)}t)$ is given by \be \label{FidtoLNE} F(\tilde{\cal
E}_t,U_t)= F(\otimes_\alpha {\cal D}^{(\alpha)},\eins) =
\left(\frac{1+p(t)}{2}\right)^m,
\ee where ${\cal D}^{(\alpha)}$ is the dephasing channel defined in
Eq. \ref{dephasing} with $p(t)=\exp(-\gamma t)$, and we used that
Hamiltonian and noise parts commute. The relation of the LNE to
the fidelity $F \equiv F({\cal E}_t,U_t)$ of a process is in this
case given by \be \gamma=-\frac{\ln(2F^{1/m}-1)}{t}. \ee For other
Hamiltonians, and depolarizing noise, where Hamiltonian and noise
parts do not commute, it is often difficult to solve the
corresponding master equation analytically. However, the LNE can
be easily determined numerically also in these cases.

\section{Noise model 2: Timing errors}\label{Noisemodel2}

As we have discussed, all of the Hamiltonian simulation methods in
this article can be thought of as being implemented via control
sequences of Hamiltonian evolutions. In these protocols we
manipulate the amount of time that a Hamiltonian evolution is
induced on a system. In this section we examine the effect of
inaccuracies in this evolution time.

Say we wish to physically implement the unitary operation
$U=e^{-iH t}$ and we give ourselves control over the amount of
time that the evolution occurs. In reality, we can only control
$t$ with finite precision. In a single instance of a controlled
evolution by $H$ for some time $t$, we assume that there will be
some Gaussian distributed random error $\delta$ with mean 0. The
corresponding completely positive map is given by,
\begin{equation}
\label{eq:timerror} \mathcal{E}(\rho) =
\int_{-\infty}^{\infty}d\delta p(\delta) e^{-iH(t+\delta)}\rho
e^{iH(t+\delta)},
\end{equation}
where $p(\delta)$ is a Gaussian distribution.

Importantly, an error such as that described by Eq.
\ref{eq:timerror} can be thought of as two distinct, commuting
operations on the the state $\rho$. That is, we can define two
super-operators $\mathcal{U}_t(\rho) \equiv e^{-iH t}\rho e^{iHt}$
and $\mathcal{T}(\rho)\equiv \int_{-\infty}^{\infty}d\delta
p(\delta) e^{-iH\delta}\rho e^{iH\delta}=\int_{-\infty}^\infty
d\delta p(\delta) \mathcal{U}_\delta (\rho)$ such that,
\begin{equation}
\label{eq:supererrorfactor} \mathcal{E}(\rho) =
\mathcal{T}\circ\mathcal{H}(\rho) =
\mathcal{H}\circ\mathcal{T}(\rho).
\end{equation}
Noting this, we are free to think of the map given in Eq.
\ref{eq:timerror} as being comprised of some ``perfect'' unitary
operation, that is the operation $\mathcal{U}$, which is followed
by a noisy operation, $\mathcal{T}$.

\subsubsection{Unitary Hamiltonians}

An evolution according to Eq. \ref{eq:timerror} has a particularly
simple form in the special case where a Hamiltonian is of the form
$H=\omega H_\omega$, where $H_\omega^2=\eins$. This property
allows us to express the unitary evolution generated by such
Hamiltonians in the following simple form,
\begin{equation}
\label{eq:unithams} e^{-iHt} = \cos (\omega t)\eins - i\sin
(\omega t) H_\omega.
\end{equation}
There are a number of interesting Hamiltonians in this class. For
instance the two-qubit Ising interaction, as well as single-qubit Pauli
rotations.

Utilizing the identity in Eq. \ref{eq:unithams}, we can express
the map $\mathcal{T}(\rho)$ as,
\begin{eqnarray}
\mathcal{T}(\rho) & = & \int^\infty_{-\infty} d\delta
p(\delta)\big[\cos^2(\omega\delta)\eins\rho \eins + \sin^2(\omega \delta)H_\omega \rho H_\omega \nonumber \\
&  & -i\sin(\omega \delta)\cos(\omega \delta)(H_\omega \rho \eins
-\eins \rho H_\omega)\big].
\end{eqnarray}
As $p(\delta)$ is an even function of $\delta$ and the product
$\sin(\omega \delta)\cos(\omega \delta)$ is an odd function, we
know that,
\begin{equation}
\int_{-\infty}^\infty p(\delta)\sin(\omega \delta)\cos(\omega
\delta) = 0.
\end{equation}
Thus we find,
\begin{equation}
\label{eq:integralformtimeerror} \mathcal{T}(\rho) =
\int^\infty_{-\infty} d\delta
p(\delta)\cos^2(\omega\delta)\eins\rho \eins + \sin^2(\omega
\delta)H_\omega \rho H_\omega.
\end{equation}
As we have assumed that the timing error is Gaussian distributed
around zero we can write the probability distribution $p(\delta)$
in the following way,
\begin{equation}
\label{eq:Gaussdist} p(\delta) =
\frac{1}{\sigma\sqrt{2\pi}}e^{-\frac{\delta^2}{2\sigma^2}},
\end{equation}
where $\sigma$ is the standard deviation of the probability
distribution. We can find an analytic expression for
$\mathcal{T}(\rho)$ by inserting Eq. \ref{eq:Gaussdist} into Eq.
\ref{eq:integralformtimeerror}, using some standard trigonometric
identities, and noting the following integral identity,
\begin{equation}
\int_{-\infty}^\infty dx e^{ax^2}\cos (bx) =
\sqrt{\frac{\pi}{4a}}e^{-\frac{b^2}{4a}}.
\end{equation}
Doing this we find that the completely positive map associated
with a random Gaussian error in the control time of a Hamiltonian
is given by,
\begin{equation}
\label{eq:noisyerrorlong} \mathcal{T} (\rho) =
\frac{1}{2}(q+1)\eins\rho \eins +\frac{1}{2}(1-q)H_\omega \rho
H_\omega
\end{equation}
where $q (\sigma,\omega)=e^{-2\sigma^2\omega^2}$. Equation
\ref{eq:noisyerrorlong} can be re-written as,
\begin{equation}
\label{eq:noisyerrorshort} \mathcal{T} (\rho) = q \rho  +
\frac{(1-q)}{2}(\rho + H_\omega \rho H_\omega).
\end{equation}
which has a similar form to that previously seen for the
de-phasing channel.

The corresponding Jamio{\l}kowski fidelity a for random timing
error on a unitary Hamiltonian would be,
\begin{equation}
\label{eq:unitaryTerror}
F(\mathcal{H},\mathcal{E})=F(\eins,\mathcal{T})=\frac{(1+q)}{2}.
\end{equation}
The corresponding Jamio{\l}kowski distance is
\begin{equation}
D(\mathcal{H},\mathcal{E})= \sqrt{\frac{(1-q)}{2}}.
\end{equation}

\textbf{Example:} Consider the two-qubit Ising Hamiltonian, $H =
\omega \sz\otimes \sz$. If we were to evolve a two-qubit system by
$H$ to a system of two qubits for some time $t$, then the map
associated with a random timing error on the evolution time would
take the form:
\begin{equation}
\label{eq:2qubitisingtimeerror}
 \mathcal{T} (\rho) = q \eins\rho \eins +
\frac{(1-q)}{2} (\rho + \sz\otimes \sz\rho \sz\otimes \sz).
\end{equation}
Such a map can be thought of as a collective dephasing channel. We
shall use this example frequently in the remainder of this paper.

\subsubsection{Sums of commuting unitary Hamiltonians}
\label{subsec:timingerrorssumofunitary}

We can extend the results of the previous section to a wider
variety of Hamiltonians. Consider the following Hamiltonian,
\begin{equation}
H = \sum_{j=1}^n \omega_j H_j,
\end{equation}
where each $H_j$ is a unitary Hamiltonian, $\omega_j$ is  real,
and any two $H_j$ commute. As each of the $H_j$ commute, an
evolution according to such a Hamiltonian can be factorized into a
sequence of $n$ unitary evolutions,
\begin{equation}
e^{-iHt}\rho e^{iHt}=e^{-i\omega_1H_1t}...e^{-i\omega_nH_nt}\rho
e^{i\omega_nH_nt}...e^{i\omega_1H_1t},
\end{equation}
or in terms of super-operators,
\begin{equation}
\mathcal{U}_t(\rho) = \mathcal{U}^1_t\circ ...\circ
\mathcal{U}^n_t(\rho).
\end{equation}
The effect of a timing error on such an evolution can be analyzed
in a fashion similar unitary Hamiltonians that we examined in the
previous subsection. We can describe an evolution of $H$ with a
timing error using the completely positive map $\mathcal{E}$ that
was defined in Eq. \ref{eq:timerror} in the previous section. Like
we saw in the previous section, we find that because all of the
$H_j$ commute $\mathcal{E}= \mathcal{U}_t\circ\mathcal{T}_H=
\mathcal{T}_H\circ\mathcal{U}_t$. Hence in order to understand the
effects of the timing error we only need to consider the nature of
the operator $\mathcal{T}_H$.

In terms of superoperators, $\mathcal{T}_H$ can be written,
\begin{equation}
\label{eq:collectiveEH} \mathcal{T}_H(\rho) =
\int_{-\infty}^\infty p(\delta)\mathcal{U}_\delta^1\circ ...\circ
\mathcal{U}_\delta^n(\rho),
\end{equation}
where the order the sequence of $\mathcal{U}_t^j$ is irrelevant.
Note that the integral in the expression for $\mathcal{T}_H$ is
only over a single variable, $\delta$. As we did in the previous
section, we can expand each of the $\mathcal{U}_\delta^j$ terms in
$\mathcal{T}_H$ using Eq. \ref{eq:unithams} and integrating out
odd terms in the expansion. That is,
\begin{eqnarray}
\label{Eq:nqubittimeing} \mathcal{T}_H(\rho) &=&
\int_{-\infty}^\infty p(\delta)\mathcal{U}_\delta^1\circ ...\circ
\mathcal{U}_\delta^{n-1}\nonumber\\
& & (\cos^2(\omega_n \delta)\eins\rho \eins + \sin^2(\omega_n
\delta)H_n \rho H_n).
\end{eqnarray}
Some algebra demonstrates that if each $\mathcal{U}_\delta^j$ is
expanded in this way, then the resulting expression becomes quite
complicated and writing it here would be unhelpful. Generally, the
expression for $\mathcal{T}_H$ involves a sum of all products of
the unitary Hamiltonians $H_j$. The nature of such an expression
depends greatly on what each of the Hamiltonians $H_j$ actually
are. This is best illustrated by an example.

\textbf{Example:} Imagine we wish to apply an Ising Hamiltonian to
3 qubits of the form:
\begin{equation}
H = \omega \sz^1 \sz^2 + \omega \sz^1 \sz^3,
\end{equation}
for a time $t$. The resulting timing error operation can be
written,
\begin{equation}
\mathcal{T}_H(\rho) = \int_{-\infty}^\infty d\delta
p(\delta)(e^{-i\omega \delta \sz^1\sz^2} e^{-i\omega \delta
\sz^1\sz^3}\rho e^{i\omega \delta \sz^1\sz^2}e^{i\omega \delta
\sz^1\sz^3}).
\end{equation}

This can be expanded in the fashion described above,
\begin{eqnarray}
\label{eq:collective3qisingerror} \mathcal{T}_H(\rho) &=&
\int_{-\infty}^\infty d\delta p(\delta)\big[\cos^4(\omega
\delta)\eins\rho \eins \nonumber \\&& + \cos^2(\omega
\delta)\sin^2(\omega \delta)(\sz^1\sz^2\rho \sz^1\sz^2 +
\sz^1\sz^3\rho \sz^1\sz^3 )\nonumber
\\ && + \sin^4(\omega \delta)\sz^2\sz^3 \rho \sz^2\sz^3\big].
\end{eqnarray}
This expression can be simplified to,
\begin{eqnarray}
\mathcal{T}_H (\rho)& = & g \eins\rho \eins+ (1-g)(\sz^1\sz^2\rho
\sz^1\sz^2+ \sz^2\sz^3\rho \sz^2\sz^3 \nonumber \\ & &+
(1-g)^2\sz^1\sz^3\rho \sz^1\sz^3,
\end{eqnarray}
where
\begin{equation}
g = \int_{-\infty}^\infty d\delta p(\delta)\cos^4(\omega \delta).
\end{equation}
Assuming that $p(\delta)$ is a Gaussian distribution with mean 0
it can be written as it is in Equation \ref{eq:Gaussdist} and we
find an analytic expression for $g$,
\begin{equation}
g = \frac{3}{2^3}
+\frac{1}{2}e^{-2\sigma^2\omega^2}+\frac{1}{2^3}(e^{-2\sigma^2\omega^2})^4.
\end{equation}
In this example, the Jamio{\l}kowski fidelity is simply,
$F(\mathcal{H},\mathcal{E})=g$.

From this example we see for any $H$ that is a sum of commuting
unitary Hamiltonians that it is possible to derive a lower bound
on the Jamio{\l}kowski fidelity given, a random timing error.
Expanding $\mathcal{T}_H$ we find that there is always a term with
the following form:
\begin{equation}
I_H=\int_{-\infty}^\infty d\delta p(\delta)
\prod_{j=1}^n\cos^2(\omega_j \delta) \eins^{\otimes n}\rho
\eins^{\otimes n}.
\end{equation}
This means that the Jamio{\l}kowski fidelty must be at least,
\begin{equation}
F(\mathcal{H},\mathcal{E})\geq \int_{-\infty}^\infty d\delta
p(\delta) \prod_{j=1}^n\cos^{2n}(\omega_j \delta).
\end{equation}
If each $\omega_j = \omega$ and $p(\delta)$ is a Gaussian, then it
is possible to find an analytic form for this lower bound by using
basic trigonometric identities,
\begin{equation}
F(\mathcal{H},\mathcal{E})\geq \frac{1}{2^{2n}}{2n\choose n} +
\frac{1}{2^{2n-1}}\sum_{k=0}^{n-1}{2n\choose
k}e^{-2(n-k)^2\sigma^2\omega^2}.
\end{equation}
As our three-qubit Ising model example demonstrates, this lower
bound on the fidelity can be saturated. Saturation of this
inequality occurs when no product of the Hamiltonians $H_j$ are
equal to the identity. That is, the only term in ${\cal T}_H$
which gives a superoperator proportional to $\eins$ is $I_H$.
There are a many Hamiltonians that have this property. Two
examples include 1-D Ising chains with open boundary conditions
and Ising interaction Hamiltonians whose interaction pattern forms
a star graph.

The Hamiltonian $H=\sum_{j=1}^n \omega_j H_j$ can be simulated by
a sequence of $n$ independent evolutions of the Hamiltonians
$H_j$. If each of these evolutions had a timing error then Eq.
\ref{eq:collectiveEH} would have $n$ independent $\delta_j$ terms
over which one would integrate. This is equivalent to applying $n$
different error maps like that given in Eq.
\ref{eq:noisyerrorshort}. That is the timing error map would be
given by a product of maps like ${\cal T}_{H_1}...{\cal T}_{H_n}$.
The fidelity of a timing error in such a case is then lower
bounded by
\begin{equation}
F(\mathcal{H},\mathcal{E})= F(\eins, {\cal T}_{H_1}...{\cal
T}_{H_n})\geq \prod_{j=1}^n\left(\frac{1+q_j}{2}\right).
\end{equation}
This bound on the fidelity is saturated for the same $H$ as the
bound derived for ${\cal T}_H$ above.

If we consider the situation where the above fidelity bounds are
saturated, we can compare the effect of timing errors on an
evolution according $H$ alone versus the effect of simulating $H =
\sum_{j=1}^{n} \omega H_j$ by $n$ evolutions according to its
constituent Hamiltonians $H_j$. For simplicity, we consider the
case where $\omega_j = \omega$ and  the standard deviation in the
timing error for implementing $H$ is the same as for implementing
each of it's constituent terms $H_j$. The resulting fidelities are
given by the following integrals;
\be
F(\eins, {\cal T}_H)=\int_{-\infty}^\infty d\delta p(\delta)
\cos^{2n}(\omega \delta)
\ee
for the case where we implement $H$ directly; and
\be
F(\eins, {\cal T}_{H_1}...{\cal
T}_{H_n})=\left(\int_{-\infty}^\infty d\delta p(\delta)
\cos^{2}(\omega \delta)\right)^n
\ee
for the case where $H$ is implemented by evolutions according to
the Hamiltonians $H_j$. We have reverted to writing these
expressions in their integral form because in this form it is
simpler to see how to apply H{\"o}lder's inequality to them.
Recall H{\"o}lder's inequality,
\be
\int_a^b|\zeta(x)\eta(x)| dx\leq
\left(\int_a^b|\zeta(x)|^rdx\right)^{\frac{1}{r}}\left(\int_a^b|\eta(x)|^wdx\right)^{\frac{1}{w}}
\nonumber \\
\ee
where $\zeta(x)$, $\eta(x)$ are integrable complex functions and
$r,w>1$ satisfying
\be
\frac{1}{r}+\frac{1}{w}=1.
\ee
We set $\zeta(\delta) = p(\delta)^{\frac{n-1}{n}}$, $\eta(\delta)
= p(\delta)^{\frac{1}{n}}\cos^2(\omega\delta)$, $r=\frac{n}{n-1}$
and $w=n$ and substitute these values into H{\"o}lder's
inequality\footnote{M.B. would like to thank Otfried G{\"u}hne for
pointing out this choice of variables.}. Then, noting that
$p(\delta)$ and $\cos^2(\omega \delta)$ are both always positive
we find that
\be
\left(\int_{-\infty}^\infty d\delta p(\delta) \cos^{2}(\omega
\delta)\right)^n\leq\int_{-\infty}^\infty d\delta p(\delta)
\cos^{2n}(\omega \delta).\nonumber \\
\ee
Thus we find that for $\delta =\delta_j$ and $\omega_j=\omega$ for
all $j$
\be
F(\eins, {\cal T}_H)\geq F(\eins, {\cal T}_{H_1}...{\cal
T}_{H_n}).
\ee

\section{Commutator method}\label{commutatorNOISE}

We will now analyze the influence of noise, described by some
master equation of Lindblad form, on the simulation of many--body
interaction Hamiltonians generated by using the commutator method.
The commutator method is based on a sequential application of
evolutions for times $\delta t$ with respect to different
Hamiltonians in such a way that, when performing a Taylor
expansion in $\delta t$, all first order terms vanish and the
second order terms --that include the commutator of the initial
Hamiltonians and hence many--body interaction terms-- remain. The
analysis is usually done at the level of operators (see Sec.
\ref{commutator}), where one obtains (up to higher order
corrections) an effective Hamiltonian given by $H_{\rm
eff}=-i/2[H_1,H_2]$ which appears in second order $\delta t$. One
can formally perform the analysis also at the level of
superoperators, as is required when considering noisy interactions
described by master equations.

\subsection{Three--body interactions}

We consider a sequence of four operations, each applied for time
$\delta t$, where we denote the Hamiltonian part of the $j^{\rm
th}$ operation by $\mathcal{H}_j$, and the Liouvillian part by
${\mathcal L}_j$. An initial state $\rho(0)$ has evolved under
this sequence of operations after a total time $t=4\delta t$ as
follows
\be \label{MEnoise}
\rho(t)=e^{({\mathcal H}_4+{\mathcal L}_4)\delta t}e^{({\mathcal H}_3+{\mathcal L}_3)\delta t}e^{({\mathcal H}_2+{\mathcal L}_2)\delta t}e^{({\mathcal H}_1+{\mathcal L}_1)\delta t}\rho(0).\nonumber\\
\ee

For small $\delta t$, Eq. \ref{MEnoise} can be Taylor expanded
using Eq. \ref{expansionME}, and one obtains
\be \label{Sop}
\rho(t)=& & \rho(0)+ \delta t \left(\sum_{j=1}^4 {\mathcal H}_j + \sum_{j=1}^4 {\mathcal L}_j\right)\rho(0) \nonumber\\
&+& \frac{\delta t^2}{2} \left( \sum_{j=1}^4 {\mathcal H}_j^2 + 2\sum_{l=2}^4 \sum_{j=1}^{l-1}{\mathcal H}_l{\mathcal H}_j \right)\rho(0)   \nonumber\\
&+& \frac{\delta t^2}{2} \sum_{l=2}^4 \sum_{j=1}^{l-1}2({\mathcal H}_l{\mathcal L}_j+{\mathcal L}_l{\mathcal H}_j+{\mathcal L}_l{\mathcal L}_j)\rho(0) \nonumber\\
&+& \frac{\delta t^2}{2} \sum_{j=1}^4 ({\mathcal L}_j^2+{\mathcal H}_j{\mathcal L}_j+{\mathcal L}_j{\mathcal H}_j)\rho(0) \nonumber\\
&+& O(\delta t^3)
\ee
As for ideal evolutions, we consider the case where ${\mathcal
H}_4 = -{\mathcal H}_2$, ${\mathcal H}_3=-{\mathcal H}_1$, i.e.
$H_4=-H_2, H_3=-H_1$. In this case, $\sum_{j=1}^4 {\mathcal H}_j
\rho(0)=0$, and the second line of Eq. \ref{Sop} simplifies to $2
{\delta t^2} \tilde{\mathcal H} \rho(0)$ with $\tilde {\mathcal
H}\rho=-i[\tilde H,\rho]$ and $\tilde H= \frac{-i}{2}[H_1,H_2]$.
Also some of the terms in lines 3 and 4 of Eq. \ref{Sop} are
simplified. Note, however, that the term $\delta t\sum_{j=1}^4
{\mathcal L}_j\rho(0)$ --which is first order in $\delta t$--
remains. While ${\mathcal H}_j$ correspond to coherent processes
that can annihilate each other, ${\mathcal L}_j$ correspond to
{\em incoherent} noise processes, where no such interference is
possible. This implies that noise appears in first order $\delta
t$, while the (desired) Hamiltonian part only appears in second
order, $\delta t^2$. We denote additional noise terms of lines 3
and 4 that appear in second order, $\delta t^2$, in Eq. \ref{Sop}
by $\frac{\delta t^2}{2}{\cal L}'$, and use the short-hand
notation $\bar{\mathcal L}= \sum_{j=1}^4 {\mathcal L}_j$. The
total time evolution can hence be written as
\be
\rho(t)&=&\rho(0) + 2 \delta t^2 \tilde {\mathcal H} + 2 \delta t^2\left(\frac{\bar{\mathcal L}}{2\delta t} + \frac{\cal L'}{4}\right) + O(\delta t^3)\nonumber \\
&=& \exp \left[ 2\delta t^2 \left(\tilde{\mathcal H} + \frac{\bar{\cal L}}{2\delta t} + \frac{\cal L'}{4}\right)\right]\rho(0) + O(\delta t^3) \nonumber \\
&=& \exp \left[ \delta t' \left(\tilde{\mathcal H} + \frac{\bar{\cal L}}{\sqrt{2\delta t'}} + \frac{\cal L'}{4}\right)\right]\rho(0) + O(\delta t'^{3/2}). \nonumber\\
\ee
where we have introduced, as in the noiseless case, a dilated time
$\delta t'=2\delta t^2$. That is, when applying the sequence of
evolutions specified in Eq. \ref{MEnoise} for a total time of
$4\delta t$, this is equivalent (up to higher order corrections),
to an evolution with respect to a Liouvillian superoperator \be
\label{effectiveL} \tilde{\mathcal H} + \frac{\bar{\cal
L}}{\sqrt{2\delta t'}} + \frac{\cal L'}{4} \ee for a time $\delta
t'=2\delta t^2$. Note that the Hamiltonian part of the evolution
is given by $\tilde H=\frac{1}{2}[H_1,H_2]$ which provides the
desired many--body interaction term, but the noise part of the
evolution, dominated by $\bar{\cal L}/\sqrt{2\delta t'}$ is
enhance by a factor of $(2\delta t')^{-1/2}$, or equivalently
$(2\delta t)^{-1}$, as compared to the two--body evolutions. This
is a significant increase of noise level, which clearly limits the
applicability of the commutator method. In particular, the
increase of noise level becomes {\em larger} for smaller times
$\delta t$, while on the other hand smaller $\delta t$ increase
the precision with which the desired many--body interaction terms
are generated. Recall that the commutator method is based on
validity of Taylor expansion, which requires small $\delta t$ and
precision increases with decreasing  $\delta t$. In other words,
the dilation factor in simulation time given by the time cost,
translates in the case of noisy interactions directly into an
increase of noise level.

{\bf Example:} We consider in the following a simple example that
illustrates the increase of noise level. We consider the case
where, independent of the interaction Hamiltonian, the Liouvillian
describing the noise part of the evolution corresponds to
single--qubit dephasing noise, i.e. is given by
\be
\label{Liovphasenoise}\sum_\alpha {\cal L}^{(\alpha)}\rho=
\sum_\alpha
\gamma_0/2(\rho-\sigma_z^{(\alpha)}\rho\sigma_z^{(\alpha)}),
\ee
for all particles $\alpha$ that are involved in an interaction.
For Hamiltonians $H_1,H_2$ given in Eq. \ref{Hamiltonians12} and
$H_3=-H_1,H_4=-H_2$, the effective evolution resulting from the
sequence of evolutions specified in Eq. \ref{MEnoise} corresponds
--in the ideal case-- to an evolution with respect to the
effective three--body Hamiltonian
\be H_{\rm eff}=\sigma_z^{\otimes 3}
\ee
for time $\delta t'=2\delta t^2$. When $H_1$ is applied the
corresponding Liouvillian is given by ${\cal L}_1={\cal
L}^{(A)}+{\cal L}^{(B)}$. If $\delta t\ll 1$ and $\gamma_0 << 1$
then the evolution ${\cal H}_1+{\cal L}_1$ corresponds to an
evolution according to the Hamiltonian $H_1$ and a single-qubit
dephasing channels acting on qubits $A$ and $B$. The resulting
fidelity is given by
\be
F_1 = \left( \frac{1+p}{2}\right)^2,
\ee
where $p=e^{-\gamma_0 \delta t}$.  Assuming the same fidelities
for each of the two-qubit operations used to simulate $H_{\rm eff}$ we
can use the simple fidelity bound derived in the previous section,
Eq. \ref{eq:simpleFbound}, to give a lower bound on the fidelity
of the simulation
\be
F \geq 1 - 16\gamma_0 \delta t.
\ee
Note that this bound is only valid for small $\delta t$ and
$\gamma_0$. In terms of the effective simulation time $\delta t'$
it reads,
\be
F \geq 1-\frac{16}{\sqrt{2}}\gamma_0\sqrt{\delta t'}.
\ee
We improve on this bound below by directly calculating the
fidelity under similar approximations.

The effective Liouvillian ${\cal L}_{\rm eff}$ describing the
noise part of this simulation is dominated by the term $\frac{\bar
{\cal L}}{\sqrt{2\delta t'}}=\frac{\sum_k {\cal
L}_k}{\sqrt{2\delta t'}}$, where ${\cal L}_1 = {\cal L}_3 ={\cal
L}^{(A)}+{\cal L}^{(B)}, {\cal L}_2={\cal L}_4={\cal
L}^{(B)}+{\cal L}^{(C)}$ as only qubits $AB$ or $BC$ are involved
in a particular operation. We thus obtain \be \label{Leffective}
{\cal L}_{\rm eff} \approx \frac{2{\cal L}^{(A)}+2{\cal
L}^{(C)}+4{\cal L}^{(B)}}{\sqrt{2\delta t'}}. \ee In principle,
the sequence of operations leading to the effective Hamiltonian
$H_{\rm eff}$ can be symmetrized by permuting the roles of qubits
$A,B,C$, either probabilistically or sequentially, which leads to
symmetrized noise in the effective Liouvillian \be {\cal L}_{\rm
eff,sym} \approx \frac{8/3 ({\cal L}^{(A)}+{\cal L}^{(B)}+{\cal
L}^{(C)})}{\sqrt{2\delta t'}}, \ee where the factor $8/3$ results
from $(2 \times 2 + 1 \times 4)/3$, i.e. in two out of three cases
we obtain a factor 2 in front of the Liouvillian term (role of
particles $A,C$), while in 1 out of 3 cases there is a factor 4
(role of particle $B$) -- see Eq. \ref{Leffective}. The LNE with
respect to dephasing noise can be directly read off from above
equation and is (approximately) given by \be \gamma \approx
\frac{8}{3\sqrt{2 \delta t'}}\gamma_0, \ee where the expressions
are only valid for $\delta t' \ll 1$. The Jamio{\l}kowski fidelity
can be estimated via Eq. \ref{FidtoLNE}, and one obtains \be F
\approx \left(\frac{1+e^{-\sqrt{\delta
t'}8\gamma_0/(3\sqrt{2})}}{2}\right)^3 \approx 1- 2\sqrt{2\delta
t'}\gamma_0. \ee
We remark that dephasing noise is in many cases not an appropriate
model, and also the assumption that dephasing noise occurs
independent of the Hamiltonian might often not be fulfilled.
Consider e.g. the case where a basic two--body Hamiltonian
$H=\sigma_z\sigma_z$ is manipulated by means of local unitary
operations to simulate a new effective two--body Hamiltonian, e.g.
$H'=\sigma_x\sigma_x$. In this case, the dephasing noise is also
transformed, and one obtains bit--flip noise for the new effective
Hamiltonian instead.

A similar calculation can be performed for depolarizing (white)
noise, i.e. the Liouvillian describing the noise part is a sum of
local terms, $\sum_\alpha {\cal L}^{(\alpha)}$, with ${\cal
L}^{(\alpha)}$ given by Eq. \ref{Lindblad} with  $s=1/2$ and
$B=C\equiv \kappa$. Again, one finds \be \label{LNEcommutatorest}
\kappa \approx \frac{8}{3\sqrt{2 \delta t'}}\kappa_0, \ee for the
LNE, while for the fidelity we can simply bound it in the limit of
small $\kappa$ and $\delta t$ using Eq. \ref{eq:simpleFbound} by
\be
F\geq 1- \frac{24}{\sqrt{2}}\kappa_0\sqrt{\delta t'}
\ee
and we can explicitly estimate the fidelity in the same limits to
be
\be \label{Fcommutatorest} F \approx
\left(\frac{1+3e^{-\sqrt{\delta
t'}\kappa_0/(8\sqrt{2})}}{4}\right)^3 \approx 1-3\sqrt{2\delta t'}
\kappa_0. \ee

\begin{figure}[ht]
\begin{picture}(230,200)
\put(0,0){\epsfxsize=230pt\epsffile[87   262   507
578]{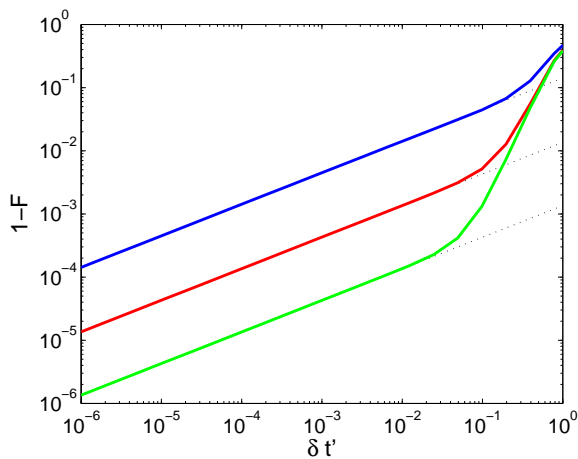}}
\end{picture}
\caption[]{\label{FigCommutator_N3} (color online) Error of
three--qubit interaction ${\rm exp}(-i\delta t' \sigma_z^{\otimes
3})$ applied for time $\delta t'$ generated by commutator method.
Plot shows deviation from ideal evolution measured by the
fidelity, $1-F$, as a function of $\delta t'$. Curves from top to
bottom correspond to $p=0.9$ (blue), $p=0.99$ (red), $p=0.999$
(green), where $p=\exp(-\pi \kappa)$. Dotted lines are analytic
estimates of Eq. \ref{Fcommutatorest}, $1-F=3\sqrt{2 \delta t'}
\kappa_0$.}
\end{figure}

\begin{figure}[ht]
\begin{picture}(230,200)
\put(0,0){\epsfxsize=230pt\epsffile[87   262   507
578]{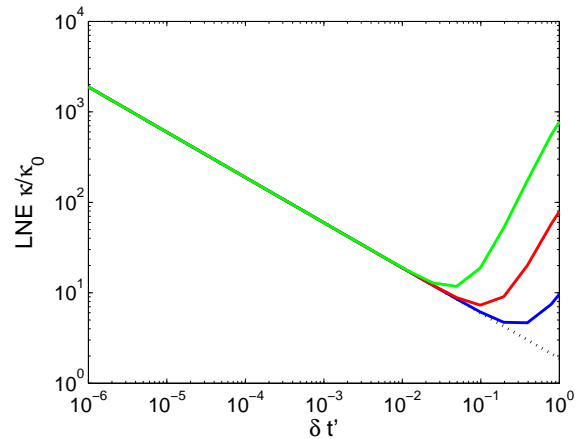}}
\end{picture}
\caption[]{\label{FigCommutator2_N3} (color online) Increase of
Local noise equivalent of three--qubit interaction ${\rm
exp}(-i\delta t' \sigma_z^{\otimes 3})$ applied for time $\delta
t'$ generated by commutator method. Plot shows
$\kappa/\kappa_{0}$, as a function of $\delta t'$. Curves from
bottom to top correspond to $p=0.9$ (blue), $p=0.99$ (red),
$p=0.999$ (green), where $p=\exp(-\pi \kappa)$. Dotted line is the
analytic estimate given in Eq. \ref{LNEcommutatorest},
$\kappa/\kappa_0 = \frac{8}{3\sqrt{2 \delta t'}}$.}
\end{figure}

Above calculations should be considered as estimates of LNE and
fidelity. A complete calculation involves also additional errors
due to Taylor expansion and noise terms occurring in order $\delta
t'$. We have thus performed numerical simulations taking all
errors into account, and considered the (often) more realistic
model of white noise (depolarizing channels). The results of these
simulations for the generation of an effective three--body
interaction are shown in Figs.
\ref{FigCommutator_N3},\ref{FigCommutator2_N3}. Plots correspond
to coupling strength $\kappa$ that would lead to single qubit
white noise channels with parameter $p=0.9; 0.99; 0.999$
respectively when applied for time $\delta t =\pi$ (see Eq.
\ref{Depol}). The results fully agree with the analytic estimates
(see Eqs. \ref{LNEcommutatorest},\ref{Fcommutatorest}) for small
$\delta t'$. For larger $\delta t'$, higher order corrections due
to Taylor expansion also play a role and further increase the
noise level (and hence increase the LNE and reduce the fidelity).
We remark that although for small $\delta t'$ the fidelity is
close to unity, the quality of the simulation is nevertheless very
low as the local noise equivalent is enhanced significantly.

\subsection{Timing errors}

In Section \ref{Noisemodel2} we derived CP maps that describe the
effect of random timing errors in the implementation of a
Hamiltonian evolution. As we saw in Section \ref{commutator} we
can simulate $H_{\rm eff} = \sz\otimes \sz \otimes \sz$ by a
control sequence involving the Hamiltonians $\pm H_1 = \pm
\sz^{(A)}\sigma_z^{(B)}$ and $\pm H_2 = \pm
\sigma_y^{(B)}\sz^{(C)}$. Both $H_1$ and $H_2$ are unitary and
Hermitian, thus they are their own inverses. In Section
\ref{Noisemodel2} (Eq. \ref{eq:unitaryTerror}) we explicitly
calculated the resulting fidelity of a timing error for such
Hamiltonians. Considering such a noise model and given that
$\delta t<< 1$ then we can use Eq. \ref{eq:simpleFbound} to
estimate a lower bound on the resulting fidelity. Assuming
additionally that the standard deviation, $\sigma$, of each timing
error is the same we find the following lower bound on the
fidelity of the simulation of $H_{\rm eff}$ to be
\be
F \geq 1-8(1-q)
\ee
where $q=e^{-2\sigma^2}$.

\subsection{Many--body interactions}

It is straightforward to perform a similar analysis for many--body
interactions involving more than three qubits. Essentially, one
finds that the time cost of simulating many--body Hamiltonians
using basic two--body interactions translates to an increase in
noise level. As shown in Sec. \ref{Manybodyinteractions}, the
desired $m$--body Hamiltonian appears in $\delta t_m=O(\delta
t^{(m-1)})$, where we have denoted by $\delta t_m$ the time for
which the effective $m$--body Hamiltonian is applied. The noise,
however, still appears in first order $\delta t$, or equivalently
$O(\delta t_m^{1/(m-1)})$. This corresponds to an increase of
noise level by a factor of order $O(\delta t^{-(m-2)})$, or
equivalently $O(\delta t_m^{-(m-2)/(m-1)})$. The total evolution
for time $\delta t_m$ is thus essentially governed by a
Liouvillian superoperator of the form \be \label{multibodyL}
\tilde{\cal H} + O(\delta t_m^{-(m-2)/(m-1)}) {\bar {\cal L}}, \ee
where $\tilde{\cal H}=-i[H_{\rm eff},\rho]$ and $H_{\rm eff}$ is
the desired $m$--body Hamiltonian, while ${\bar {\cal L}}$ is
given by an appropriate sum of Liouvillians describing noise in
two--body interactions involved in the process. This noise is
enhanced by a factor of $O(\delta t_m^{-(m-2)/(m-1)})$.

The LNE also increases accordingly, i.e. \be \gamma \approx
O(\delta t_m^{-(m-2)/(m-1)})\gamma_0, \ee while the fidelity is
approximately given by \be F \approx 1-m\gamma_0 \times O(\delta
t_m^{-(m-2)/(m-1)}). \ee In above expressions, we have again
assumed a simple model with only dephasing noise specified by a
coupling constant $\gamma_0$ [$\gamma$] respectively. For large
$m$, the dilation factor approaches $\delta t_m^{-1}$ and can be
significant. In particular, the resulting CPM may no longer be
capable of creating entanglement for all times.

We have also performed a full numerical analysis taking all errors
and imperfections into account. We find that qualitative scaling for small
simulation time $\delta t_m$ is essentially as described above.
For larger $\delta t_m$ additional errors due to Taylor expansion
and/or noise terms in higher order $\delta t$ further increase the
noise level.

\subsection{Simulation of strongly entangling multi--qubit gate}
\label{twoHams}

\begin{figure}[ht]
\begin{picture}(230,200)
\put(0,0){\epsfxsize=230pt\epsffile[87   262   507
578]{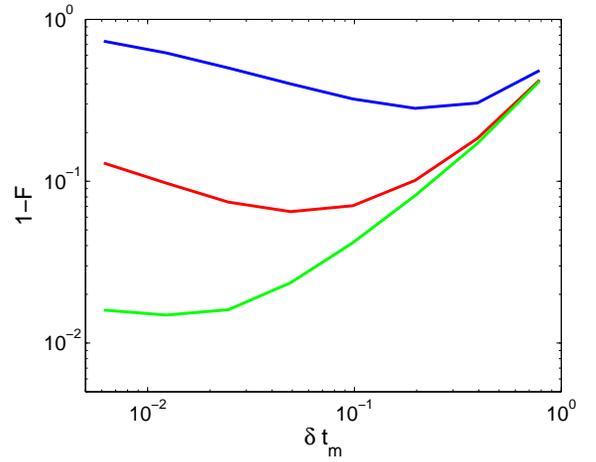}}
\end{picture}
\caption[]{\label{FigGateCommutator_N3} (color online) Error of
three--qubit gate ${\rm exp}(-i\pi/4 H)$ with
$H_2=\sigma_z^{(1)}\sigma_z^{(2)}\sigma_z^{(3)}+\sum_{k=1}^{3}\sigma_x^{(k)}$
generated by sequential applications of $\exp(-i \delta t_m \sigma_z^{(1)}\sigma_z^{(2)}\sigma_z^{(3)})$ and $\exp( - i\delta
t_m \sum_{k=1}^{3}\sigma_x^{(k)})$. The three--qubit interaction
is generated by the commutator method, and the figure shows the
derivation from ideal gate as measured by the fidelity, $1-F$,
plotted against different values of $\delta t_m$. Curves from top
to bottom correspond to $p=0.9$ (blue), $p=0.99$ (red), $p=0.999$
(green), where $p=\exp(-\pi \kappa)$. Notice that there is an optimal choice for $\delta t_m$ to maximize the gate fidelity (see main text for details).}
\end{figure}

Above considerations are concerned with the generation of a basic
$m$--body Hamiltonian of the form $H_{\rm eff}=\sigma_z^{\otimes m}$
for time $\delta t_m$. However, such a basic $m$--body Hamiltonian
only serves as a building block to generate --via standard
Hamiltonian simulation techniques-- other many--body Hamiltonians,
or multi--qubit gates. To illustrate the accuracy of such
simulation in the presence of noise, we consider two examples:
\begin{itemize}
\item[(i)] The simulation of the three--body Hamiltonian
$H_1=\sigma_z^{(1)}\sigma_z^{(2)}\sigma_z^{(3)}$ for time
$t=\pi/4$; \item[(ii)] The simulation of the three--body
Hamiltonian
$H_2=\sigma_z^{(1)}\sigma_z^{(2)}\sigma_z^{(3)}+\sum_{k=1}^{3}\sigma_x^{(k)}$
for time $t=\pi/4$;
\end{itemize}
Example (i) only contains the basic Hamiltonian, while example
(ii)  also includes single--qubit terms and the application of
Hamiltonian simulation techniques to generate sequentially the two
non--commuting terms $\sigma_z^{\otimes 3}$ and
$\sum_{k=1}^{3}\sigma_x^{(k)}$.
Again, we determine fidelity and LNE as a function of the time
$\delta t_m$ for which the basic two--body interaction Hamiltonian
is generated. Small times $\delta t_m$ require several application
(and additional intermediate local unitary operations) of the
basic 3--body Hamiltonian to achieve an evolution for time
$t=\pi/4$, while larger times $\delta t_m$ require fewer
applications but involve larger Taylor expansion errors in both
the generation of the basic 3--body Hamiltonian and the simulation
of the desired Hamiltonian $H$ from $H_{\rm eff}$. The results of
the numerical simulation for the Hamiltonian (ii) are shown in
Fig. \ref{FigGateCommutator_N3} for different values of the white
noise parameter $p$, where $p=\exp(-\pi \kappa)$. That is, the
corresponding coupling strength to the thermal bath $\kappa$ is
such that the evolution with respect to the noise part of the
Liouvillian for time $t=\pi$ leads to a depolarizing map (see Eq.
\ref{Depol}) with parameter $p$. One observes that there exists an
optimal time $\delta t_m$ that leads to a maximal gate fidelity,
and the optimal value of $\delta t_m$ depends on $p$. Recall that the LNE to generate the basic 3--body hamiltonian increases with decreasing $\delta t_m$ (see Fig. \ref{FigCommutator2_N3}), while $\delta t_m$ cannot be chosen too big to limit Taylor expansion errors. The optimal value of $\delta t_m$ is hence a compromise between the two competing requirements. The results
for simulation of the Hamiltonian (i) are very similar and are not
shown here.

\subsection{Simulation of quDits using qudits}\label{quDits_Commutator}

We have illustrated the commutator method for qubits, however the
results can without further effort be directly adopted to quDits.
When using pairwise interacting $d$--level systems to simulate an
arbitrary Hamiltonian of a single $D$--level system with $D > d$,
an effective $m$--body interaction Hamiltonian is required, where
\be m = \left\lceil \frac{\log D}{\log d} \right\rceil, \ee and
$\lceil x \rceil$ denotes the next largest integer. This
immediately follows from the fact that the Hilbert space of $m$
interacting qudits has dimension $d^m$, and $D \geq d^m$ is
required that the $m$ qudits can simulate a single quDit, i.e. a
$D$--dimensional system. The increase of noise level (dilation
factor), the LNE and the fidelity are given by exactly the same
expressions as in the case of qubits (see Eqs.
\ref{effectiveL},\ref{multibodyL}), where the definition of LNE
has to be adopted accordingly (local white noise of individual
qudits rather than qubits).

Similarly, the simulation of pairwise interacting $D$--level
systems requires $m$--body interactions of qudits, where \be m =
\left\lceil 2\frac{\log D}{\log d} \right\rceil. \ee The
simulation of many--body interactions of quDits, say $n$--body
interactions, requires effective $m$--body interactions of qudits
with $m = \left\lceil n \log D/\log d \right\rceil$.

\section{Unitary conjugation and graph state encoding}\label{GSENOISE}

We now turn to the second method to generate many--body
Hamiltonians, unitary conjugation and graph state encoding (see
Sec. \ref{GSE}). In this case, certain two--qubit gates $U$,
$U^{\dagger}$ (corresponding to a graph state encoding) are
applied before and after the evolution with respect to a single--
or two--body Hamiltonian, which leads to an effective evolution
with respect to a many--body Hamiltonian. The choice of the
encoding determines the many--body interaction that is generated.
In this method, errors occur due to
\begin{itemize}
\item[(i)] imperfect application of
two qubit gates $U,U^{\dagger}$
\item[(ii)] imperfect application of
single and two--body Hamiltonians.
\end{itemize}
While errors do not appear in
higher order $\delta t'$ as in the commutator method, they can
still be significantly amplified. In particular, any simulation
involves application of gates $U$ and $U^{\dagger}$ and hence any
error associated with these gates. As these gates require the
application of a basic Hamiltonian $H=\sigma_z\otimes\sigma_z$ for
time $t=\pi/4$, errors will be of order $\exp (-\kappa_0 \pi/4)$,
even if $\delta t' \ll 1$.

\subsection{Fixed graph state encoding}

We will first consider a fixed graph state encoding, i.e. the
operations $U,U^{\dagger}$ only have to be applied at the
beginning and end of simulation process. We start by analytically
estimating the effect of noise in this case, and then consider
exact numerical simulations which confirm our estimates.

\subsubsection{Three-body interaction}

The easiest way to generate a three--body interaction using a
graph state encoding is to consider a three qubit system and an
encoding operation \be U=U_{PG}^{(12)} U_{PG}^{(13)}, \ee where
$U_{PG} = diag(1,1,1,-1)$ is a phase gate. With this encoding,
single qubit operations of the form $\exp(-i\delta t'
\sigma_x^{(1)})$ allow one to generate an evolution with respect
to the three--body Hamiltonian $H=\sigma_x^{(1)}\otimes
\sigma_z^{(2)}\otimes \sigma_z^{(3)}$ for time $t=\delta t'$, i.e.
\be \label{Utalpha} U_t(\delta t') = U e^{-i\delta t'
\sigma_x^{(1)}} U^{\dagger} = e^{-i \delta t'
\sigma_x^{(1)}\otimes \sigma_z^{(2)}\otimes \sigma_z^{(3)}}. \ee

{\em Phase noise: } We now turn to a simple example to illustrate
how noise effects this process. We consider a basic evolution that
is governed by the Hamiltonian $H=\sigma_z\otimes \sigma_z$, and
where the noise part is described by a Liouvillian that
corresponds to single qubit dephasing, i.e. is given by Eq.
\ref{Liovphasenoise}. Since noise part and Hamiltonian part
commute, the solution of the corresponding master equation is
given by
\be \label{simplecommute} \rho(t)= {\cal D}^{(1)}(p){\cal
D}^{(2)}(p) U \rho U^{\dagger},
\ee
where $U=\exp(-i t \sigma_z^{(1)} \otimes \sigma_z^{(2)})$ and
${\cal D}^{(k)}_t$ is a dephasing map with parameter
$p=p(t)=\exp(-\gamma_0 t)$ (see Eq. \ref{dephasing}). In the
remainder of this section we will drop the $t$ subscript from the
dephasing and depolarizing maps, so, unless otherwise stated,
${\cal D}^{(k)}_t\equiv {\cal D}^{(k)}$. Applying this noisy
evolution for time $t=\pi/4$, together with local rotations among
the $z$--axis (i.e. operations of the form $\exp(- \delta t
\sigma_z)$, a noisy phase gate can be generated
\be
{\cal E}_{PG} \rho= {\cal D}^{(1)}(p){\cal D}^{(2)}(p) U_{PG} \rho
U_{PG}^{\dagger},
\ee
where $p=p(t)=\exp(-\gamma_0 \pi/4)$. The
the total (noisy) process to generate the evolution with respect
to the three-body Hamiltonian $H$ for time $t=\delta t'$ is thus
given by
\be
{\cal E}\rho ={\cal E}_{PG}^{(12)}{\cal E}_{PG}^{(13)}{\hat
U}_x^{(1)}(\delta t'){\cal E}_{PG}^{(12)}{\cal E}_{PG}^{(13)}
\rho,
\ee
where ${\hat U}_x^{(1)}(\delta t') \rho = U_x(\delta t') \rho
U_x(\delta t')^\dagger$ with $U_x(\delta t')=\exp(-i\delta t'
\sigma_x)$. Using that the noise process commutes with $\sigma_z$,
and that ${\cal D}(p){\cal D}(p) = {\cal D}(p^2)$, one can
simplify this expression and obtains
\be \label{eq:graphdephase}
{\cal E}\rho = {\cal D}^{(1)}(p^2) {\hat U_t(\delta t')} {\cal
D}^{(1)}(p^2)  {\cal D}^{(2)}(p^2){\cal D}^{(3)}(p^2) \rho,
\ee
where $U_t(\delta t')$ is given in Eq. \ref{Utalpha}. Notice that
${\cal D}^{(1)}$ and $U_t(\delta t')$ do not commute. Using Eq.
\ref{eq:graphdephase} and the chaining inequality (Eq.
\ref{eq:chaining}) we can derive an exact bound on the
Jamio{\l}kowski distance and the corresponding fidelity. The
distance between ${\cal E}$ and $\hat{U}_t(\delta t')$ can be
expressed as,
\be
&& D({\cal E}, \hat{U}_t(\delta t')) \nonumber\\ &&=D({\cal
D}^{(1)}(p^2) {\hat U_t(\delta t')} {\cal D}^{(1)}(p^2)  {\cal
D}^{(2)}(p^2){\cal D}^{(3)}(p^2), \hat{U}_t(\delta
t'))\nonumber\\.
\ee
Applying the chaining property of the Jamio{\l}kowski distance we
find,
\be
&&D({\cal E}, \hat{U}_t(\delta t'))\nonumber \\
&&\leq D({\cal D}^{(1)}(p^2) {\hat U_t(\delta
t')},\hat{U}_t(\delta t')) \nonumber \\&&+ D({\cal D}^{(1)}(p^2)
{\cal D}^{(2)}(p^2){\cal
D}^{(3)}(p^2), \eins) \nonumber \\
&& = D({\cal D}^{(1)}(p^2), \eins)+D({\cal D}^{(1)}(p^2)  {\cal
D}^{(2)}(p^2){\cal D}^{(3)}(p^2), \eins). \nonumber \\
\ee
Noting the fidelity of a one qubit dephasing channel with the
identity is easily calculated and one finds that,
\be
F({\cal D}^{(1)}(p^2), \eins) = \frac{1+p^2}{2}
\ee
and
\be
F({\cal D}^{(1)}(p^2)  {\cal D}^{(2)}(p^2){\cal D}^{(3)}(p^2),
\eins) = \left(\frac{1+p^2}{2}\right)^3.
\ee
Which yields the corresponding distances,
\be
D({\cal D}^{(1)}(p^2), \eins) = \sqrt{1-\frac{1+p^2}{2}},
\ee
\be
D({\cal D}^{(1)}(p^2)  {\cal D}^{(2)}(p^2){\cal D}^{(3)}(p^2),
\eins) = \sqrt{1- \left(\frac{1+p^2}{2}\right)^3}.
\ee

Thus, $D({\cal E}, \hat{U}_t(\delta t'))$ has the upper bound,
\be
&&D({\cal E}, \hat{U}_t(\delta t')) \nonumber \\
&&\leq \sqrt{1-\frac{1+p^2}{2}} +\sqrt{1-
\left(\frac{1+p^2}{2}\right)^3},
\nonumber \\
\ee
which provides an lower bound on the fidelity, $D({\cal E},
\hat{U}_t(\delta t'))$ by
\be \label{eq:Fbound3qGSE}
&&F({\cal E}, \hat{U}_t(\delta t'))\geq 1 - \nonumber \\
&&\left( \sqrt{1-\frac{1+p^2}{2}} +\sqrt{1-
\left(\frac{1+p^2}{2}\right)^3}\right)^2.
\nonumber \\
\ee
For small $\delta t'$ a much simpler expression for the fidelity
can be found that is tighter than the above bound. Using that
${\cal D}(p) {\hat U_x(\delta t')} = {\hat U_x(\delta t')} \tilde
{\cal D}(p)$, where
\be
\tilde {\cal D}(p)\rho = p \rho +
\frac{1-p}{2} ( \rho + A \rho A^{\dagger}),
\ee
and $A= \hat{U}_t(\delta t')^{\dagger} \sigma_z \hat{U}_t(\delta
t')$. It follows that the total process can also be written as,
\be \label{phasenoisem3}
{\cal E}\rho = {\hat U_t(\delta t')} \tilde {\cal D}^{(1)}(p^2)
{\cal D}^{(1)}(p^2)  {\cal D}^{(2)}(p^2){\cal D}^{(3)}(p^2) \rho,
\ee
i.e. a sequence of certain local noise processes followed by the
ideal evolution. The Jamio{\l}kowski fidelity of the total map
${\cal E}$ with respect to the desired evolution $U_t(\delta t')$,
$F({\cal E},U_t(\delta t'))$ is thus the same as the one of the
map $\tilde {\cal E}\rho = \tilde {\cal D}^{(1)}(p^2) {\cal
D}^{(1)}(p^2) {\cal D}^{(2)}(p^2){\cal D}^{(3)}(p^2) \rho$ with
respect to the identity, $F(\tilde {\cal E}, \eins)$. For small
$\delta t'$, one has that $\tilde{\cal D}\approx {\cal D}$, and hence
\be
\tilde {\cal E}\rho \approx {\cal D}^{(1)}(p^4){\cal
D}^{(2)}(p^2){\cal D}^{(3)}(p^2) \rho.
\ee
The fidelity $F(\tilde {\cal E}, \eins)$, and hence $F({\cal
E},U_t(\delta t'))$ can be easily calculated, and one obtains
\be
F({\cal E},U_t(\delta t'))&\approx& \left (\frac{1+p^4}{2}\right )\left (\frac{1+p^2}{2}\right )^2 \nonumber\\
&\approx& 1- \pi \gamma_0,
\ee
where $p=\exp(-\gamma_0 \pi/4)$ and the last simplification only
holds for $\gamma_0 \pi/4 \ll 1$. Under the same approximation the
exact lower bound on the fidelity that we found in Eq.
\ref{eq:Fbound3qGSE} can be approximated to
\be
F({\cal E},U_t(\delta t'))\gtrsim  1-
\left(1+\frac{\sqrt{3}}{2}\right)\pi \gamma_0,
\ee
demonstrating that our bound provides a relatively good
approximation to the achievable fidelity in this parameter regime.

The local noise equivalent (see Sec. \ref{subsec:LNE} for
definition) can be evaluated under the previous assumptions. We
find that $F = \left(\frac{1+\exp(-\delta t' \gamma)}{2}\right)^3
\approx 1- \frac{3}{2}\delta t' \gamma$, and hence the local noise
equivalent $\gamma$ is given by \be \label{GSE_LNE_phase} \gamma
\approx \frac{2\pi}{3\delta t'} \gamma_0. \ee

{\em White noise: } Similar arguments can be used to estimate the
effect of white noise, i.e. when the Liouvillian describing the
noise process in the basic two--qubit interaction is a sum of
local terms described by Eq. \ref{Lindblad} with $s=1/2$ and
$B=C\equiv \kappa_0$. The solution of this master equation (when
taking interaction into account) does in general not lead a simple
form such as Eq. \ref{simplecommute}, where the ideal evolution is
simply composed by noise channels. However, one may still use such
a form, and still obtains a good approximation, in particular if
$\kappa_0$ is relatively small.

In particular, we have
\be \label{simplecommute}
\rho(t) \approx {\cal M}^{(1)}(p){\cal M}^{(2)}(p) U \rho
U^{\dagger},
\ee
where $U=\exp(-i t \sigma_z^{(1)} \otimes \sigma_z^{(2)})$ and
${\cal M}$ is a depolarizing map with parameter
$p=p(t)=\exp(-\kappa_0 t)$ (see Eq. \ref{Depol}). As the
depolarizing channel does not commute with $U$ the best analytic
bound that we can find on the Jamio{\l}kowski distance is
that given by the simple bound derived in Eq.
\ref{eq:simpledistbound}. Noting that,
\be
&&F({\cal M}^{(1)}(p){\cal M}^{(2)}(p) U, U) =F({\cal
M}^{(1)}(p){\cal
M}^{(2)}(p),\eins)\nonumber \\
&& = \left(\frac{1+3p}{4}\right)^2
\ee
and that in order to simulate $U_t(\delta t')$ we must use $U$
four times we find,
\be
D({\cal E},U_t(\delta t'))\leq
4\sqrt{1-\left(\frac{1+3p}{4}\right)^2}.
\ee
Thus, the corresponding Jamio{\l}kowski fidelity is lower bounded
by:
\be
\label{eq:FboundWN3GSE} F({\cal E},U_t(\delta t'))\geq 1 -
16\left(1-\left(\frac{1+3p}{4}\right)^2\right).
\ee

Following the same line of argument as for the dephasing
calculation above, we can find a simple approximation that beats
this bound when $\delta t'$ is small. One finds that the total
(noisy) process is approximately described by
\be
\label{caledepol} \tilde {\cal E}\rho \approx {\cal
M}^{(1)}(p^4){\cal M}^{(2)}(p^2){\cal M}^{(3)}(p^2) \rho.
\ee
This leads to a final fidelity of the total process given by
\be
\label{estimatewhite}
F({\cal E},U_t(\delta t'))&\approx& \left (\frac{1+3 p^4}{4}\right )\left (\frac{1+3 p^2}{4}\right )^2 \nonumber\\
&\approx& 1- \frac{3 \pi}{2} \kappa_0
\ee
where here $p=\exp(-\kappa_0 \pi/4)$ and the last simplification
only holds for $\kappa_0 \pi/4 \ll 1$. If we alternately used the
exact error bound of Eq. \ref{eq:FboundWN3GSE} in the same
parameter regime we find
\be
F({\cal E},U_t(\delta t')) \gtrsim 1 - 3\pi \kappa_0,
\ee
which is close to the approximation of Eq. \ref{estimatewhite}.

Again, the local noise equivalent can be evaluated, where in the
case of depolarizing noise one can estimate $F \approx \left(
\frac{1+3 \exp(-\kappa \delta t')}{4}\right )^3 \approx 1-
\frac{9}{4} \delta t' \kappa$, which leads to a local noise
equivalent
\be
\label{LNEmultipartiteGSE} \kappa \approx \frac{2\pi}{3\delta t'}
\kappa_0.
\ee

Notice in particular that the final expression for the fidelity is
{\em independent} of $\delta t'$, the time for which the
three--body interaction should be applied. Such a constant
fidelity in turn implies that the local noise equivalent grows
with decreasing $\delta t'$. This can easily be understood by
recalling that the reliability parameter $p$ for dephasing or
depolarizing noise decreases with time, and hence a constant
fidelity actually means a larger noise level for short time
evolutions.

These observations are in contrast to the commutator method, where
the fidelity $F$ essentially decreases for larger $\delta t'$. A
direct comparison shows that for small values of $\delta t'$, the
commutator method leads much better fidelities. However, when
considering evolutions for longer times $\delta t'$, e.g. the
generation of gates or the simulation of the evolution with
respect to a more complicated Hamiltonian generated by Hamiltonian
simulation techniques, the graph state encoding method performs
significantly better (see table \ref{table1_1}). In particular when
considering time evolution over several full cycles (multiples of
$2 \pi$), the advantage of graph state encoding method becomes
obvious. In this case, the local noise equivalent can even be {\em
smaller} than $\kappa_0$. The latter processes appear, for
instance, when simulating an evolution with respect to a time
dependent Hamiltonian in such a way that adiabatic passage to the
ground state of the final Hamiltonian occurs, and hence the ground
state of this Hamiltonian is generated \cite{Jane03,Murg04}.

We remark that in our estimation of the total fidelity we have
assumed small values $\delta t'$. It turns out, however, that even
when taking all errors into account and performing numerical
simulations of the total process, there is essentially no
dependence on $\delta t'$. In addition, our estimates turn out to
be very accurate (see table \ref{table1_1}). This table shows the
resulting fidelities when using graph state encoding techniques as
described above to generate unitary evolutions with respect to
Hamiltonian (i) $H_1=\sigma_z^{(1)}\sigma_z^{(2)}\sigma_z^{(3)}$
for time $t=\pi/4$. In this case, the graph state encoding $G_1$
corresponds to $U=U_{PG}^{(12)}U_{PG}^{(13)}$ and single--qubit
operations of the form $\exp(-i\delta t' \sigma_x^{(1)})$ are used
to generate the desired three--body Hamiltonian.  Also the usage
of an alternative graph state encoding $G_2$ using only
$U=U_{PG}^{(12)}$ together with the application of two--qubit
operators $\exp(-i\delta t' \sigma_x^{(2)}\sigma_z^{(3)})$ to
generate the desired three--qubit Hamiltonian is shown.

The latter graph state encoding $G_2$ is also used to generate
time evolution with respect to the Hamiltonian (ii)
$H_2=\sigma_z^{(1)}\sigma_z^{(2)}\sigma_z^{(3)}+\sum_{k=1}^{3}\sigma_x^{(k)}$
for time $t=\pi/4$. In this case, Hamiltonian simulation
techniques need to be applied to generate evolutions with respect
to the non--commuting terms
$\sigma_z^{(1)}\sigma_z^{(2)}\sigma_z^{(3)}$ and
$\sum_{k=1}^{3}\sigma_x^{(k)}$ for short times $\delta t_m$
sequentially. Up to an additional local basis change (a Hadamard
operation ${\rm Had}$ on qubit 2), the first term is generated by the
two--qubit Hamiltonian $H_1'= \sigma_x^{(2)}\sigma_z^{(3)}$, while
the second term is produced from the Hamiltonian $H_2'=
\sigma_x^{(1)}\sigma_z^{(2)}+\sigma_z^{(2)} + \sigma_x^{(3)}$. The
total procedure thus involves the application of
$U_{PG}^{(12)}{\rm Had}^{(2)}$, followed by sequential application of
$H_1'$, $H_2'$ for short times $\delta t_m$ a total of
$(\pi/4)/\delta t_m$ times, and a final application of
${\rm Had}^{(2)}U_{PG}^{(12)}$. Notice that the same examples are
considered in the case of the commutator method in Sec.
\ref{twoHams}.

\begin{table}
\begin{tabular}{|c||c|c|c|c|c|}
  \hline
    & $H_1$, anal. & $H_1, {G}_1$ & $H_1, G_{2}$ & $H_1$, $C$ & $H_2, G_{2}$\\
  \hline\hline
  p=0.9 & F=0.8552 & F=0.8545 & F=0.8887  & F=0.7062 & F=0.8545\\
  \hline
  p=0.99 & F=0.9851 & F=0.9850 & F=0.9888  & F=0.9305 & F=0.9850\\
  \hline
  p=0.999 & F=0.9985 & F=0.9985 & F=0.9989 & F=0.9841 & F=0.9985 \\
  \hline
\end{tabular}
\caption{Table shows fidelities to generate gate $U_1=\exp(-i\pi/4
H_1)$ using graph state encoding (column 1: analytic estimate
using graph state encoding $G_1$, columns two and three: numerical
simulation using graph state encoding $G_1$ and $G_2$
respectively) or commutator method with optimal time $\delta t_m$
(column four). The last column gives the fidelity to generate the
time evolution $U_2=\exp(-i\pi/4 H_2)$ using graph state encoding
$G_2$ (see text for details). A white noise parameter
$p=\exp(-\kappa_0 \pi)$ is assumed that corresponds to a coupling
strength $\kappa_0$ to the thermal bath that would lead to
depolarizing map with parameter $p$ when applied for time $t=\pi$.
Notice that this corresponds to $p'=p^{1/4}$ when using the
formula for analytic estimate of fidelity, Eq.
\ref{estimatewhite}, where noisy evolution is only applied for
time $t=\pi/4$.}
\label{table1_1}
\end{table}

\subsubsection{Many-body interaction}

It is straightforward to obtain methods to generate a $m$--body
Hamiltonian using graph state encoding, and analytic estimates for
the influence of noise. In particular, the graph state encoding
\be U=\prod_{k=2}^{m} U_{PG}^{(1k)}, \ee together with the
evolution with respect to the single--qubit Hamiltonian
$H=\sigma_x^{(1)}$ for time $t=\delta t'$ allows one to generate
the $m$--qubit gate $\exp(-i\delta t'
\otimes_{k=2}^{m}\sigma_z^{(k)} \otimes \sigma_x^{(1)})$, i.e.
\be
U \exp(-i\delta t' \sigma_x^{(1)}) U^{\dagger} = \exp[-i\delta t'
(\otimes_{k=2}^{m}\sigma_z^{(k)}) \otimes \sigma_x^{(1)}]
\ee

Following precisely the same line of reasoning as in the case of
three--qubit interaction, $m=3$, one obtains in the case of phase
noise that the total noisy evolution is described by (compare with
Eq. \ref{phasenoisem3})
\be {\cal E}\rho = {\hat U_t(\delta t')}
\tilde {\cal D}^{(1)}(p^{m-1})  {\cal D}^{(1)}(p^{m-1})
\prod_{k=2}^{m}{\cal D}^{(k)}(p^2)  \rho.
\ee
We find that the analytic bound on the fidelity of this operation
given to us by the chaining property of the Jamio{\l}kowski
distance to be,
\be
&&F({\cal E}, \hat{U}_t(\delta t'))\geq 1 - \nonumber \\
&&\left( \sqrt{1-\frac{1+p^{m-1}}{2}} +\sqrt{1-
\frac{1+p^{m-1}}{2}\left(\frac{1+p^2}{2}\right)^{m-1}}\right)^2.
\nonumber \\
\ee
When $\delta t'$ is small a simple estimation of fidelity can be
made
\be
F({\cal E},U_t(\delta t'))&\approx& \left (\frac{1+p^{2(m-1)}}{2}\right )\left (\frac{1+p^2}{2}\right )^{m-1} \nonumber\\
&\approx& 1- 2 (m-1) \frac{\pi}{4} \gamma_0
\ee
where $p=\exp(-\gamma_0 \pi/4)$ and the last simplification is
valid for $\pi/4 \gamma_0 \ll 1$. For the local noise equivalent
we find $\gamma \approx  \frac{(m-1)\pi}{m\delta t'} \gamma_0$.

Similarly, in the case of depolarizing (or white) noise, one finds
that the fidelity can be lower bounded by
\be
F({\cal E},U_t(\delta t'))\geq 1 -
(m-1)^2\left(1-\left(\frac{1+3p}{4}\right)^2\right).
\ee
We can also directly evaluate the fidelity in the small $\delta
t'$ limit to be
\be
F({\cal E},U_t(\delta t')) &\approx& \left (\frac{1+3 p^{2(m-1)}}{4}\right )\left (\frac{1+3 p^2}{4}\right )^{m-1} \nonumber\\
&\approx& 1- 3 (m-1) \frac{\pi}{4} \kappa_0
\ee
where here $p=\exp(-\kappa_0 \pi/4)$ and the last simplification
only holds for $\kappa_0 \pi/4 \ll 1$. Again, the final
expressions are independent of the total time $t=\delta t'$. The
local noise equivalent can be determined to be $\kappa \approx
\frac{(m-1)\pi}{m\delta t'} \kappa_0$.

\subsubsection{Timing errors}

In this subsection we consider the effect of timing errors on the
fixed graph state encoding method. Recall that in Subsection
\ref{Noisemodel2} we introduced the idea of a random timing error.
Such an error arises due to our inability to apply any Hamiltonian
for an absolutely precise amount of time.

Consider the 3-qubit simulation protocol of the previous
subsection as given by Eq. \ref{Utalpha}
\be
U_t(\delta t')& = &U_{PG}^{(12)}U_{PG}^{(13)} e^{-i\delta t'
\sigma_x^{(1)}} U_{PG}^{(12)}U_{PG}^{(13)} \nonumber \\&= & e^{-i
\delta t' \sigma_x^{(1)}\otimes \sigma_z^{(2)}\otimes
\sigma_z^{(3)}}. \nonumber
\ee
Each of the phase gates, $U_{PG}$, can be generated by evolving a
two qubit Ising interaction Hamiltonian, $H=\sz\otimes\sz$, for a
time $t=\pi/4$ and applying single qubit $\sz$ rotation to each
qubit. If we have a timing error on an Ising Hamiltonian that
generates a phase gate, then we can represent this noisy process
by the following CP map
\be
{\cal E}_{PG} (\rho) = {\cal T} U_{PG}\rho U_{PG}
\ee
where ${\cal T}$ is the CP map given in Eq.
\ref{eq:2qubitisingtimeerror}. The map ${\cal T}$ is a correlated
two-qubit dephasing channel. The noisy version this simulation
protocol is expressed by
\be
{\cal E} (\rho) = {\cal T}^{(12)}{\cal T}^{(13)}\hat{U}_t(\delta
t'){\cal T}^{(12)}{\cal T}^{(13)}(\rho).
\ee
using the chaining property of the Jamio{\l}kowski distance it is
possible to bound the fidelity of this operation. Noting that neither
${\cal T}_{12}$ nor ${\cal T}_{13}$ commute with $\hat{U}_t(\delta
t')$ and applying the chaining inequality we find,
\be
D({\cal E}, U_t(\delta t'))\leq 2 D({\cal T}^{(12)}{\cal
T}^{(13)}, \eins).
\ee
Assuming that the timing error was Gaussian distributed with a
standard deviation of $\sigma$ (as was done in Subsection
\ref{Noisemodel2}) we find
\be
F({\cal T}^{(12)}{\cal T}^{(13)}, \eins) =
\left(\frac{1+q}{2}\right)^2
\ee
where $q = e^{-2\sigma^2}$. The resulting lower bound on the
fidelity is,
\be
F({\cal E}, U_t(\delta t')) \geq
1-4\left(1-\left(\frac{1+q}{2}\right)^2\right).
\ee
If we assume that the timing error is very small, that is
$\sigma^2\ll 1$ then we can find a simpler form for this bound.
Noting that for small $\sigma^2$
\be
F({\cal T}^{(12)}{\cal T}^{(13)}, \eins) \approx 1-2\sigma^2,
\ee
then it follows that
\be
F({\cal E}, U_t(\delta t'))\gtrsim 1 - 8\sigma^2.
\ee

The above equations can be simply generalized to the $m$ qubit
case. Recall that in the $m$-qubit case that we must apply the
following encoding operation,
\be U_e=\prod_{k=2}^{m} U_{PG}^{(1k)}. \ee
The result of timing errors on the phase gates in $U_e$ is given
by the following map,
\be
{\cal E}_e (\rho) =\prod_{k=2}^m{\cal T}^{(1k)}U_e\rho U_e
\ee

The fidelity, $F({\cal E}_e, U_e)=F(\prod_{k=2}^m{\cal
T}^{(1k)},\eins)$, and is given by:
\be
F({\cal E}_e, U_e)=F(\prod_{k=2}^m{\cal
T}^{(1k)},\eins)=\left(\frac{1+q}{2}\right)^{m-1}.
\ee
This results in the following error bound for the entire
simulation protocol,
\be
F({\cal E}, U_t(\delta t')) &\geq &1 - 4(1 - F({\cal E}_e, U_e)) \nonumber \\
& = &1-4\left(1-\left(\frac{1+q}{2}\right)^{m-1}\right).
\ee
When $\sigma^2\ll 1$ the fidelity of performing an encoding
operation is given by,
\be
F({\cal E}_e, U_e)=F(\prod_{k=2}^m{\cal T}^{(1k)},\eins) \approx
1-(m-1)\sigma^2
\ee
which results in the following bound on the entire simulation
protocol,
\be
F({\cal E}, U_t(\delta t'))\gtrsim 1-4(m-1)\sigma^2.
\ee

So far we have considered the situation where the operation
$U_{PG}^{(12)}U_{PG}^{(13)}$ is generated by single qubit $\sz$
rotations and evolutions of the Hamiltonians
$H^{(12)}=\sz^{(1)}\sz^{(2)}$ and $H^{(13)}=\sz^{(1)}\sz^{(3)}$.
We could have in principle generated the product of these phase
gates by applying single-qubit $\sz$ rotations and evolving by the
collective Hamiltonian $H^{(123)} = H^{(12)} + H^{(13)}$. Such an
evolution would give rise to the following CP map,
\be
{\cal E}^{(123)} (\rho) = {\cal T}^{(123)}
U_{PG}^{(12)}U_{PG}^{(13)}\rho U_{PG}^{(12)}U_{PG}^{(13)}.
\ee
Where the CP map ${\cal T}^{(123)}$ is a result of timing error on
the Hamiltonian $H_{123}$. The form of ${\cal T}^{(123)}$ was
derived in Subsection \ref{Noisemodel2} for the case where the
timing error is Gaussian distributed with a standard deviation of
$\sigma$ (this is given in Eq. \ref{eq:collective3qisingerror}).
Noting that
\be
F({\cal E}^{(123)},U_{PG}^{(12)}U_{PG}^{(13)}) &=& F({\cal
E}^{(123)},e^{-i\frac{\pi}{4}H^{(123)}})  \nonumber\\ &=& F({\cal T}^{(123)},
\eins)
\ee
the fidelity of this channel was derived and we found it to be
\be
F({\cal T}^{(123)}, \eins) & = & \frac{3}{2^3}
+\frac{1}{2}e^{-2\sigma^2}+\frac{1}{2^3}(e^{-2\sigma^2})^4.
\nonumber \\
\ee

Using similar methods to that which we have seen above, we can
calculate a bound on the fidelity $F({\cal E}, U_t(\delta t'))$
given that we used a collective Ising Hamiltonian $H^{(123)}$ as
our entangling resource,
\be
F({\cal E}, U_t(\delta t')) \geq 1-4\left(1-F({\cal T}^{(123)},
\eins)\right).
\ee
When $\sigma^2\ll 1$, we see that
\be
F({\cal T}^{(123)}, \eins) \approx 1-2\sigma^2
\ee
which results in a lower bound for $F({\cal E}, U_t(\delta t'))$,
\be
F({\cal E}, U_t(\delta t')) \gtrsim 1 - 8\sigma^2.
\ee
Interestingly, we see that this is the same fidelity that we
derived in the case where our entangling operations were generated
by two-qubit Hamiltonian interactions.

Generalizing to the $m$-qubit case, we recall that we must apply
an $m$-qubit encoding operation, $U_e = \prod_{k=2}^{m}
U_{PG}^{(1k)}$. This $m$-qubit encoding operation can be performed
by single qubit $\sigma_z$ operations and an evolution for a time
$\pi/4$ of the Hamiltonian,
\be
H_e = \sum_{k=2}^m \sz^{(1)}\sz^{(k)}.
\ee
$H_e$ is a sum of commuting unitary operations, in Subsection
\ref{subsec:timingerrorssumofunitary} we analyzed the effect of
timing errors on such Hamiltonians. Because there are only $m-1$
$\sz^{(1)}\sz^{(k)}$ terms in $H_e$, and because of the geometry
of the interaction graph, it is possible to find an analytic
expression for a timing error in the application of the
Hamiltonian $H_{e}$. We found the resulting fidelity to be,
\be
&& F({\cal E}_e,U_e)=F({\cal T}_{e}, \eins) \nonumber \\ &&=
\frac{1}{2^{2n}}{2n\choose n} +
\frac{1}{2^{2n-1}}\sum_{k=0}^{n-1}{2n\choose
k}e^{-2(n-k)^2\sigma^2}, \nonumber \\
\ee
where $n=m-1$. As we saw above, this allows us to bound the
fidelity of the entire simulation protocol to be:
\be
F({\cal E}, U_t(\delta t')) \geq 1-4\left(1-F({\cal T}^{(12\dots
m)}, \eins)\right).
\ee

\begin{figure}
 \includegraphics[trim=100 0 0 0,scale=.24]{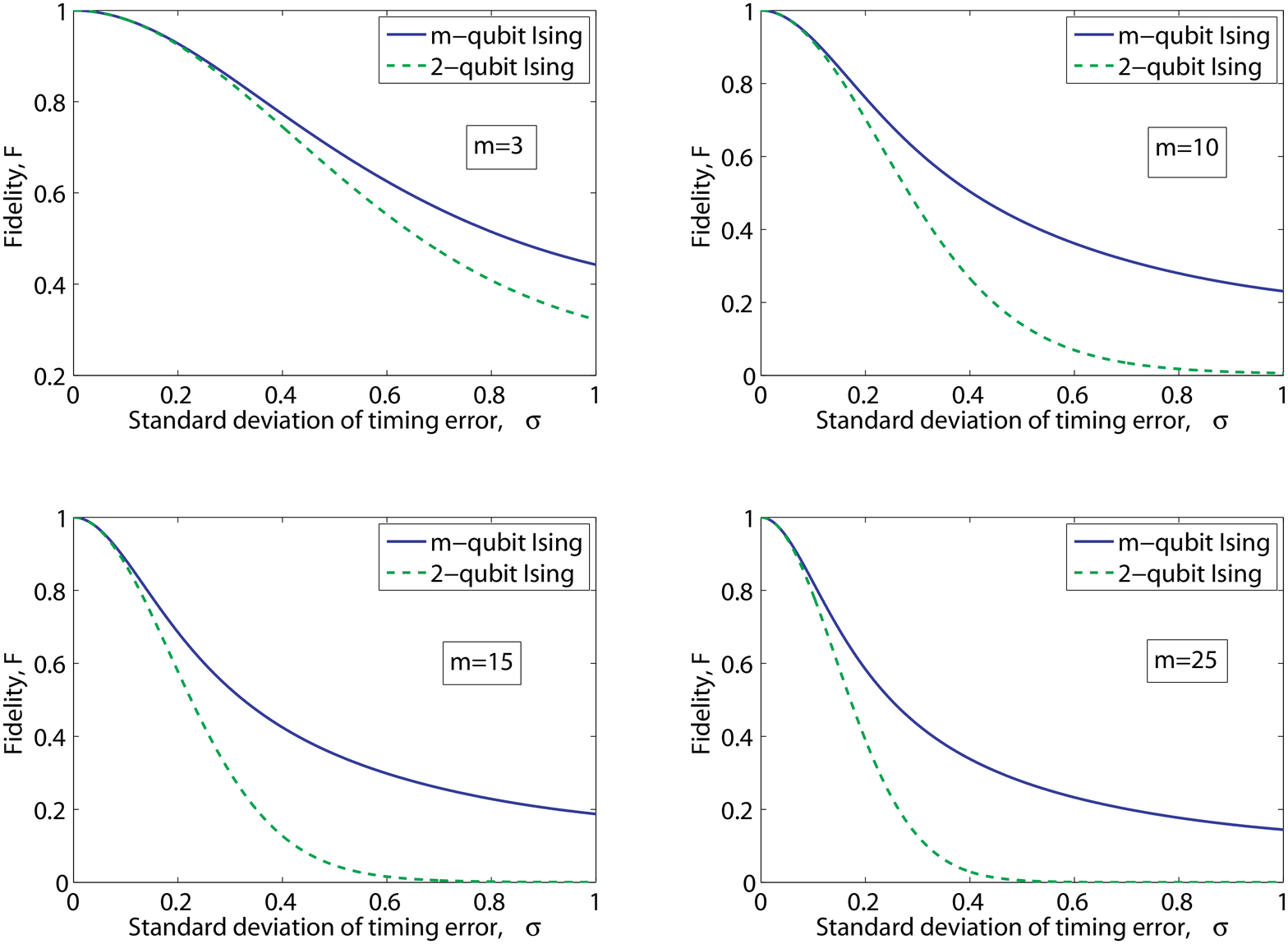}\\
  \caption{(Color online) In these figure we see how the fidelity of performing the encoding
operation $U_e$ changes with the standard deviation of a timing
error, $\sigma$. We have examined how this fidelity changes with
the number of qubits, $m$, encoded by $U_e$. The solid line
represents the fidelity of performing the encoding operation with
a single collective Ising Hamiltonian $H_e$. The dashed line
represents the fidelity as a result of using a number of two-qubit
Ising Hamiltonians $H_{1k}$ to generate $U_e$. We see that as
$\sigma \rightarrow 0$ the two methods give the same fidelity.
However, it appears for larger $\sigma$ the collective method is
superior. This effect becomes more apparent as $m$
increases.}\label{fig:collectivevs2qubit2}
\end{figure}

All of the noise that appears in our analysis of the graph state
encoding protocol appears within the timing error of the encoding
operation, $U_e$. As we saw, the success of the protocol depends
on the fidelity, $F({\cal E}_e, U_e)$. The higher this fidelity,
the higher the fidelity of the overall protocol. Our analysis begs
the obvious question; \emph{in the presence of timing errors is it
better to use a collective Ising Hamiltonian acting on many qubits
or many applications of two-qubit Ising Hamiltonians?} We
addressed a more general version of this question in Subsection
\ref{subsec:timingerrorssumofunitary}. Applying the results of
Subsection \ref{subsec:timingerrorssumofunitary} to the case of
applying the encoding operation $U_e$ we find
\be
F(\prod_{k=2}^m {\cal T}^{(1k)},\eins)\leq F({\cal T}^{(12...m)},
\eins),
\ee
where we assuming that the standard deviation of the timing errors
in the two-qubit Hamiltonians is the same as that for the
collective Hamiltonian $H_e$. Under these assumptions using a
collective Ising Hamiltonian results in a larger fidelity for the
simulation. In Figure \ref{fig:collectivevs2qubit2} we have
plotted the fidelity of performing the encoding operation $U_e$
using a collective $m-qubit$ Ising Hamiltonian and using two-qubit
Ising Hamiltonians. We see that for $\sigma\rightarrow 0$ that
both methods converge to give a fidelity of one. As $\sigma$
increases we see that the fidelity of the two methods diverges and
that this effect becomes more pronounced as the number of qubits
coupled by $U_e$ increases.

\subsection{Variable graph state encoding}

In our previous consideration, we have assumed that a fixed graph
state encoding is sufficient to generate the desired many--body
Hamiltonian using only single- and two--qubit interactions. In
Sec. \ref{GSE} we have discussed several examples where this is
indeed the case, and the strength of the graph state encoding
method is certainly in such situations. In particular, if the
total time for which an evolution with respect to a many--body
Hamiltonian should be generated is large, the graph state encoding
method performs significantly better than the commutator method.

In certain situations, however, a fixed graph state encoding is
not sufficient and the encoding has to be changed frequently. Any
change of encoding requires the application of phase gates, i.e.
applying the two--body interaction described by the Hamiltonain
$H=\sigma_z\otimes \sigma_z$ for time $t=\pi/4$. Such a situation
is discussed in detail in Sec. \ref{variableGSE}. Recall that for
a given interaction Hamiltonian, it can be that not all required
(non--commuting) interaction terms $H_1, \ldots , H_n$ can be
generated using a fixed graph state encoding. Hence Hamiltonian
simulation techniques need to be applied, requiring the sequential
evolution with respect to Hamiltonians  $H_1, \ldots , H_n$ for
short times $\delta t'$ that are generated with help of certain
graph state encodings $G_1, G_2 , \ldots ,G_n$. As discussed in
detail above, any change of graph state encoding introduces a
certain (fixed) amount of noise. This is reflected in a constant
fidelity of the simulation process using a fixed graph state
encoding, or alternatively by an increase of the local noise
equivalent that is essentially given by \be \kappa/\kappa_0
\approx \pi/\delta t'. \ee

On the one hand, $\delta t'$ needs to be chosen sufficiently small
to ensure that errors due to taylor expansion in the simulation
process remain small. On the other hand, smaller values of $\delta
t'$ lead a larger local noise equivalent due to frequent change of
graph state encoding. Similar as in the commutator method, a
compromise between these two competing requirements need to be
chosen, such that for the simulation with respect to a certain
Hamiltonian for time $t$ the resulting fidelity is maximized.

\section{Teleportation based methods}\label{TBMNOISE}

We now turn to teleportation based methods to generate many--body
Hamiltonians. Recall that the basic idea is to use two--body
interactions to generate certain entangled multi--party states
$|\psi_E\rangle$. These states are then used as a resource to
generate by means of (joint) measurements the desired evolution
with respect to a many--body Hamiltonian for time $\delta t'$.
Depending on the chosen scheme, the states $|\psi_E\rangle$ are
(i) weakly entangled and sequences of probabilistic teleportation
processes together with preparation of states with increasing
amount of entanglement are required to make the process
deterministic; or (ii) maximally entangled and involve also an
ancilla particle, which is used to chose the evolution time
$\delta t'$ and to make the process deterministic.

If two--body interactions are noisy, then the generated
multi--party states are mixed. The fidelity of the state is
determined by the quality of the two--body interaction. In turn,
the fidelity of the multi--party states determines the noise level
of the many--body interaction. We remark that for the approach
(i), it is not always clear how to use two--body interactions for
{\em short times} to generate the desired state. Even if this is
possible, it turns out that approach (i) yields smaller fidelities
than approach (ii). We will thus only consider approach (ii) in
the following, and assume that maximally entangled states of
several particles (including ancilla particles) are generated by
two--body interactions, and processed by measurements. As in the
case of graph state encoding, here we are again faced with a {\em
fixed} amount of noise, independent of the desired interaction
time $\delta t'$. On the one hand, this results from the
generation of the maximally entangled state $|\psi_E\rangle$,
which involves application of two--body interactions for times
$\delta t \approx O(\pi)$. On the other hand, the teleportation
process consists of joint (Bell) measurements, which may also
require the application of entangling operations that further
increase the noise level.

In the following we will discuss a
scenario where higher dimensional systems are used to simulate
qubit systems (or more generally lower dimensional systems).
Extra dimensions are used to store the particles corresponding to
the state $|\psi_E\rangle$ and at the same time the system
degrees of freedom which should be processed. In this case, the
measurements in the teleportation process are {\em local}
as they involve only single systems, and one can argue that the
noise for such local processes is different (and lower) than for
two--system interactions. Hence the dominating part of noise will
come from the preparation of the entangled state $|\psi_E\rangle$.
We will elaborate further on this scenario, and will also consider
the case where additional auxiliary dimensions can be used to
perform {\em entanglement purification} to reduce the amount of
noise, or in the case when single system operations are perfect,
to eliminate noise completely.

\subsection{Simulation of low dimensional systems using high dimensional systems}

In Sec. \ref{quDits_Commutator}, we have argued that low
dimensional systems of dimension $d$ can also be used to simulate
interacting high dimensional systems of dimension $D$. In this
case, many body Hamiltonians are required, even to achieve
single-system operations on the simulated $D$--dimensional system.
Here we consider the opposite approach, where {\em high}
dimensional systems of dimension $D$ are used to simulate low
dimensional systems of dimension $d$, and we concentrate on $d=2$,
i.e. qubits. We will, however, not embed the qubits into the full
Hilbert space of the $D$--dimensional systems, but rather associate each
$D$--level system with a {\em single} qubit, i.e. we use only two
of the $D$ dimensions to represent and store the quantum
information. The remaining levels are used as auxiliary levels in
such a way that operations on the the qubit systems can be
performed in an simplified way, or to increase the fidelity of
such operations and interactions.

To be more precise, we consider the case where the dimension $D$
is some power of two, $D=2^m$. In this case, the $D$--dimensional
system can be considered as a system consisting of $m$ {\em
virtual} subsystems (virtual qubits), where one of these qubits,
the storage qubit $A_1$, is used for storage of quantum
information, while the remaining $m-1$ qubits $A_2, \ldots ,A_m$
are auxiliary systems. We remark that this simply corresponds to a
labelling of the $D$--dimensional Hilbert space which we choose
for notational convenience and to clearly distinguish between
levels that are used for storage purposes and auxiliary levels.
These virtual qubits do not have any true physical relevance.

Interactions between the auxiliary qubits and the storage qubit,
as well as interactions between the auxiliary qubits, correspond
to {\em single system operations} on the $D$--dimensional system.
We also consider two--body interactions between two such
$D$--dimensional systems labelled by $A$ and $B$. Depending on the
type of interaction, such a two--body interaction can couple the
two storage qubits $A_1$ and $B_1$, yielding a pairwise
interaction, or two auxiliary qubits $A_k$ and $B_k$ (or
combinations of these processes). We will assume that single
system operations, which we also call {\em local operations}, can
be performed with a high fidelity. The noise of such local
processes is described by a noise parameter $p_l$ (when
considering gates), or by coupling strengthes $\gamma_l,\kappa_l$
when considering single system interactions (i.e. interactions
between auxiliary qubits and the storage qubit) that are described
by a master equation. Interactions between different
$D$--dimensional systems $A$ and $B$, which we also call {\em
non--local} operations, can be performed with lower fidelity. We
will denote the noise parameter by $p_{0}$, or the coupling
strengthes by $\gamma_{0},\kappa_{0}$. We remark that these
assumptions are natural in the sense that often single system
operations, i.e. interactions between different levels of the same
system, are easier to perform than controlled interactions between
two (spatially separated) systems. For example, atoms or ions have
many potentially usable internal states (levels) which can be
easily coupled via Raman-- or Microwave transitions, while
interactions between two such atoms or ions involves more
complicated processes (e.g. induced dipole couplings, sequences of
Laser pulses and couplings via motional states etc.). Note that a
similar approach was considered in the context of quantum
computation in Ref. \cite{Dur03}, where the auxiliary levels where
used to prepare entangled states shared between systems, which
where then used as resource to implement non--local gates. In
\cite{Dur03}, the exact requirements that such a scheme can be
applied were formulated, and in the following we assume that these
requirements are met. In particular, the schemes we consider
requires the realizability of the following operations: (i)
(non-local) two--system interactions that act only on specific
virtual subsystems of each particle, e.g., $A_2, B_2$, without
affecting other virtual subsystems; (ii) single-system measurement
on one virtual subsystem without affecting other virtual
subsystems; (iii) arbitrary unitary operations on one virtual
subsystem; and (iv) (local) interactions between arbitrary virtual
subsystems $A_j, A_k$. While (i) requires a specific type of
two--body interactions between two systems $A$ and $B$,
requirements (ii)-(iv) are concerned with the ability to locally
manipulate a single system.

Similarly as in the context of quantum computation \cite{Dur03},
here we consider the case that auxiliary qubits are used to
generate and store the entangled states $|\psi_E\rangle$, and
teleportation is used to generate time evolutions governed by a
many--body Hamiltonian for time $\delta t'$ on the storage qubits.
Note that the involved measurements and manipulations are {\em
local}, and only the generation of $|\psi_E\rangle$ requires
non--local operations. In addition, if enough auxiliary levels $D
\approx 8 -32$ are available, {\em entanglement purification} can
be applied to increase the fidelity of the noisy entangled states
generated by noisy two--system interactions. The states
$|\psi_E\rangle$ are essentially GHZ states (or graph states
that are locally equivalent to GHZ states), and known multiparty
entanglement purification protocols \cite{Aschauer05a,Aschauer05} can be
used to purify these states and to increase their fidelity.
Consequently, many-body interaction Hamiltonians acting on storage
qubits with increased fidelity can be generated.

\subsection{Perfect single--system operations}

In this section we assume that local operations are perfect, i.e.
$p_l=1$ or $\gamma_l=\kappa_l=0$, and investigate the influence of
noise resulting from two--system interactions in the generation of
$|\psi_E\rangle$. In the case of $n$--body interactions that
should be applied on (virtual) qubits $A_1,B_1, \ldots ,N_1$, the
state $|\psi_E\rangle$ is a $2n+1$ qubit state given by \be
|\psi_E\rangle=1/\sqrt{2}(|\phi^+\rangle^{\otimes n}|0\rangle_E +
|\phi^-\rangle^{\otimes n}|1\rangle_E), \ee where the last qubit
corresponds to an additional auxiliary system $E$ and the
$|\phi^+\rangle$ states corresponding to virtual qubits two and
three of the same system, $X_2X_3$, $X\in\{A,B,\ldots,N\}$. It
follows that $|\psi_E\rangle$, when considered as a $n+1$ system
state, is up to local, single system operations (that include also
local CNOT or phase gates which we however consider to be
noiseless in this section), equivalent to a GHZ--type graph state
\be
|\tilde \psi_E\rangle= [1/\sqrt{2}(|+\rangle^{\otimes n}|0\rangle_E
+  |-\rangle^{\otimes n}|1\rangle_E)]\otimes|0\rangle^{\otimes n}. \ee The state $|\tilde
\psi_E\rangle$ can be generated by applying phase gates $U_{PG}$
between the auxiliary particle $E$ and all other parties,
$\prod_{X \in \{A,B, \ldots ,N\}} U_{PG}^{(EX)}$. The Bell
measurements involved in the teleportation process act on virtual
qubits one and two of the same system, and are hence local, single
system operation (which are considered to be noiseless in this
section). Hence the only noise in the process comes from imperfect
preparation of the state $|\tilde \psi_E\rangle$ which involves
$n$ two--system gates.

An alternative scheme that uses a reduced number of virtual qubits
per system (two rather than three) is the following. We use the
$n+1$ qubit state $|\tilde \psi_E\rangle$ directly, and replace
the (local) Bell measurements by local phase gates that couple the
storage qubits $A_1,B_1,\ldots N_1$ to the qubits $A_2,B_2,\ldots
N_2$ of the state $|\tilde \psi_E\rangle$, and measurements in the
$\sigma_x$-basis on the storage qubits. Together, this again leads
to transfer of quantum information to the system $A_2,\ldots
,N_2$. An appropriate measurement on the ancilla qubit $E$, where
the measurement direction again depends on the outcome of all
previous $\sigma_x$ measurement, finally allows one to implement
the operation $e^{-\alpha  \sigma_z^{\otimes n}}$ for arbitrary
$\alpha$ in the storage particles (now held in $A_2,\ldots ,N_2$).
Additional Pauli operations --depending on the results of the
measurements-- on the final state are required to adjust the local
basis. This scheme has the advantage that fewer virtual qubits are
involved. In fact, this procedure is equivalent to the way of
obtaining $n$--qubit phase gates in one-way quantum computation
put forward by Browne and Briegel in Ref. \cite{Browne06}.

We remark that the auxiliary system $E$ may also be thought of
being an additional virtual qubit of one of the systems particles,
e.g. $A_3$ of system $A$. In this case, fewer two--system gates
are required, and the influence of noise is lower, yielding to
higher fidelities. However one of the systems needs to have twice
as many levels as the other systems, and the process becomes
non--symmetric. In the following, we will not consider this
situation, but rather the symmetric situation were the auxiliary
qubit is given by an additional, independent system particle, and
we use the second scheme involving fewer virtual qubits for
simplicity.


\subsubsection{Three--body interactions}
We start by considering three--body interactions,  $U(\delta
t')=e^{-\delta t'  \sigma_z^{\otimes 3}}$, i.e. $\alpha =\delta
t'$. We consider first the case of phase noise, and then the case
of white noise in two--system interactions.

{\em Phase noise:}

We consider a basic evolution coupling two virtual qubits of two
different systems that is governed by the Hamiltonian
$H=\sigma_z\otimes \sigma_z$, and where the noise part is
described by a Liouvillian that corresponds to single qubit
dephasing, i.e. is given by Eq. \ref{Liovphasenoise}. Noise part
and Hamiltonian part commute, and thus the solution of the
corresponding master equation is simple. A noisy phase gate can be
obtained using additional single--qubit $z$--rotations, and one
finds ${\cal E}_{PG} \rho= {\cal D}^{(1)}(p){\cal D}^{(2)}(p)
U_{PG} \rho U_{PG}^{\dagger}$, where $p=p(t)=\exp(-\gamma_0
\pi/4)$ and ${\cal D}$ is a single--qubit dephasing map specified
in Eq. \ref{dephasing}.

Generation of the state $|\tilde \psi_E\rangle$ using such noisy
gates leads to a mixed state \be \rho_E= {\cal D}^{(A_2)}(p){\cal
D}^{(B_2)}(p){\cal D}^{(C_2)}(p){\cal D}^{(E)}(p^3) |\tilde
\psi_E\rangle\langle \tilde \psi_E|. \ee Notice that $|\tilde
\psi_E\rangle =
U_{PG}^{(A_2E)}U_{PG}^{(B_2E)}U_{PG}^{(C_2E)}|++++\rangle$, and
the phase gates commute with the noise processes. Also the phase
gates $U_{PG}^{(A_1A_2)},U_{PG}^{(B_1B_2)},U_{PG}^{(C_1C_2)}$ used
to couple the state $|\tilde \psi_E\rangle$ to the storage qubits,
and the $\sigma_x$ measurements on storage qubits $A_1,B_1,C_1$,
commute with the noise. Noise acting on the ancilla system $E$
transforms after the measurement to {\em correlated} phase noise
acting on the remaining qubits $A_2,B_2,C_2$. That is, for an
arbitrary input state $\rho$, the output state after the sequences
of measurements and corresponding local correction operations is
given by \be \label{Teleportation_phase_map} {\cal E} \rho ={\cal
D}^{(A_2)}(p){\cal D}^{(B_2)}(p){\cal D}^{(C_2)}(p){\cal \tilde
D}^{(A_2B_2C_2)}(p^3) U(\delta t') \rho U(\delta
t')^\dagger,\nonumber \ee where \be {\cal \tilde D}_{ABC}(p)\rho =
p\rho + \frac{1-p}{2} (\rho +
\sigma_z^{(A)}\sigma_z^{(B)}\sigma_z^{(C)} \rho
\sigma_z^{(A)}\sigma_z^{(B)}\sigma_z^{(C)}), \nonumber \ee and
$U(\delta t')=e^{-\delta t'  \sigma_z^{\otimes 3}}$.

The fidelity of the noisy map ${\cal E}$ with respect to the ideal
process $\hat U(\alpha)$ (where $\hat U \rho = U \rho U^\dagger$),
$F({\cal E}, \hat U(\delta t'))$, can easily be calculated and one
obtains
\be
F({\cal E,} \hat U(\delta t'))&=& \left (\frac{1+p}{2} \right)^3 \frac{1+p^3}{2} + \left (\frac{1-p}{2} \right)^3 \frac{1-p^3}{2} \nonumber \\
 &=&\left ( \frac{1+p^2}{2} \right )^3 \nonumber \\
 &\approx& 1-  3\pi/4 \gamma_0 ,
\ee
where the last equality only holds for $\gamma_0 \pi/4 \ll 1$.
The local noise equivalent $\gamma$ can be determined under the
previous assumptions, and one finds \be \gamma \approx
\frac{\pi}{2 \delta t'} \gamma_0. \ee Note that the local noise
equivalent is of the same order of magnitude as in the case of
graph state encoding (see Eq. \ref{GSE_LNE_phase}), and the
fidelity is constant, independent of $\delta t'$.


{\em White noise:}

A similar analysis can be performed for white noise. In this case,
noise part and unitary part do not commute, and the influence of
noise on the ancilla qubit leads to a slightly different noise
process on the final system. We find \be {\cal E} \rho ={\cal
\tilde M}^{(A_2B_2C_2)}(p^3) {\cal M}^{(A_2)}(p){\cal
M}^{(B_2)}(p){\cal M}^{(C_2)}(p)  \rho,\nonumber \ee where \be
{\cal \tilde M}^{(ABC)}(p)\rho &=& p U(\delta t') \rho U(\delta
t')^\dagger \nonumber\\ &+& \frac{1-p}{2} (\rho +
\sigma_z^{(A)}\sigma_z^{(B)}\sigma_z^{(C)} \rho
\sigma_z^{(A)}\sigma_z^{(B)}\sigma_z^{(C)}), \nonumber \ee with
$U(\delta t')=e^{-\delta t'  \sigma_z^{\otimes 3}}$, and ${\cal
M}(p)$ is a depolarizing map specified in Eq. \ref{Depol} with
$p=\exp(-\kappa_0 \pi/4)$. It follows that the fidelity of the
process with respect to the ideal operation $\hat U(\delta t')$
can be estimated to be \be
F({\cal E,} \hat U(\delta t'))&=& \left (\frac{1+3p}{4} \right)^3 \frac{1+p^3}{2} + \left (\frac{1-p}{4} \right)^3 \frac{1-p^3}{2} \nonumber \\
 &\approx& 1-  \frac{15 \pi}{16} \kappa_0,
\ee and for the local noise equivalent $\kappa$ one finds for
$\delta t' \pi/4 \ll 1$ \be \kappa \approx \frac{5\pi}{12 \delta
t'} \kappa_0 \ee


\subsubsection{Many--body interactions}
It is straightforward to perform the analysis for $n$--qubit
operations $U(\delta t')=e^{-\delta t'  \sigma_z^{\otimes n}}$. In
this case, a $n+1$ qubit GHZ state is generated by noisy two--body
interactions, and processed by performing local operations and
measurements. In the case of phase noise, one finds a fidelity \be
F({\cal E}, \hat U(\delta t'))&=& \left (\frac{1+p}{2} \right)^n \frac{1+p^n}{2} + \left (\frac{1-p}{2} \right)^n \frac{1-p^n}{2} \nonumber \\
  &\approx& 1-  n\pi/4 \gamma_0 ,
\ee where the last equality only holds for $\gamma_0 \pi/4 \ll 1$.
For the local noise equivalent one finds $\gamma \approx
\frac{\pi}{2 \delta t'} \gamma_0$.

The corresponding expressions for white noise, where $p=\exp
(-\kappa_0 \pi/4)$, are \be
F({\cal E,} \hat U(\delta t'))&=& \left (\frac{1+3p}{4} \right)^n \frac{1+p^n}{2} + \left (\frac{1-p}{4} \right)^n \frac{1-p^n}{2} \nonumber \\
 &\approx& 1-  \frac{5n \pi}{16 } \kappa_0,
\ee and the local noise equivalent is given by $\kappa = \frac{5
\pi}{12 \delta t'} \kappa_0$.

\subsubsection{Timing errors}
\label{subsubsec:teleptimerrors}

Analyzing timing errors for this teleportation protocol is
particularly easy and has many similarities to the dephasing case.
Like the dephasing case, we will examine the situation where there
are no local errors and that all noise in the system comes from
the generation of the GHZ state $|\tilde{\psi}_E\rangle$, which is
generated via phase gates. As we saw above, these phase gates can
be generated by a two qubit Ising interaction Hamiltonian,
$H=\sz\otimes \sz$. Here we examine the effect of timing errors on
this entangling Hamiltonian. The action of our noisy phase gate on
an arbitrary state is defined as,
\be
\mathcal{E}_{PG} \rho = \mathcal{T}U_{PG}\rho U_{PG}^\dagger,
\ee
where $\mathcal{T}$ is defined in Eq.
\ref{eq:2qubitisingtimeerror}. The map $\mathcal{T}$ is a
correlated two-qubit dephasing channel. We note that $\mathcal{T}$
will commute with the coherent processes that generate
$|\tilde{\psi}\rangle$, this sibilantly simplifies our analysis.
For clarity, we will first consider the generation of three qubit
interaction. If we attempt to generate $|\tilde{\psi}_E\rangle$
with these gates the resulting state will be,
\be
\rho_E =
\mathcal{T}^{(A_2E)}\mathcal{T}^{(B_2E)}\mathcal{T}^{(C_2E)}|\tilde{\psi}_E\rangle\langle\tilde{\psi}_E|.
\ee
Now, what is the effect of attempting our teleportation-based
Hamiltonian simulation protocol using $\rho_E$ as a resource? It
is easy to see that noise process commutes with the $\sigma_x$
measurements on the storage qubits $A_1, B_1,$ and $C_1$. The only
non-trivial transformation is as a result of the projective
measurement on qubit $E$.   We saw above that, the effect of
measuring a $E$ in the basis that we have chosen causes a
dephasing channel acting on $E$ to transform into a three-qubit
dephasing channel acting on qubits $A_2, B_2,$ and $C_2$. If we
instead had a correlated phase error between $E$ and $A_2$ before
applying the necessary measurement on $E$ we will have applied a
phase flip on qubit $A_2$ followed by the same three-qubit
correlated error as above. The effect of this operation on our
protocol is to create a two qubit correlated dephasing operation
on qubits $B_2$ and $C_2$. That is, the timing noise map
$\mathcal{T}^{(A_2E)}$ gets transformed under this teleportation
protocol into $\mathcal{T}^{(B_2C_2)}$. Considering this, the
operation induced by the teleportation protocol on an arbitrary
input $\rho$ is given by,
\be
\mathcal{E} \rho =
\mathcal{T}^{(A_2B_2)}\mathcal{T}^{(A_2C_2)}\mathcal{T}^{(B_2C_2)}\hat{U}(\delta
t')\rho.
\ee
Because the noise process
$\mathcal{T}^{(A_2B_2)}\mathcal{T}^{(A_2C_2)}\mathcal{T}^{(B_2C_2)}$
commutes with $\hat{U}(\delta t')$,
\be
D(\mathcal{E},\hat{U}(\delta
t'))=D(\mathcal{T}^{(A_2B_2)}\mathcal{T}^{(A_2C_2)}\mathcal{T}^{(B_2C_2)},
\eins),
\ee
and
\be
F(\mathcal{E},\hat{U}(\delta
t'))=F(\mathcal{T}^{(A_2B_2)}\mathcal{T}^{(A_2C_2)}\mathcal{T}^{(B_2C_2)})
\ee
Thus If we now make the assumption that the timing error is
gaussian distributed, as we did in Section \ref{Noisemodel2} we
can calculate the resulting fidelity to be
\be
F(\mathcal{E},\hat{U}(\delta t')) = \left(\frac{1+q}{2}\right)^3
+\left(\frac{1-q}{2}\right)^3.
\ee

Generalizing this analysis for the generation of the $n$-qubit
operations like $U(\delta t')=e^{-i'\delta t' \sz^{\otimes n}}$ is
straightforward. As a resource we consider an $n+1$ qubit GHZ
state that is generated by phase gates as in the previous section.
Given an arbitrary input $\rho$ the effect of performing the
teleportation protocol ideally is to output the state
$\hat{U}(\delta t')\rho$ at qubits $A_2B_2...N_2$. Performing the
same analysis as above and noting that of measuring $E$ is to
transform operations like $\mathcal{T}_{A_2E}$ into operations
like $\mathcal{T}_{B_2C_2... N_2}$ we find that the effect of
timing errors is to induce the following map
\be
\mathcal{E} \rho =\mathcal{T}^{(A_2B_2...
N_2)}\mathcal{T}^{(A_2C_2... N_2)}\mathcal{T}^{(B_2C_2...
N_2)}...\hat{U}{\delta t'}\rho.
\ee
The resulting fidelity is,
\be
F(\mathcal{E},\hat{U}(\delta t'))=\left(\frac{1+q}{2}\right)^n
+\left(\frac{1-q}{2}\right)^n.
\ee

\subsection{Entanglement purification and the influence of local noise}

In a similar way, one can also take the influence of noise in
local (i.e. single system) operations into account. In this case,
the coupling of system particles to the auxiliary GHZ state is
also noisy, leading to a further reduction in the fidelity of the
final process. The corresponding noise processes are described by
similar master equations as we use for modelling of two--system
interactions, however the coupling parameters $\kappa_l$ or
$\gamma_l$ describing the noise level are different. As indicated
in the discussion at the beginning of this section, we will assume
that single system noise is {\em smaller} than noise in two system
operations, i.e. $\kappa_l \ll \kappa_0$ and $\gamma_l \ll
\gamma_0$.

In the case of phase noise, we have shown that the final state is
given by Eq. \ref{Teleportation_phase_map} when assuming that
local single system operations are perfect, while the influence of
noisy single system operations is essentially covered by
additional noise maps maps ${\cal D}^{(A_2)}(p_l^2)$,${\cal
D}^{(B_2)}(p_l^2)$,${\cal D}^{(C_2)}(p_l^2)$ with
$p_l=\exp(-\gamma_l \pi/4)$. Under the assumption that $\gamma_l
\ll \gamma_0$, the final fidelity and local noise equivalent are
essentially determined by $\gamma_0$, with small corrections due
to additional single system noise. A similar situation is
encountered in the case of depolarizing noise.

We will not further go into details about the influence of local
noise in the standard protocol, but rather consider the case where
multipartite {\em entanglement purification} \cite{Aschauer05a,Aschauer05a} is used to produce high--fidelity
resource states $|\tilde \psi_E\rangle$ and hence to generate
many--body interactions with high fidelity. That is, we assume
that systems with additional auxiliary levels ---which are used for
storage and purification of entangled resource states--- are
available. In this case, noise in local operations determines the
achievable fidelity of the resource states, and hence the final
fidelity of the many--body interaction. In particular, the
resulting fidelity is {\em independent} of amount of noise in
two--system operations and hence independent of $\kappa_0,
\gamma_0$, provided the noise is sufficiently small such that
entanglement purification can be successfully applied. For perfect
single system operations, the tolerable amount of noise in the two
system operations is up to 66 \% \cite{Dur03} (or slightly smaller
when using direct multi-particle entanglement purification) when
considering depolarizing noise. Non--zero amount of noise in
single system operations leads to a slightly smaller error
tolerance, however as long as $p_l = \exp(- \kappa_l \pi/4)$ is
sufficiently close to one (errors at the order of percent, i.e.
$p_l \approx 0.99$) we find that noise in two--system operations
of the order of several tens of percent can still be tolerated,
without influencing the final achievable fidelity of the purified
state and the resulting many--body interaction.

An exact analytic treatment of the total process is difficult due
to difficulties in analytically describing multiparticle
entanglement purification protocols and the reachable fidelities
of such schemes. We have thus performed a numerical simulation
taking all errors due to imperfect operations into account. The
results are presented in the following.

\subsubsection{Three--body interactions}

\begin{figure}[ht]
\begin{picture}(230,200)
\put(0,0){\epsfxsize=230pt\epsffile[87   262   507
578]{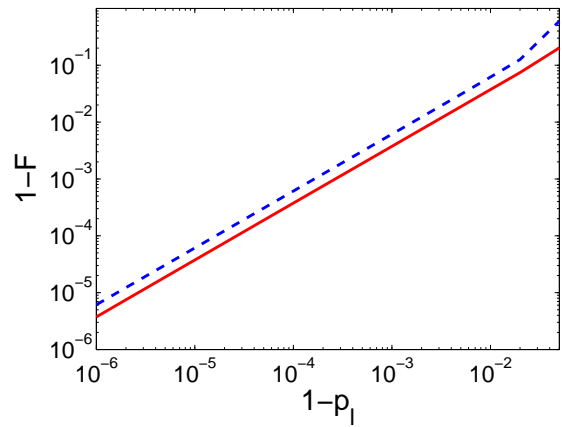}}
\end{picture}
\caption[]{Fidelity of three--body interaction $U=\exp(-\delta t'
\sigma_z^{\otimes 3})$ for time $\delta t'=\pi/2^{15} \approx
10^{-4}$ (red, solid) and $\delta t'=\pi/4$ (blue, dashed) using
teleportation based method. One minus fidelity is plotted as a
function of noise in local operation $1-p_l$, where $p_l =
\exp(-\kappa_l \pi)$ and $1-p_l = 0$ corresponds to perfect local
operations leading to unit fidelity. }
\label{Fidelity_Purification}
\end{figure}

For three--body interactions, the required resource state is a GHZ
state of four qubits, which can be purified using the entanglement
purification protocol introduced in Ref. \cite{Aschauer05a,Aschauer05}. The
fidelity of the resulting three--body interaction $\exp(-\delta t'
\sigma_z^{\otimes 3})$ is plotted in Fig.
\ref{Fidelity_Purification} for $\delta t' \ll 1$ ($\delta
t'=\pi/2^{15}$ to be precise) and for a three--qubit gate with
$\delta t'=\pi/4$. As can be seen from the plot, the resulting
fidelity is almost identical for the two cases, i.e. to a large
extend independent of $\delta t'$. The plot shows the dependence
of (one minus) the fidelity on the noise in local operation
$1-p_l$, where $p_l = \exp(-\kappa_l \pi)$ here. Notice the linear
scaling in the log-log plot.

We remark the local noise equivalent has no useful meaning in this
context, as the resulting fidelity of the many--body interaction
is independent of noise in two--system operations (the reachable
fidelity of entanglement purification only depends on noise in
local single system operations). More precisely, for any noise parameter
$\kappa_0$ (describing coupling strength to environment for
two--system operations) that is sufficiently small such that
entanglement can be generated by the resulting two--system
operation, we obtain the {\em same} local noise equivalent
$\kappa$ which only depends on $\kappa_l$, the noise parameter for
local single system operations. Notice that $\kappa$ can be
significantly {\em smaller} than the initial noise parameter
$\kappa_0$. For perfect local control operations and sufficiently
many auxiliary levels, we can even obtain $\kappa=0$.

\subsubsection{Many--body interactions}

\begin{figure}[ht]
\begin{picture}(230,200)
\put(0,0){\epsfxsize=230pt\epsffile[87   262   507
578]{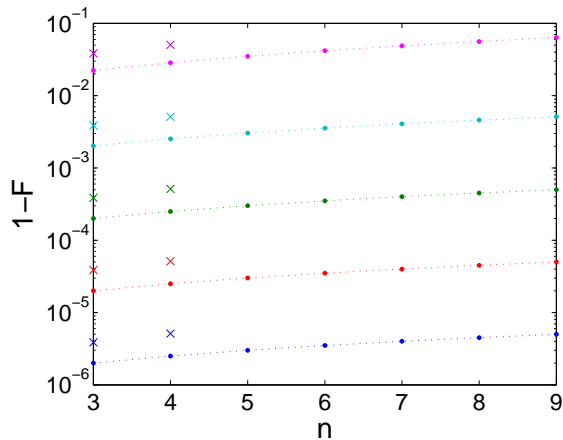}}
\end{picture}
\caption[]{Reachable fidelity to generate $n+1$ qubit GHZ states
using multiparticle entanglement purification (dots), and to
generate $n$--body interactions of the form $\exp(-\delta t'
\sigma_z^{\otimes n})$ from the GHZ states ($\times$) following
the protocol outlined in main text. One minus fidelity is plotted
against $n$. Data points/curves from top to bottom correspond to
noise parameter for local operation $(1-p_l) =
10^{-2},10^{-3},10^{-4},10^{-5},10^{-6}$ respectively, where $p_l
= \exp(-\kappa_l \pi)$. }
\label{Fidelity_Purification_multipartite}
\end{figure}

For $n$--body interactions, the required resource state is a GHZ
state of $n$ qubits. Fig. \ref{Fidelity_Purification_multipartite}
shows the reachable fidelity of the $n+1$--qubit GHZ states for
different values of noise in local control operation $1-p_l$. For
$n=3,4$ also the fidelity of resulting $n$--qubit interaction
$\exp(-\delta t' \sigma_z^{\otimes n})$ for $\delta t'=\pi/2^{15}$
is shown, which is of the same order of magnitude as the fidelity
of the purified GHZ state.

\section{Comparison of methods}\label{compare}

In this section, we will recall and compare the main properties of
the different methods to generate many--body interactions. We
consider many--body interactions $\exp(-i\delta t'
\sigma_z^{\otimes n})$ generated by the commutator method, graph
state encoding or teleportation based method with entanglement
purification.

In the commutator method, the fidelity increases for smaller
$\delta t'$, even though the local noise equivalent increases.
When implementing a gate (i.e. a certain interaction for time $t =
O(1)$) by sequentially applying interactions $\exp(-i\delta t'
\sigma_z^{\otimes n})$ generated this way, there is an optimal
$\delta t'$. The fidelity of such a gate is significantly lower
than the fidelity of a many--body interaction applied for short
time $\delta t' \ll 1$.

For graph state encoding, in contrast, the fidelity is constant
and independent of $\delta t'$. The fidelity is determined by
applications of noisy two--body interactions for times $t=O(1)$
(the graph state encoding), leading to a significantly lower
fidelity for small $\delta t'$ as compared to the commutator
method. The local noise equivalent also increases with decreasing
$\delta t'$, faster as for the commutator method. However, also
gates or evolutions for times $t \gg 1$ can be simulated with same
accuracy, leading to an advantage of graph state encoding method
over commutator method in such situations.

The teleportation based method making use of auxiliary local
degrees of freedom also leads to a constant fidelity, almost
independent of $\delta t'$, and hence also gates with $t=O(1)$ can
be implemented with same fidelity. However, for the teleportation
based method using entanglement purification, the fidelity is
determined by the noise in local single system operations rather
than noise in two--system interactions, possibly leading to much
better accuracies than reachable with graph state encoding or the
commutator method. Only for $\delta t' \ll 1$ the commutator
method may be more accurate, and for $\delta t' \gg 1$ (fixed)
graph state encoding may be favorable. The advantage of the
teleportation based method seems to increase when considering the
implementation of $n$--body interactions with increasing $n$. It
follows that the teleportation based method, together with
entanglement purification, provides a possibility to successfully
simulate interacting high--dimensional quantum system or systems
with many--body interactions with high accuracy. This can still be
seen as an intermediate stage to full scale, universal fault
tolerant quantum computation, and should be significantly easier
to implement than the latter.

\section{Summary and conclusions}\label{conclusions}

We have introduced and investigated several methods to generate
many--body interactions from two--body interactions. We reviewed
the standard commutator method which is based on usage of higher
order terms in the Trotter--Suzuki expansion. We have shown how to use
unitary conjugation or graph state encoding to generate large
classes of many--body interaction Hamiltonians when using a fixed
encoding, and arbitrary Hamiltonians when allowing for a variable
encoding. Several examples, including three--body Hamiltonians
with phase transition, simulation of interacting $d$--level
systems and generation of plaquette interactions have been put
forward. Furthermore, we have investigated the usage of maximally
entangled states produced by two--body interactions to generate
many--body interactions by means of teleportation, i.e. using
teleportation based methods.

We have studied the influence of noise, described by generic
error models, on the simulation process for the three schemes
discussed above. For long--time simulations, the fixed graph state
encoding method is favorable. 

However, if one has access to auxiliary systems (e.g. higher dimensional systems), then one may use them to reduce noise. In particular, one can use higher dimensional systems to simulate lower dimensional systems, and use auxiliary levels to perform entanglement purification.  
This makes the teleportation based method a
possible solution to overcome, or at least significantly reduce,
the influence of noise on a simulation process. While this
method makes use, to a certain extent, of elements of measurement
based quantum computation, it still does not make use of the whole
blown up machinery of a full fault tolerant quantum computer. We
believe that our considerations are useful with respect to the
ongoing effort to design and eventually use quantum simulators
based on well controllable, but nevertheless noisy quantum systems.

\section*{Acknowledgements}
M.J.B. would like to thank Otfried G\"uhne, Gavin Brennen, Dan
Browne, Ken Brown, Chris Dawson, and Henry Haselgrove for helpful
discussions.

This work was supported by the Austrian Science Foundation (FWF),
the European Union (OLAQUI,SCALA,QICS) and the Austrian Academy of
Sciences (\"OAW) through project APART (W.D.).

%




\end{document}